\RequirePackage{rotating}
\documentclass[useAMS,usenatbib]{mnras}
\pdfoutput=1
\usepackage{psfig}
\usepackage{amsmath}
\usepackage{amssymb}
\usepackage{upgreek}
\usepackage{longtable}
\usepackage[mathscr]{eucal}
\usepackage{appendix}
\usepackage{graphicx}
\usepackage{float}
\usepackage{afterpage}
\usepackage{subcaption}
\usepackage[export]{adjustbox}
\usepackage[usenames, dvipsnames]{color}
\usepackage{titlesec}
\usepackage{csvsimple}
\titlespacing*{\section}{0pt}{1.1\baselineskip}{\baselineskip}
\titlespacing*{\subsection}{0pt}{1.1\baselineskip}{\baselineskip}

\title[The 6D kinematics of Upper Scorpius]{Investigating the Upper Scorpius OB association with HERMES. I. The spectroscopic sample and 6D kinematics}
\author
[Armstrong et al.]
{Joseph J. Armstrong$^{1}$, Jonathan C. Tan$^{1}$, Nicholas J. Wright$^{2}$, R.D. Jeffries$^{2}$, Janez Kos$^{3}$, \newauthor 
E. Fiorellino$^{4}$, Sven Buder$^{5, 6, 7}$, D. Barrios López$^{8}$ 
\\
$^{1}$Department of Space, Earth \& Environment, Chalmers University of Technology, SE-412 96 Gothenburg, Sweden \\
$^{2}$Astrophysics Group, Keele University, Keele, ST5 5BG, UK \\ 
$^{3}$Faculty of mathematics and physics, University of Ljubljana, Jadranska 19, 1000 Ljubljana, Slovenia \\
$^{4}$INAF – Osservatorio Astronomico di Trieste, via Tiepolo 11, I-34143 Trieste\\
$^{5}$Research School of Astronomy and Astrophysics, Australian National University, Canberra, ACT 2611, Australia \\
$^{6}$ARC Centre of Excellence for All Sky Astrophysics in 3 Dimensions (ASTRO 3D), Australia \\
$^{7}$ACCESS-NRI, Australian National University, Canberra, ACT2601, Australia \\
$^{8}$Instituto de Estudios Astrofísicos, Facultad de Ingeniería y Ciencias, Universidad Diego Portales, Av. Ejército Libertador 441, \\
Santiago, Chile\\
}

\begin{document}
\maketitle

\begin{abstract}
OB associations are large, unbound groups of young stars, which typically exhibit great complexity in spatial, kinematic and age structure, hinting at 
formation scenarios involving an intricate interplay of molecular cloud turbulent kinematics and stellar feedback over extended periods. The kinematic properties of the numerous low-mass populations within OB associations can provide valuable constraints on their initial configurations, and thus the dominant mechanisms driving star formation and their dispersal into the field. 
We present results from a large spectroscopic survey of the nearest young association to the Sun, Upper Scorpius, conducted using 2dF/HERMES on the Anglo-Australian Telescope. We use spectroscopic youth criteria such as Li-equivalent widths to identify $>$1000 pre-main sequence (PMS) members across the region and measure radial velocities, combining these with Gaia EDR3 5-parameter astrometry to obtain 6D kinematic information. We separate confirmed PMS association members into distinct kinematic groups and measure expansion and rotation trends in each. We also trace the past motion of these groups using an epicycle approximation and estimate the time since their most compact configuration. These kinematic properties are compared to literature ages and the star formation history of Upper Scorpius is discussed. We find evidence that a scenario in which star formation in the subgroups of Upper Scorpius proceeded independently, either by self-instability or external feedback from Upper Centaurus-Lupus, is more likely than a recently proposed ``cluster chain'' scenario in which these subgroups have triggered each other.
\end{abstract}
	
	
\section{Introduction}
Most stars form in dense, gravitationally bound clusters or sparse substructured associations \citep[e.g.,][]{lada03,wright23}. Apart from the brightest O \& B type stars, by which OB associations were first identified \citep{ambartsumian47,blaauw64,dezeeuw99}, the sparse distributions of low-mass young stellar objects (YSOs) that accompany them have historically been harder to detect. In addition, they tend to be unbound and thus disperse into the Galactic field over timescales of a few tens of Myrs \citep{wright20}.

Recent investigations, particularly using high precision astrometry from \textit{Gaia} \citep{gaia16,Gaiaedr3}, have been able to identify the sparse and highly substructured distributions of nearby associations \citep[e.g.,][]{wright16,wright18,cantatgaudin19a,zari19,Armstrong22}. Associations have also been revealed to contain significant age gradients \citep{kounkel18,Kos19,cantatgaudin19,damiani19,ratzenbock23}, which, along with significant kinematic structure \citep{wright16,wright18,Armstrong22,miret-roig22,kerr25}, indicates complex star formation histories involving the impact of feedback processes and turbulence within molecular clouds to produce initially unbound distributions of young stars in large volumes. This is in contrast to the monolithic cluster formation model where stars form in initially compact configurations, but disperse after residual gas expulsion \citep{tutukov78,hills80,kroupa01b}, which would imply that OB associations are an intermediate stage of cluster dissolution.

However, it has yet to be determined what are the dominant mechanisms driving star formation in these complexes. It has been suggested that the age gradients found in OB associations point toward multiple epochs of star formation, with the initiation of star formation in one epoch being triggered by the impact of feedback from massive stars formed in the previous epoch \citep[e.g.,][]{elmegreen77,posch24,kerr25}. There is also the suggestion that star formation could be initiated by collisions between giant molecular clouds (GMCs) \citep[e.g.,][]{1986ApJ...310L..77S}, with a model in which collisions are driven by galactic shearing motions of orbiting GMCs \citep{tan00,2009ApJ...700..358T,2018PASJ...70S..56L} being potentially relevant to explaining observed large-scale star formation properties of disk galaxies \citep[][]{2010ApJ...710L..88T,2014ApJ...787...68S,2024ApJ...977L...6F}. Simulations of star formation resulting from individual GMC-GMC collisions, with relative cloud velocities of $\sim 10\:{\rm km\:s}^{-1}$ can produce spatial structure and disturbed kinematics among the newly formed stars similar to those observed in nearby regions \citep[e.g.,][]{wu17,2020ApJ...891..168W}. Stellar feedback driven cloud collisions have also been proposed as the main driver of new star formation in galaxies \citep{2015A&A...580A..49I}. Other possibilities include models of star formation regulated by turbulence in GMCs \citep[e.g.,][]{2005ApJ...630..250K,2011ApJ...730...40P,2011ApJ...743L..29H}. Combinations of the above models are also possible, e.g., if shear-driven GMC collisions are the main driver of turbulence in GMCs \citep[][]{2013IAUS..292...19T,2018PASJ...70S..57W}.

\begin{figure} 
    \includegraphics[width=\columnwidth]{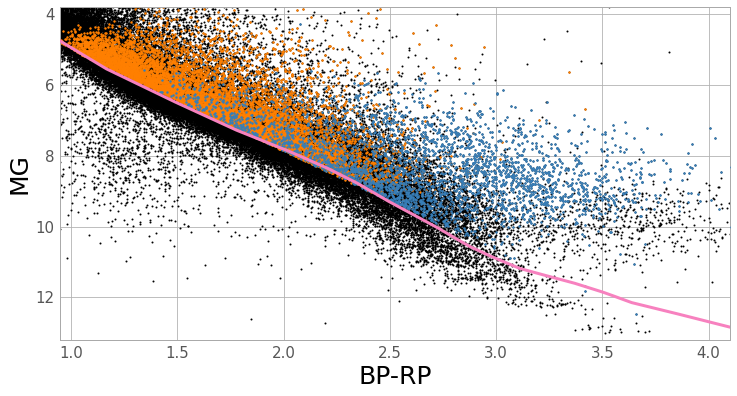}
    \setlength{\belowcaptionskip}{-10pt}
    \setlength{\textfloatsep}{0pt}
    \caption{Gaia DR3 BP-RP colour vs. absolute G magnitude for sources in the Upper Sco region with $13<Gmag<16.5$ and $\varpi<5$ (blue) or $\varpi<2.5$ (orange). Absolute magnitudes are calculated using distance estimates from \citet{bailerjones21} and average $A_G$ and $E$(BP-RP) values from Gaia DR2 per $0.25\deg^2$ square by 50 pc los-distance volume, similar to the method employed by \citet{zari19}. Overplotted is the 20 Myr PMS isochrone (pink) from \citet{baraffe15}.}
    \label{CMD}%
\end{figure}

\begin{figure*} 
    \includegraphics[width=\textwidth]{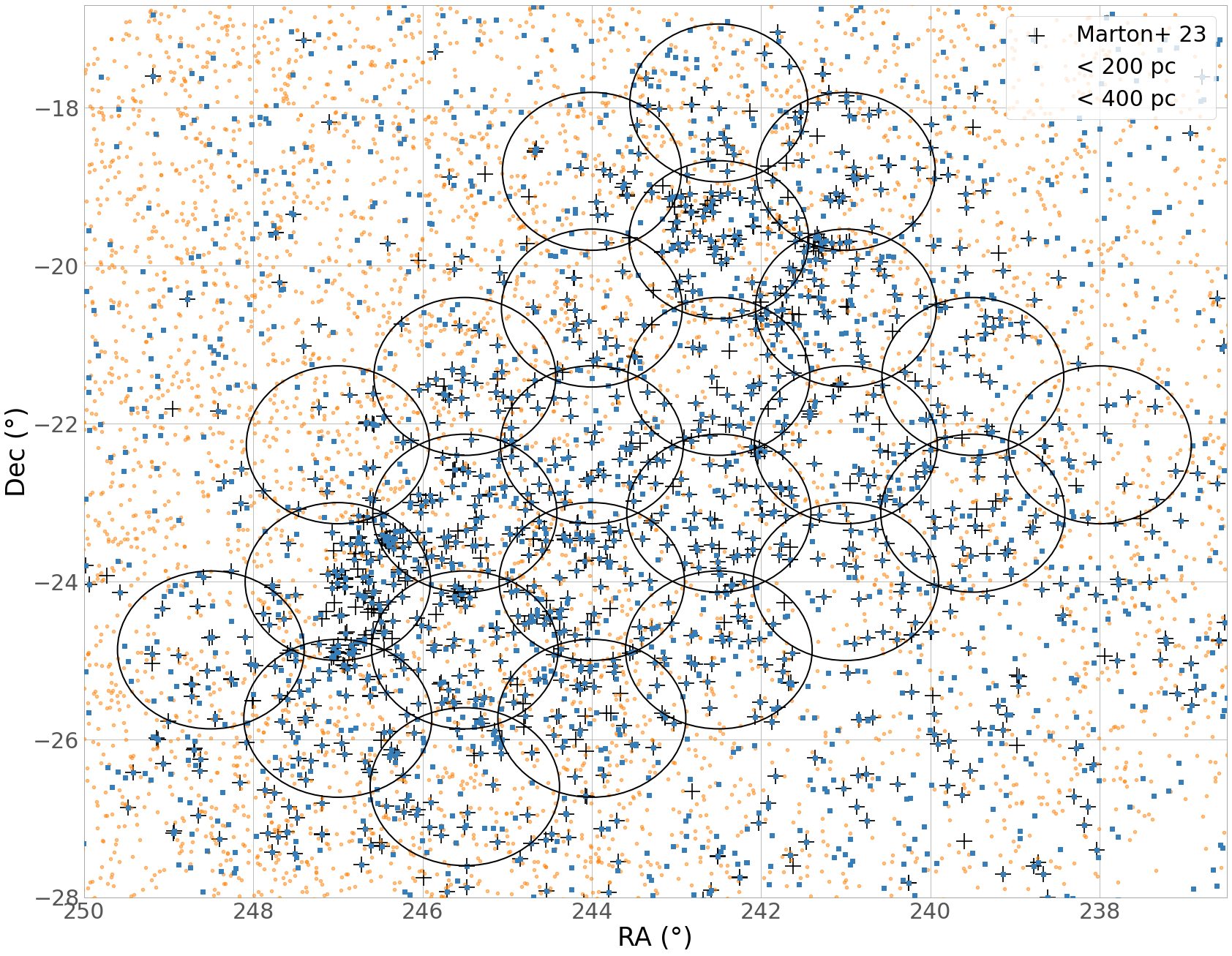}
    \setlength{\belowcaptionskip}{-10pt}
    \setlength{\textfloatsep}{0pt}
    \caption{Sky positions of photometrically selected targets colour-coded by their priority (Fig.~\ref{CMD}). Sources identified as YSOs by variability in Gaia DR3 \citep{marton23} are indicated with plus symbols. The 2$^\circ$-diameter circles indicate the individual fields of observation of the AAT within which targets were selected. The fields were distributed to effectively cover the region of Upper Scorpius with the greatest density of likely YSOs. }
    \label{FieldPlot}%
\end{figure*}

Upper Scorpius is the youngest \citep[$\approx$ 11 Myr;][]{pecaut16} subgroup within Scorpius-Centaurus, the nearest OB association to the Sun \citep[100-150 pc;][]{dezeeuw99}, containing $>100$ OB stars spread over $\approx$ 200 square degrees. Its proximity, youth and low extinction make it a prime target for studies of the low-mass population of OB associations, even to the brown-dwarf regime \citep{wright18,damiani19,luhman20}. The low extinction for association members indicates that the association has dispersed from its primordial molecular cloud. Recent kinematic studies show that Upper Sco is highly kinematically substructured and almost certainly unbound \citep{wright18,squicciarini21}. The full extent of the low-mass population is currently unknown, but thousands of low-mass counterparts to the OB population should exist for a normal initial mass function \citep[IMF; e.g.,][]{kroupa01,chabrier03}. The velocity dispersion \citep[$\approx$ 3.2 km/s;][]{wright18} and age \citep[$\approx$ 11 Myr;][]{pecaut16} of the known PMS and OB stars, and total spatial extent \citep[$\approx$ 40 pc;][]{wright18} of Upper Sco seem incompatible with the simple expansion of a monolithic cluster after gas expulsion, and in fact \citet{wright18} could not find evidence that the association had expanded from a single compact cluster. Instead, a hypothesis is that Upper Sco formed hierarchically at various sites, perhaps in a sequential way, so that there is a widespread population of unbound PMS stars dispersed across the association, while the massive stars may coincide with bound subclusters with unique kinematics \citep{damiani19}.

Recently the clustering and kinematics of Upper Scorpius have been investigated using high-precision astrometry from Gaia \citep{Gaiaedr3} with the finding that Upper Scorpius can indeed be divided into multiple subgroups \citep{squicciarini21,miret-roig22,ratzenbock23}, which differ in their kinematics and ages, with older subgroups being centrally located within the Upper Scorpius region and younger subgroups located at the outskirts. This has led to the proposed ``cluster chain'' formation scenario \citep{posch24} which hypothesises that the formation of the younger subgroups was triggered by feedback from the older subgroups in sequence. This would produce a significant age gradient across the region, as well as the large-scale ``acceleration'' observed between these subgroups. 

However, these previous kinematic studies have gathered the majority of their radial velocity information from public surveys, such as APOGEE \citep{ahumada20} and GALAH \citep{Buder21}, which vary in radial velocity precision and in coverage across the region, meaning that the internal 3D kinematics of much of the low-mass population has not yet been fully analysed. Also, in these studies, membership for the bulk of the low-mass population of Upper Sco has so far been determined using astro-photometric or kinematic criteria, which risks biasing sample selection and analysis away from sources and structures with kinematics distinct from the bulk of the region. Ideally, membership determination and target selection for further observations should be kinematically unbiased in order to investigate the full complexity of the region. In particular, spectroscopic observations are necessary to provide both the missing radial velocities and firm youth criteria, such as Li-equivalent widths, enabling us to perform membership determination for young stars in the region independent of the kinematics we wish to analyse.

In this study we present results from a large-scale spectroscopic survey across the Upper Scorpius region with the goal of identifying young stars and studying their kinematics.    Section \ref{section_data} outlines the data used and the spectroscopic observations performed. Section \ref{section_overview} provides an overview of the region and dissects the young stellar population into distinct kinematic groups, with Section \ref{section_kinematics} going on to analyse their internal dynamics, focusing on expansion trends, kinematic anisotropies and kinematic age estimates via dynamical traceback. In particular, we look for evidence that the subgroups in the region are unbound and were likely in more compact configurations in the past, with the goal of assessing whether they could have formed as a sequence of dense clusters, or whether it is more likely that they formed in sparse distributions. In Section \ref{section_discussion} we discuss our results in light of recent literature and infer the likely star formation scenario of the Upper Scorpius association.
 
\section{The data}
\label{section_data}

In this section we describe the target selection for spectroscopic observations with 2dF/AAT, and derivation of spectroscopic parameters, which we combined with astrometric and photometric data from Gaia to produce our sample of young stars with complete kinematic information.

\subsection{AAT observations}

We selected candidate PMS stars in the Upper Sco region for spectroscopic follow-up using a Gaia DR3 G$_{BP}$-G$_{RP}$ versus M$_{G}$ colour-absolute magnitude diagram (using distance estimates from \citealt{bailerjones21}) and selecting sources located above a 20 Myr PMS isochrones \citep[see Fig.~\ref{CMD}]{baraffe15}, with $13 < G < 16.5$ and $\varpi > 2.5$ mas ($<$400 pc) to filter out background contaminants.

We selected targets in 25 fields (Fig.~\ref{FieldPlot}; listed in Table~\ref{field_table}), which cover the densest groups of likely young stars in the region, as indicated by our photometric selection (Fig.~\ref{CMD}) and by Gaia DR3 variability criteria \citep{marton23}, and effectively sample the distinct populations present in the region \citep{luhman20,miret-roig22,ratzenbock23}.

Observations were made on April 16th, May 21st - 22nd \& June 10th - 12th 2023 with the 2-degree field \citep[2dF]{2dF} fibre positioner and the high-efficiency and resolution multi-element spectrograph \citep[HERMES]{HERMES}, which provides a resolution of $R \sim$ 28,000 in four optical bands, at the Anglo-Australian Telescope (AAT) as part of program 2023A/30. The red band of HERMES covers the wavelength range 6478–6737 \AA\ within which both the H$\alpha$ 6562.8 \AA\ and Li 6707.8 \AA\ lines are found. For each field we selected $>300$ targets based on colour-absolute magnitude diagram (CaMD) position, which totaled 8061 targets for all fields with overlap, 7039 of which were unique sources. Multiple 1800s exposure were taken for each field, the total time for which, along with numbers of targets and numbers of spectroscopically confirmed PMS stars (Section~\ref{YSOs}), are given in Table~\ref{field_table}. Calibration frames, including dark frames and multifibre flat fields were taken for each field and 25-40 fibres were used to measure the sky spectrum for each field.

\begin{figure}
\begin{center}
\includegraphics[width=\columnwidth]{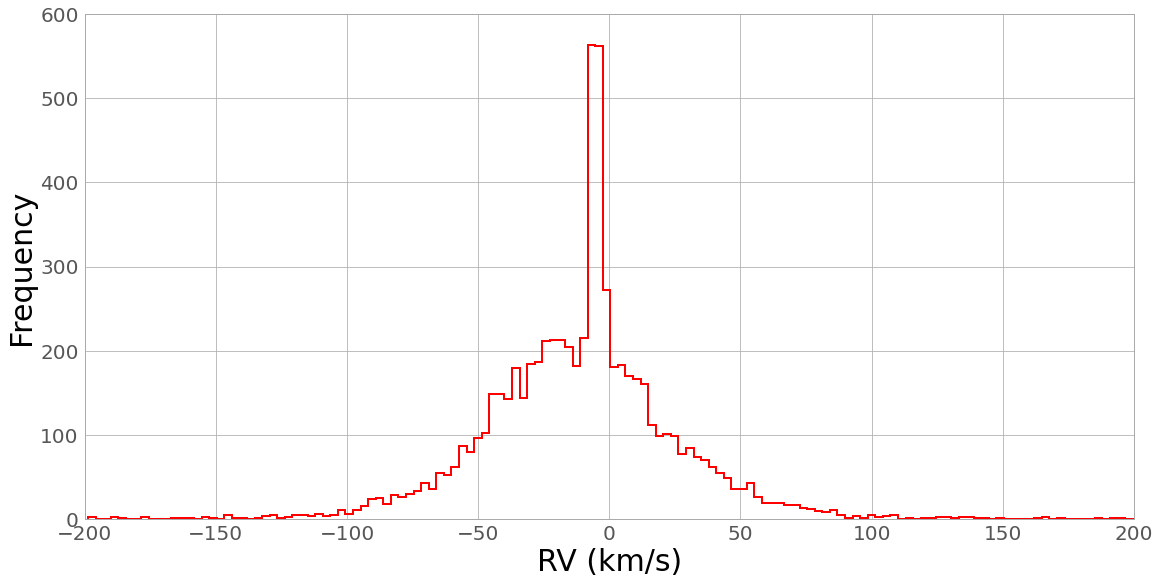}
\caption{Histogram of calibrated radial velocities (RVs) for 6727 sources in our sample of HERMES targets. }.
\label{RVs}
\end{center}
\end{figure}

\subsubsection{Data reduction and analysis}

The spectroscopic data were calibrated and reduced using the 2dF Data Reduction (2DFDR) software tool \citep{2dfdr}. Measurement of spectroscopic parameters (equivalent widths of the Li 6707.8 \AA\ line; EW(Li)s) from reduced spectra was done using the same approach as described in \citet{Armstrong22}, following the procedures of \citet{jeffries21}. We produced synthetic spectra using the MOOG spectral synthesis code \citep{Sneden2012a}, with \citet{Kurucz1992a} solar-metallicity model atmospheres, for log$g$ = 4.5 and down to $T_{\rm eff}$ = 3500 K in 100k steps, with rotational broadening and instrument resolution accounted for in the extraction profile.

\subsubsection{Radial Velocities}
\label{radial velocities}

Radial velocities are the last component of complete 6D position-velocity information missing from Gaia for most sources. With high-resolution spectra from HERMES we can measure radial velocities up to a typical precision of $0.1\:{\rm km\:s}^{-1}$ \citep{zwitter18}, which is sufficient to investigate the internal kinematics of young clusters and star-forming regions.


The spectra we obtained in these observations have been included in the GALAH survey data release 4 \citep{galahdr4}. Among the many parameters derived are RVs, which we incorporate into our data sample for kinematic analysis. However, not all candidate YSO members of Upper Sco we observed necessarily have RVs in GALAH DR4, due to the various quality controls imposed in the compilation of the catalogue. Therefore, we also derive RVs and uncertainties from our own reduced spectra using the same approach as described in \citet{Armstrong22}.

Reduced target spectra were cross-correlated with their best-fitting synthetic spectra and RVs were determined from the position of the peak in the cross-correlation function (CCF) by fitting a Gaussian function. Targets for which a Gaussian function cannot be satisfactorily fitted to the CCF are not given valid RVs and are rejected from our sample.

RV uncertainties were also determined using the same approach as described in \citet{Armstrong22}, by measuring the change in RV between $n$ separate exposures of the same target ($E_{RV} = \Delta RV/\sqrt{n}$), normalised per field using the scaling function \citep[equation 1;][]{Armstrong22}. 

We also cross-match our sources with the \textit{Survey of surveys} catalogue \citep{tsantaki22}, which contains calibrated RVs for over 11,000,000 Gaia sources, compiled from multiple spectroscopic surveys. We find RVs for 874 of our targets; 605 with RVs from APOGEE \citep{apogee}, 38 with RVs from the Gaia-ESO survey \citep{gilmore12} and 295 with RVs from Gaia, with overlap.

We find 4006 of our targets with RVs $< |200|$ km s$^{-1}$ and SNR $>10$ from both the GALAH DR4 catalogue and our analysis, with a median offset of 0.084  km s$^{-1}$. We also find 386 targets with RVs $< |200|$ km s$^{-1}$ from both GALAH DR4 (SNR $>10$) and the \textit{Survey of surveys} catalogue \citep{tsantaki22}, with a median offset of 0.459 km s$^{-1}$. We calibrate the compiled RVs by applying these offsets to RVs from the relevant source catalogues. For the targets with multiple RVs we take the weighted mean RVs, weighted by the square of the inverse measurement uncertainties in each case, with the aforementioned RV offsets applied.

In total we have RVs available for 6727 of our 7039 unique targets, the distribution of which is plotted in Fig.~\ref{RVs}.







\begin{table*}
\begin{center}
{\renewcommand{\arraystretch}{1.5}
\begin{tabular}{|p{2cm}|p{1.5cm}|p{1.5cm}|p{2.5cm}|p{1.8cm}|p{3cm}| }
\hline
Field number & RA ($^\circ$) & Dec ($^\circ$) & Exposure time (s) & Targets & Confirmed YSOs \\
\hline
1 & 244.0 & -24.0 & 5400 & 321 & 88 \\
2 & 244.0 & -22.268 & 5400 & 328 & 59 \\
3 & 245.5 & -23.134 & 6210 & 325 & 107 \\
4 & 245.5 & -24.866 & 9000 & 321 & 108 \\
5 & 244.0 & -25.732 & 9000 & 314 & 57 \\
6 & 242.5 & -24.866 & 5400 & 321 & 67 \\
7 & 242.5 & -23.134 & 5400 & 314 & 71 \\
8 & 244.0 & -20.536 & 7200 & 321 & 33 \\
9 & 245.5 & -21.402 & 5400 & 314 & 38 \\
10 & 247.0 & -22.268 & 5400 & 321 & 25 \\
11 & 247.0 & -24.0 & 5400 & 313 & 133 \\
12 & 247.0 & -25.732 & 3600 & 326 & 96 \\
13 & 245.5 & -26.598 & 9000 & 324 & 47 \\
14 & 241.0 & -24.0 & 5400 & 328 & 41 \\
15 & 241.0 & -22.268 & 5400 & 324 & 65 \\
16 & 242.5 & -21.402 & 5400 & 324 & 62 \\
17 & 244.0 & -18.804 & 5400 & 324 & 31 \\
18 & 248.5 & -24.866 & 5400 & 328 & 47 \\
19 & 239.5 & -23.134 & 5400 & 328 & 56 \\
20 & 239.5 & -21.402 & 5400 & 324 & 30 \\
21 & 241.0 & -20.536 & 11520 & 327 & 67 \\
22 & 242.5 & -19.670 & 5400 & 324 & 86 \\
23 & 241.0 & -18.804 & 5400 & 324 & 60 \\
24 & 242.5 & -17.938 & 10800 & 328 & 24 \\
25 & 238.0 & -22.268 & 5400 & 324 & 24 \\
\hline
Total &  &  &  & 8061 (\textbf{7039}) & 1522 (\textbf{1204})\\
\hline
\end{tabular}}
\end{center}
\setlength{\belowcaptionskip}{-10pt}
\setlength{\textfloatsep}{0pt}
\caption{Details of the  fields targeted and observed, listing the central coordinates, total exposure time (s), number of science targets and number of confirmed PMS stars per field. As there are some targets included in overlapping fields, the total number of unique targets are given in bold.}
\label{field_table}
\end{table*}

\begin{figure*} 
    \includegraphics[width=500pt]{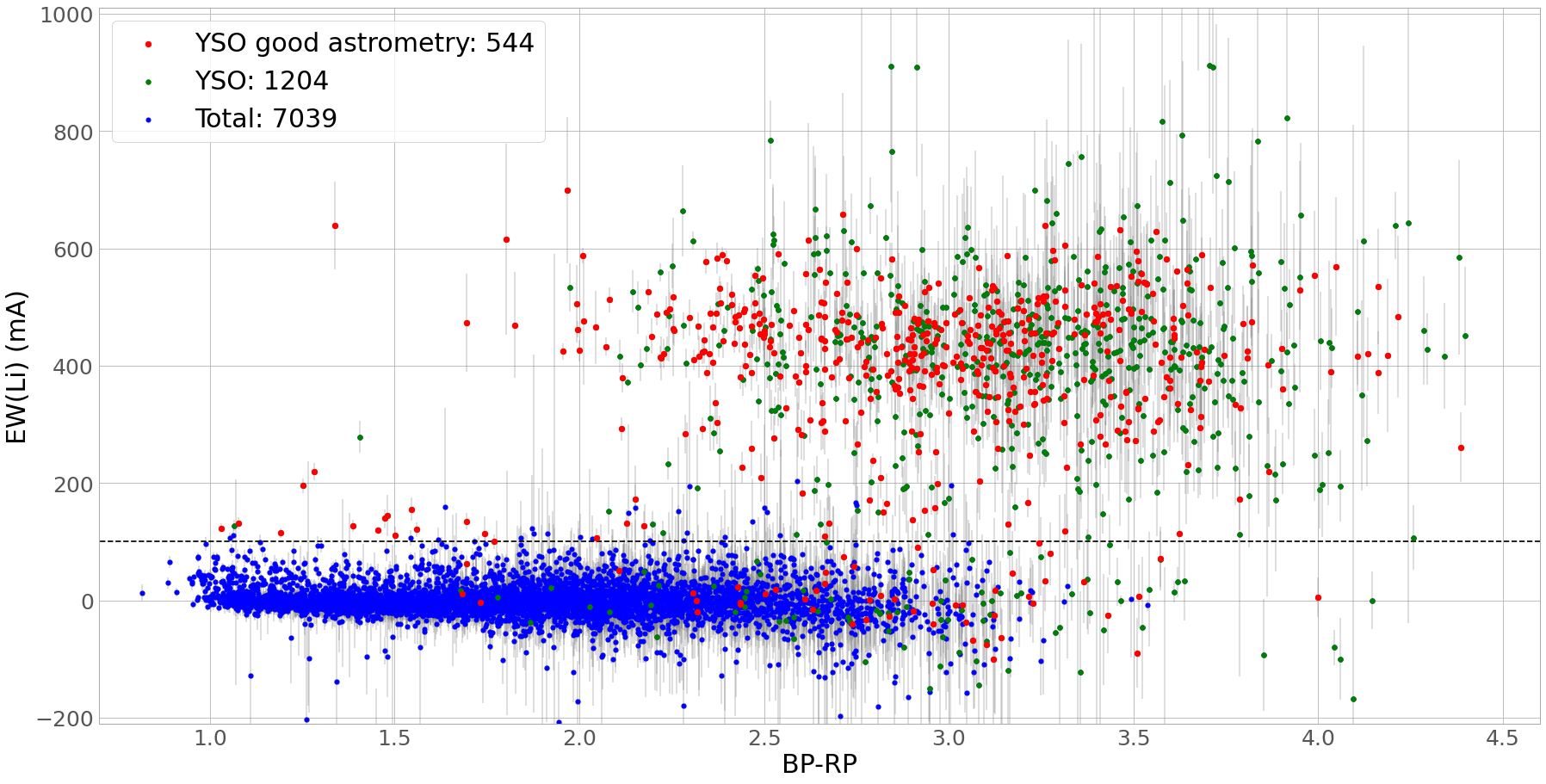}
    \setlength{\belowcaptionskip}{-10pt}
    \setlength{\textfloatsep}{0pt}
    \caption{Equivalent width of the lithium 6708 \AA\ line plotted against the Gaia DR3 BP - RP colour. Sources that pass one or more of our YSO criteria are plotted in green, sources that pass one or more of our YSO criteria and also pass our astrometric quality criteria are plotted in red, sources that meet no YSO criteria are plotted in blue (likely non-YSOs). The EW(Li) threshold value for selecting YSOs (100 m\AA) is indicated with a dashed line. }
    \label{EWLi}%
\end{figure*}

\subsubsection{Identifying YSOs}\label{YSOs}

The presence of Li is an effective indicator of youth in low-mass young stars \citep{soderblom10}. We measured the equivalent width of the Li $6707.8$ \AA\ line (EW(Li)) following the same approach as described in \citet{Armstrong22}, by subtracting the synthetic spectrum from the target spectrum and then integrating under the relevant profile. EW(Li) uncertainties are taken as the RMS value of the EWs measured using the same procedure with the Gaussian profile of the CCF centred at five wavelengths either side of the Li $6707.8$ \AA\ line \citep{jackson18}. 


We obtained values of EW(Li) and EW(Li) uncertainties for 6937 (98.6\%) of our 7039 unique targets, the distribution of which is illustrated in Fig.~\ref{EWLi}, and we identify 999 YSOs that pass the youth criteria of EW(Li)-EW(Li)\_error $>100$ m\AA\ .

Due to veiling, EW(Li)s can be underestimated for YSOs with high mass accretion rates \citep[$\dot M_{acc}>10^{-9} M_{\odot}$yr$^{-1}$,][]{palla05,frasca17}, so we also measure spectroscopic signatures of accretion to identify such YSOs, in particular, excess emission of the H$\alpha$ line at 6562.8\AA\ . 








As in \citet{Armstrong22} we use both the equivalent-width of H$\alpha$ and the spectral index $\alpha_w$ of the wing components of the H$\alpha$ emission profile, as described by \citet{damiani14}, to select likely Classical T Tauris (CTTs). We adopt thresholds of EW(H$\alpha$) - EW(H$\alpha$)\_error $>$ 10 \AA\ and $\alpha_w$ $>$1.1 (for sources with $SNR>5$). We identify 230 YSOs (likely CTTs) in our sample via these H$\alpha$ criteria, 62 of which did not meet the EW(Li) criteria of EW(Li)-EW(Li)\_error $>100$ m\AA\ .


We also cross-match spectroscopic targets with YSOs from the Gaia DR3 variability catalogue \citep{marton23}. One of the main reasons why YSOs are expected to exhibit optical variability is as a result of fluctuating accretion activity while there remains a circumstellar disk. We identify 986 likely YSOs via this variability criteria, 176 of which we identify as CTTs by our H$\alpha$ criteria and 831 which are also confirmed as YSOs via EW(Li).

Thus, in total, we identify 1204 YSOs.

\subsection{Gaia astrometry}\label{Gaia}
Spectroscopic targets are matched to the Gaia DR3 catalogue \citep{Gaiaedr3} to obtain positions, proper motions and parallaxes. Combined with our RVs, this gives us full 6D kinematic information for all of our 1204 confirmed YSOs.
We filter targets with poor Gaia astrometric quality following the recommended criteria of \citet{fabricius21}; RUWE $>$ 1.4, \textit{ipd\_frac\_multi\_peak} $>$ 3 and \textit{ipd\_gof\_harmonic\_amplitude} $>$ 0.1, in order to remove non-single stars. The velocity contribution of binary orbits, in particular, could bias our kinematic analysis. We also remove targets with RV$> |200|$ km s$^{-1}$ or with RV uncertainty $> 1$ km s$^{-1}$, in order to further filter sources with unreliable RVs and possible binaries. Out of our 1204 confirmed YSOs, 79 fail the \textit{ipd\_gof\_harmonic\_amplitude} criterion, 92 fail the \textit{ipd\_frac\_multi\_peak} criterion, 251 fail the RUWE criterion and 482 fail the RV criteria, with overlap. After applying these astrometric quality filters we are left with 544 confirmed YSOs with quality 6D kinematic information.

\subsection{Deriving 3D positions and velocities}\label{cartesian}

We obtain Cartesian positions XYZ and velocities UVW in the Galactic system following the same approach as \citet{Armstrong22}, using the coordinate transformation matrices from \citet{johnson87}. This is done via Bayesian inference where we perform 1000 iterations with 100 walkers in an unconstrained parameter space. We sample the posterior distribution using the MCMC sampler \textit{emcee} \citep{emcee}, reporting the median, 16th and 84th percentiles as the best fit and 1 $\sigma$ uncertainties respectively, after discarding the first 500 iterations as a burn in.

When deriving 3D positions we use Gaia parallaxes rather than distance estimates from \citet{bailerjones21}. As described in \citet{fiorellino24}, the priors involved in the probabilistic estimation of distances by \citet{bailerjones21} are not optimised for YSOs, nor for sources in clustered regions. We apply the parallax zero point correction to our sources, as prescribed in \citet{lindegren21}, as a function of ecliptic latitude, magnitude and colour. We also apply the magnitude dependent correction to parallax uncertainties prescribed by \citet{elbadry21}. 


\subsection{Summary of the data}

After reducing the spectra of our AAT sources, we have 6646 unique sources with spectroscopic RV and EW(Li) values spread across 25 fields over the Upper Sco region, with 5-parameter astrometry from Gaia EDR3, with which we calculate Cartesian XYZ positions and UVW velocities. 1204 of these sources are identified as YSOs with significant EW(Li)s, EW(H$\alpha$)s or Gaia variability flag. The median uncertainties on parallax, distance, proper motion and RV for the confirmed PMS stars are 0.05 mas, 0.9 pc, 0.05 mas yr$^{-1}$ and 0.56 km s$^{-1}$ respectively, with a median SNR $= 16.1$ of their HERMES spectra. 544 of these confirmed PMS stars pass our filters for astrometric quality.

 
\section{Overview of the sample}
\label{section_overview}

\begin{figure*} 
    \includegraphics[width=\textwidth]{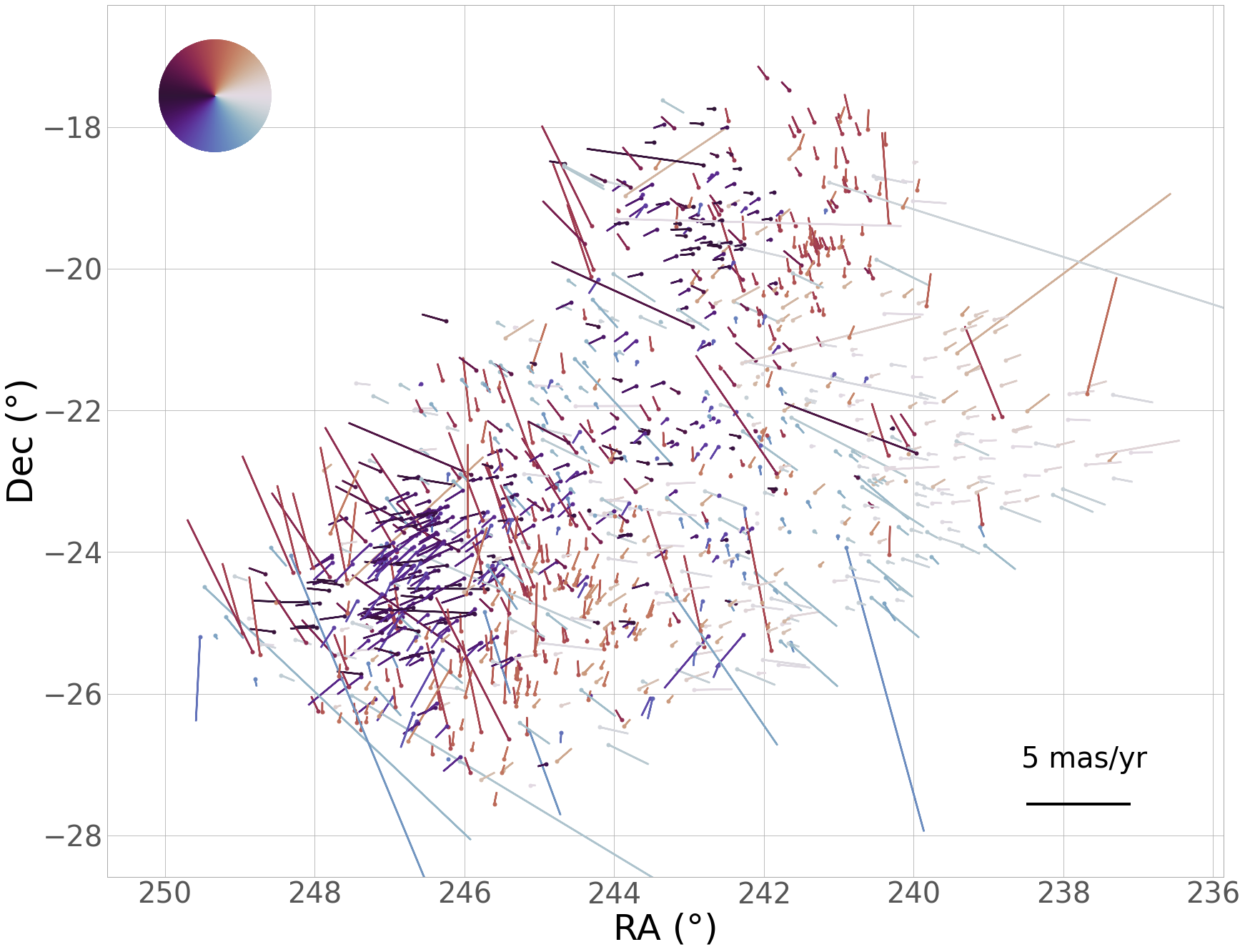}
    \setlength{\belowcaptionskip}{-10pt}
    \setlength{\textfloatsep}{0pt}
    \caption{Sky positions of confirmed YSOs observed with the AAT. Vectors indicate the proper motion of each source relative to the median of all Upper Sco YSOs in the sample, colour-coded based on the position angle of the proper motion (see the colour wheel in the top left as a key). The magnitude scale (mas yr$^{-1}$) of proper motion vectors is indicated by the scale bar in the bottom right.  }
    \label{PMvectorPlot}%
\end{figure*}

\begin{figure} 
    \subfloat{{\includegraphics[width=240pt]{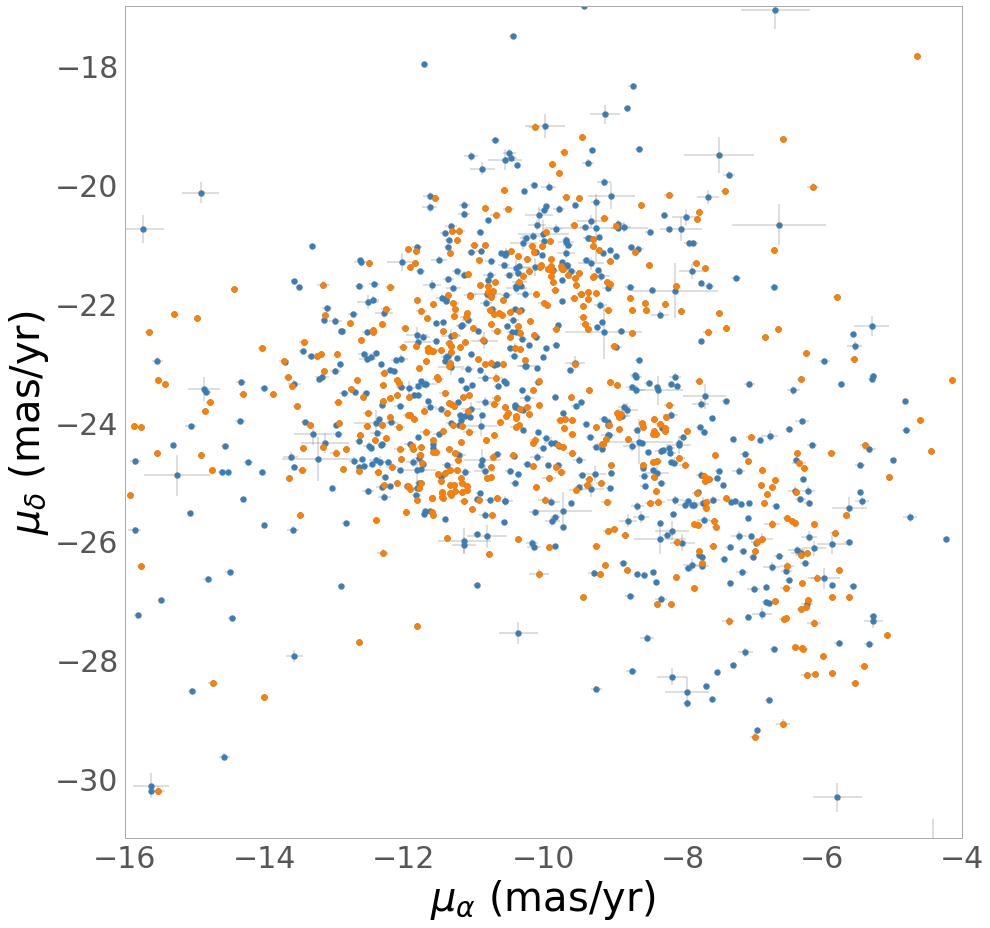}}}%
    \qquad
    \subfloat{{\includegraphics[width=240pt]{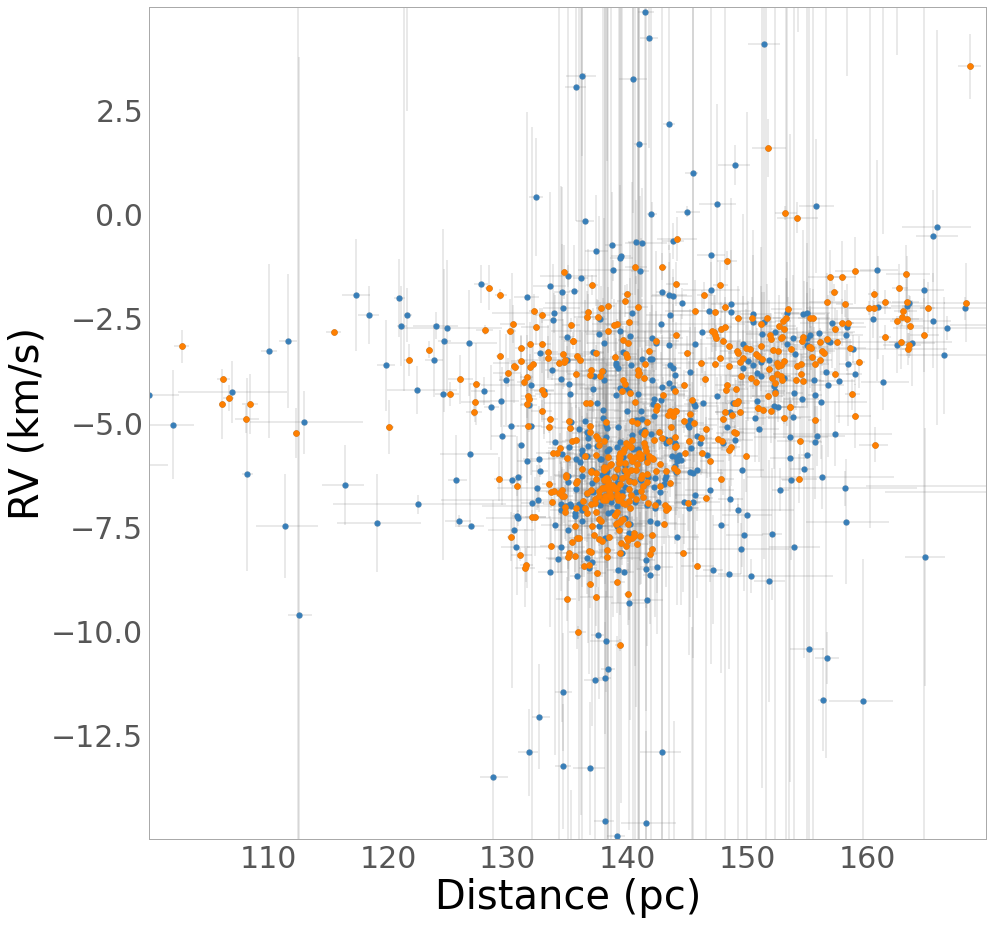}}}%
    \setlength{\belowcaptionskip}{-10pt}
    \setlength{\textfloatsep}{0pt}
    \caption{\textit{Top:} Proper motion space and \textit{Bottom:} \citet{bailerjones21} distance versus RV of confirmed YSOs with quality 6D kinematic information (red) and without (blue).}
    \label{Astrometry}%
\end{figure}

Fig. \ref{PMvectorPlot} shows the spatial distribution of the 1204 confirmed YSOs with their proper motion vectors colour-coded according to their relative proper motion (relative to the median of the sample). It is immediately apparent that there are multiple subgroups among the YSOs with distinct bulk proper motions, as well as a number of high proper motion YSOs seemingly unassociated with any of the major groups. The distribution of YSOs in proper motion space and line-of-sight (LOS) distance versus RV space is plotted in Fig.~\ref{Astrometry}. There is also apparent spatial and velocity structure along the LOS, with the bulk of the YSO population located at $\sim140$ pc, but also a smaller concentration at $\sim155\:$pc, as well as a sparse distribution up to $\sim40\:$pc in the foreground.

We compare our sample of confirmed YSOs to previously compiled catalogues of candidate YSOs in Upper Sco to assess their relative completeness and to highlight the advantages of different methods of YSO identification.

\begin{figure} 
    {\includegraphics[width=240pt]{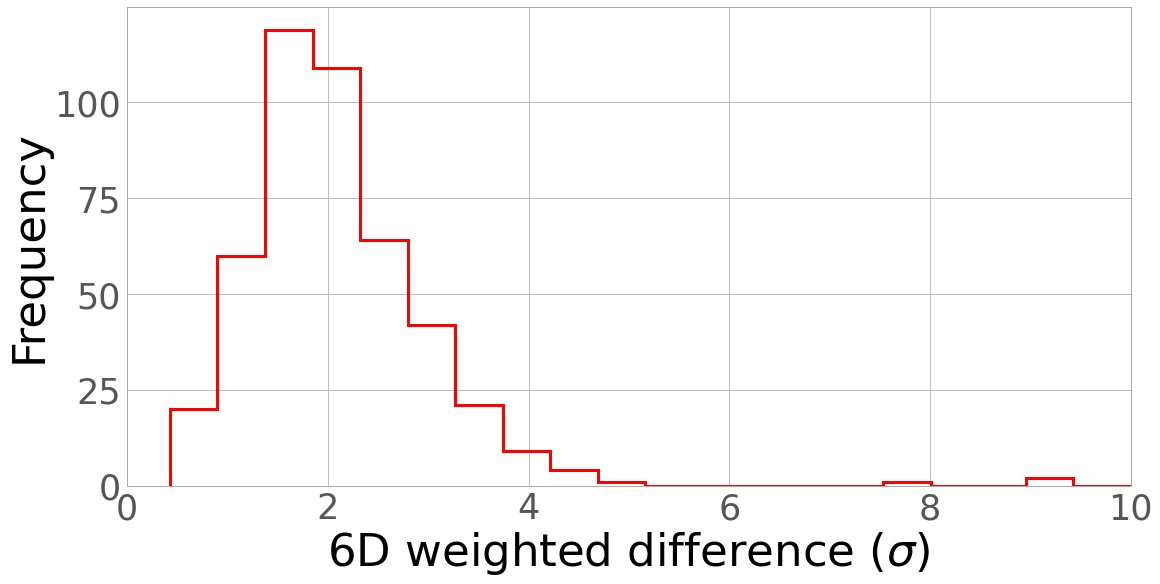}}
    \setlength{\belowcaptionskip}{-10pt}
    \setlength{\textfloatsep}{0pt}
    \caption{Histogram of 6D weighted distance between YSOs not belonging to a \protect\citet{miret-roig22} subgroup and their nearest subgroup.}
    \label{Distance_to_groups}%
\end{figure}

\subsection{Cross-match with \protect\citet{miret-roig22}}

\citet{miret-roig22} used a Gaussian mixture model to separate the Upper Sco and Ophiuchus YSO candidate list of \citet{miret-roig22a} into kinematically distinct subgroups. Out of our full target sample we find 1065 matches with the members of Upper Sco identified by \citet{miret-roig22} using the HDBSCAN algorithm. 228 belong to $\alpha$ Sco, 98 belong to $\beta$ Sco, 199 belong to $\delta$ Sco, 119 belong to $\pi$ Sco, 85 belong to $\nu$ Sco, 197  belong to $\rho$ Oph, 139 belong to $\sigma$ Sco. Also, 946 of these are identified as YSOs in the Gaia variability catalogue \citep{marton23}, and 946 have RVs in our sample.

Considering our 1204 confirmed YSOs, we find 1030 matches with the members of Upper Sco identified by \citet{miret-roig22}. 224 belong to $\alpha$ Sco, 98 belong to $\beta$ Sco, 198 belong to $\delta$ Sco, 94 belong to $\pi$ Sco, 84 belong to $\nu$ Sco, 197 belong to $\rho$ Oph, 135 belong to $\sigma$ Sco. The remaining 174 YSOs were not identified as members of any subgroup by \citet{miret-roig22}, and we discuss their membership in Section~\ref{clustering}. There are also 5 subgroup members of \citet{miret-roig22} for which we have spectra of SNR$>20$, but do not meet any of our YSO criteria. The number of false-negatives in  particular highlights the importance of spectroscopic youth indicators for identifying YSOs with kinematics distinct from the major subgroups.

Out of our 544 confirmed YSOs with quality 6D kinematic information we find 493 matches with the members of Upper Sco identified by \citet{miret-roig22}. 111 belong to $\alpha$ Sco, 53 belong to $\beta$ Sco, 110 belong to $\delta$ Sco, 35 belong to $\pi$ Sco, 43 belong to $\nu$ Sco, 93 belong to $\rho$ Oph, 48 belong to $\sigma$ Sco. The remaining 51 YSOs were not allocated to any subgroup by \citet{miret-roig22}, and we discuss their membership in Section~\ref{clustering}. 

\begin{figure*} 
    \includegraphics[width=\textwidth]{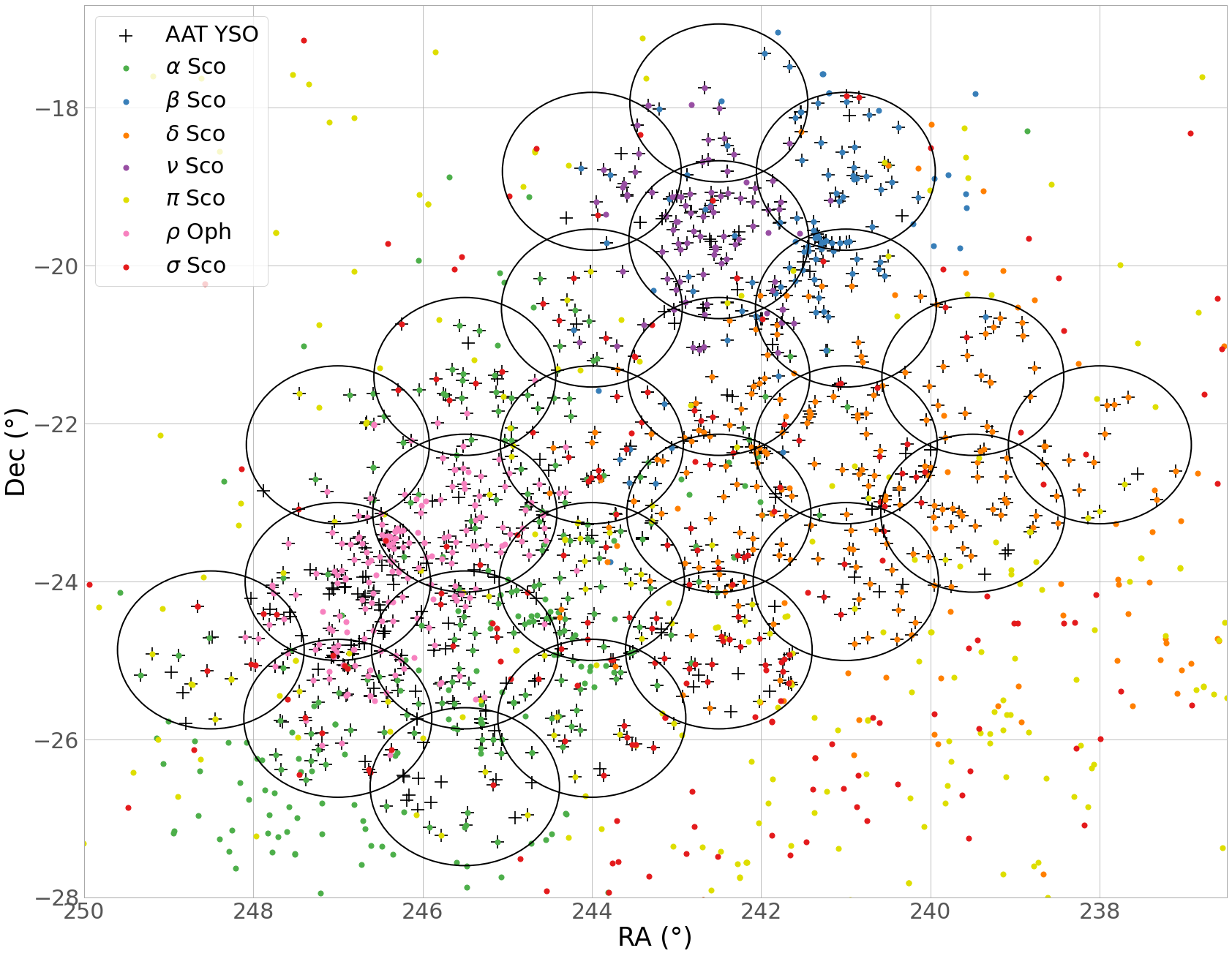}
    \setlength{\belowcaptionskip}{-10pt}
    \setlength{\textfloatsep}{0pt}
    \caption{Sky positions of spectroscopically confirmed YSOs observed with the AAT, indicated with plus symbols, as well as the positions of Upper Sco subgroup members from \protect\citet{miret-roig22} within the same magnitude range as our observed sample ($13 < G < 16.5$), colour-coded according to their subgroup membership. The 2$^\circ$-diameter circles indicate the individual fields of observation of the AAT within which targets were selected.   }
    \label{MiretRoigFields}%
\end{figure*}

Fig. \ref{MiretRoigFields} shows the spatial distribution of the 1204 confirmed PMS stars indicated with plus symbols, as well as the positions of 1580 Upper Sco subgroup members of \citet{miret-roig22} within the same magnitude range as our observed sample ($13 < G < 16.5$), colour-coded according to their subgroup membership. The AAT FOVs are indicated by the overlaid circles.

It is important to note that the FOVs coverage of each subgroup varies. $\beta$, $\nu$ Sco and $\rho$ Oph are almost completely covered, while $\delta$, $\pi$ and $\sigma$ Sco have many members that lie outside the observed region, towards Lupus and Upper Centaurus-Lupus (UCL). Therefore, the central positions of these subgroups in our sample will differ from those given by \citet{miret-roig22}, but our sample should still be sufficient to measure kinematic properties.

\begin{figure} 
    \subfloat{{\includegraphics[width=240pt]{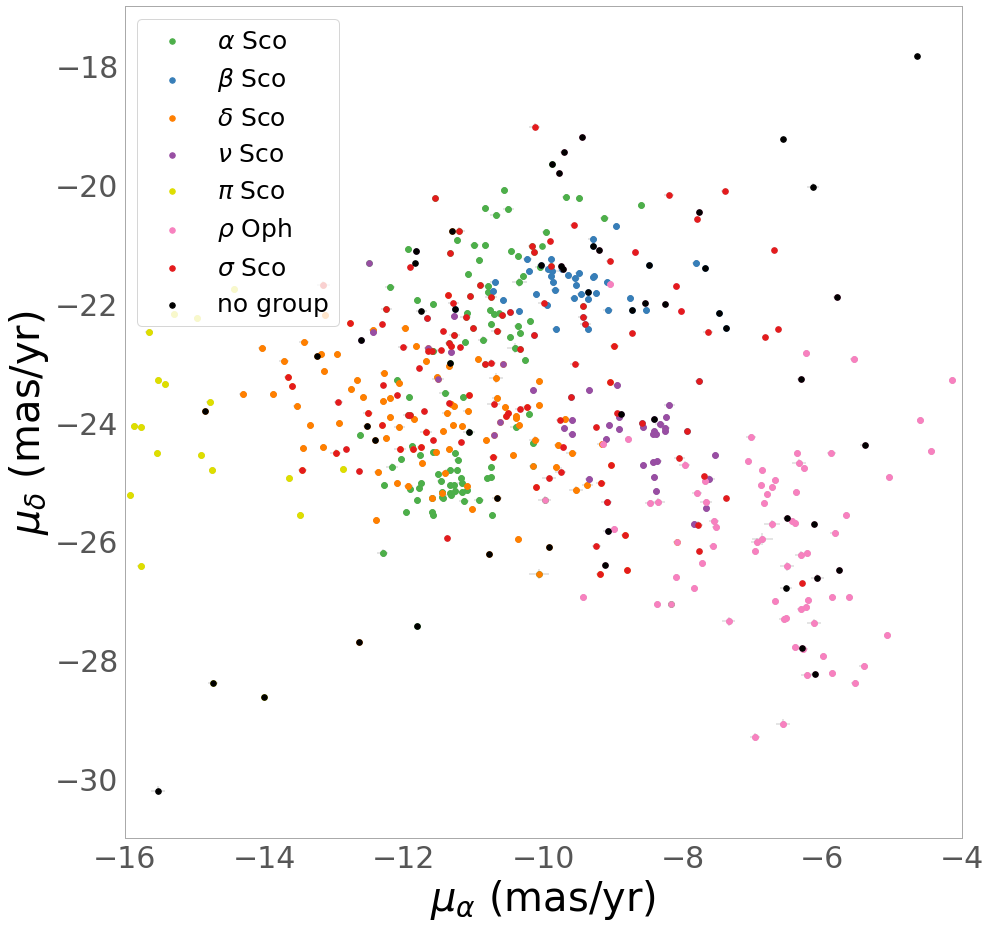}}}%
    \qquad
    \subfloat{{\includegraphics[width=240pt]{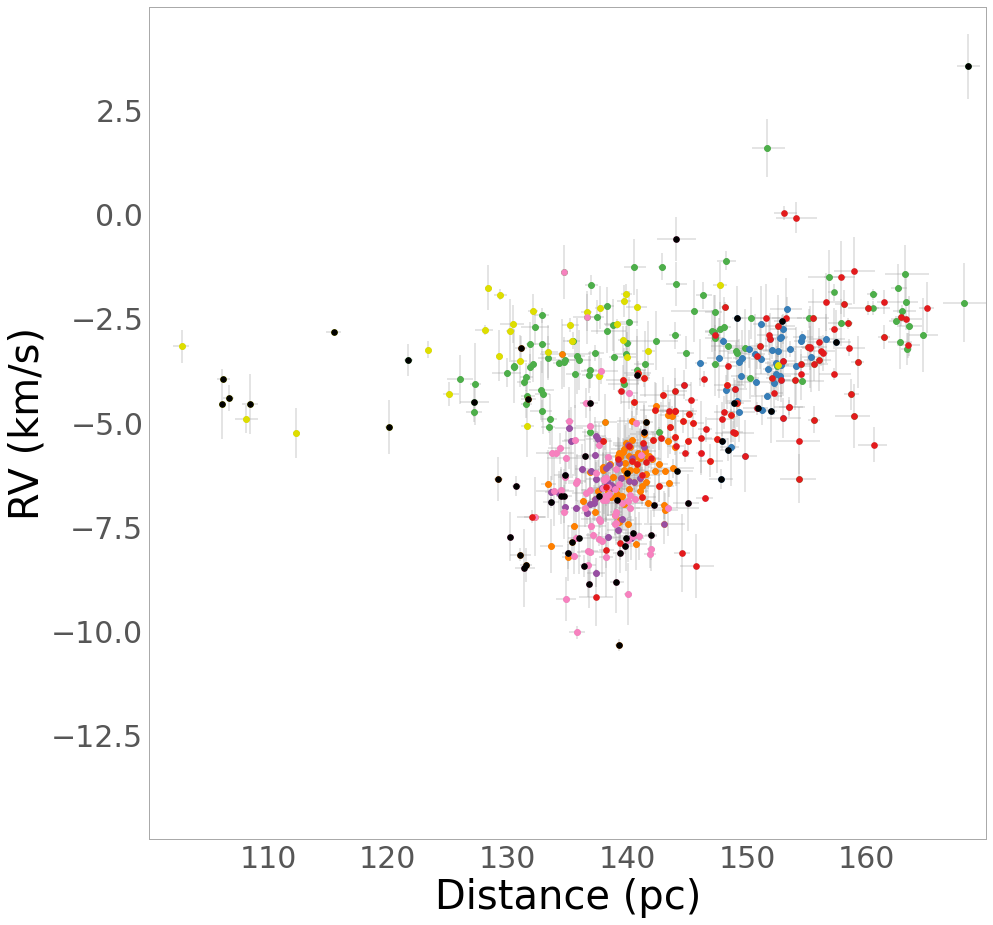}}}%
    \setlength{\belowcaptionskip}{-10pt}
    \setlength{\textfloatsep}{0pt}
    \caption{\textit{Top:} Proper motion space and \textit{Bottom:} \citet{bailerjones21} distance versus RV of confirmed YSOs colour-coded according to the kinematic subgroup they have been allocated to.}
    \label{Astrometry_groups}%
\end{figure}

\subsection{Cross-match with \protect\citet{ratzenbock23}}

\citet{ratzenbock23} selected a large sample of candidate YSOs across Sco-Cen, including Upper Sco, from the Gaia DR3 catalogue using photometric cuts and astrometric quality criteria. They then identify kinematically distinct subgroups using the SIGMA clustering algorithm \citep{ratzenbock23a}.

Out of our full target sample we find 1121 matches with the members of Sco-Cen (including Upper Sco) identified by \citet{ratzenbock23} using the SIGMA algorithm. 217 belong to $\rho$ Oph, 79 belong to $\nu$ Sco, 276 belong to $\delta$ Sco, 121 belong to $\beta$ Sco, 165 belong to $\sigma$ Sco, 157 belong to Antares, 54 belong to $\rho$ Sco, 52 belong to US-foreground. Also, 960 of these are identified as YSOs in the Gaia variability catalogue \citep{marton23}, and 994 have RVs in our sample.

Out of our 1204 confirmed YSOs we find 1085 matches with the members of Sco-Cen identified by \citet{ratzenbock23}. 216 belong to $\rho$ Oph, 78 belong to $\nu$ Sco, 272 belong to $\delta$ Sco, 121 belong to $\beta$ Sco, 161 belong to $\sigma$ Sco, 154 belong to Antares, 52 belong to $\rho$ Sco, 31 belong to US-foreground. The remaining 119 YSOs were not identified as members of any subgroup by \citet{ratzenbock23}, likely because their kinematics are too distinct from the subgroups for the SIGMA algorithm to allocate them. There are also 7 subgroup members of \citet{ratzenbock23} for which we have spectra of SNR$>20$, but do not meet any of our YSO criteria. 29 more of our targets with matches to the list of \citet{ratzenbock23} we do not confirm as YSOs, but these have spectra of SNR$<5$ and so their EW(Li) measurements are relatively uncertain. Again, the number of false-negatives highlights the importance of spectroscopic youth indicators for identifying YSOs independently of their kinematics.

Out of our 544 confirmed YSOs with quality 6D kinematic information we find 495 matches with the members of Sco-Cen identified by \citet{ratzenbock23}. 92 belong to $\rho$ Oph, 36 belong to $\nu$ Sco, 138 belong to $\delta$ Sco, 55 belong to $\beta$ Sco, 71 belong to $\sigma$ Sco, 70 belong to Antares, 24 belong to $\rho$ Sco, 9 belong to US-foreground. The remaining 49 YSOs were not identified as members of any subgroup by \citet{ratzenbock23}.


\subsection{Cross-match with \protect\citet{luhman22}}

\citet{luhman22} selected likely members of Sco-Cen using Gaia DR3 photometric and astrometric criteria. Out of our total observed sample we find 1009 matches with the members of Sco-Cen identified by \citet{luhman22}, 1007 of which match with our 1204 confirmed YSOs, and 2, for which we have spectra of SNR$>20$, do not meet any of our YSO criteria.

	
\section{Clustering}
\label{clustering}

Considering the limitations involved in target selection, and positional bias introduced by the fibre configuration for observations, we do not perform our own clustering analysis of Upper Sco with our spectroscopic sample. Rather, we adopt the subgroups identified by \citet{miret-roig22}.

For the YSOs that we have identified which were not included in the subgroup membership list of \citet{miret-roig22}, we allocate to the group to which its 6D kinematics are most similar. We determine which group a YSO is most similar to using the distance in 6D between the YSO and the group central Cartesian positions (X,Y,Z) and velocities (U,V,W) weighted by the inverse of the group's standard deviation in these coordinates (Fig.~\ref{Distance_to_groups}). The total number of YSO members in each subgroup and the number of new YSOs, either with or without quality 6D kinematic information, are given in Table~\ref{groups_table}. 

We then recalculate the new median and dispersion of each Cartesian coordinate per subgroup using the total number of YSOs with quality 6D kinematic information and present these with their associated uncertainties in Table~\ref{groups_table}. In subsequent figures members of these subgroups are plotted in green for $\alpha$ Sco, blue for $\beta$ Sco, orange for $\delta$ Sco, purple for $\nu$ Sco, yellow for $\pi$ Sco, pink for $\rho$ Oph \& red for $\sigma$ Sco.

For YSOs with significantly distinct kinematics from any group, i.e., 6D weighted difference $> 3\sigma$ (Fig.~\ref{Distance_to_groups}), we keep separately in a group of kinematic outliers, which we investigate later to determine if any might be dynamically ejected YSOs either from Upper Sco or from another nearby young cluster. We find 84 YSOs that we categorise as kinematic outliers. In subsequent figures these are plotted in black.

We re-plot the proper motion distribution and LOS distance versus RV distribution of YSOs in Fig.~\ref{Astrometry_groups} colour-coded according to the kinematic subgroup they have been allocated to. Most subgroups appear sparse and have significant overlap in proper motion space, but clearer separation is seen in LOS distance versus RV between $\alpha$ Sco (green) and $\pi$ Sco (yellow) which extend further into the foreground and have greater RVs than $\delta$ Sco (orange), $\nu$ Sco (purple) and $\sigma$ Sco (red).

\begin{figure*} 
    \subfloat{{\includegraphics[width=490pt]{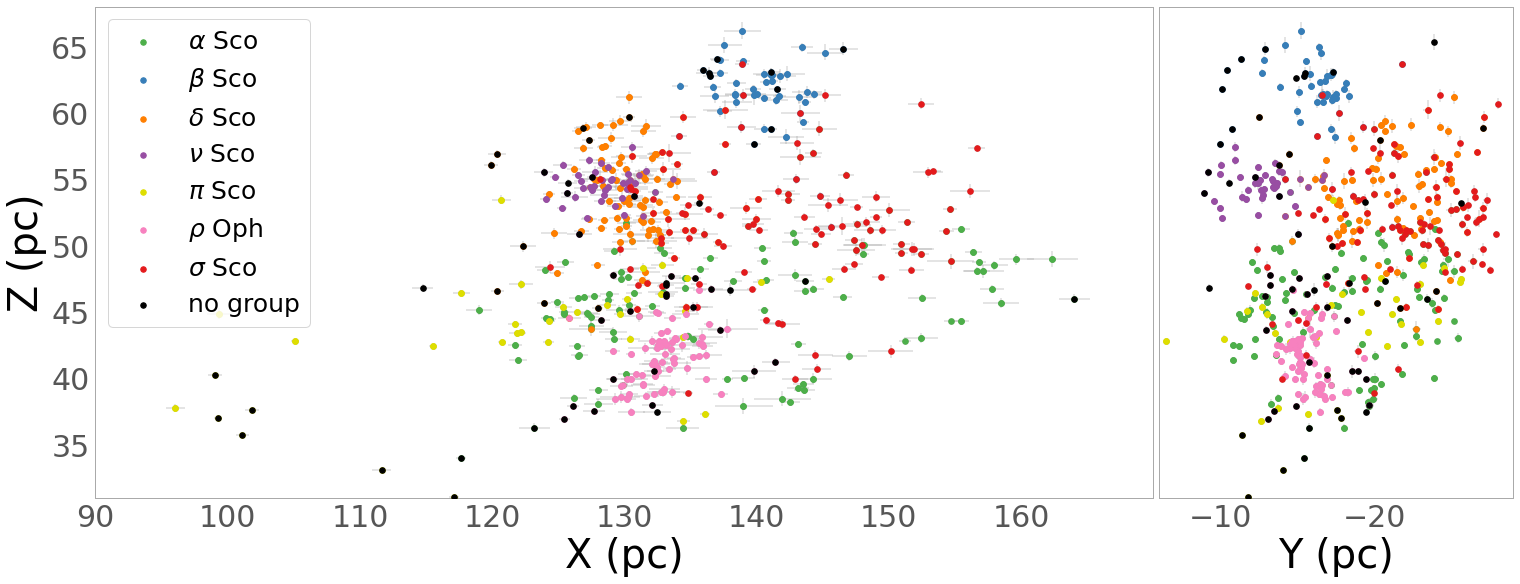}}}%
    \qquad
    \subfloat{{\includegraphics[width=490pt]{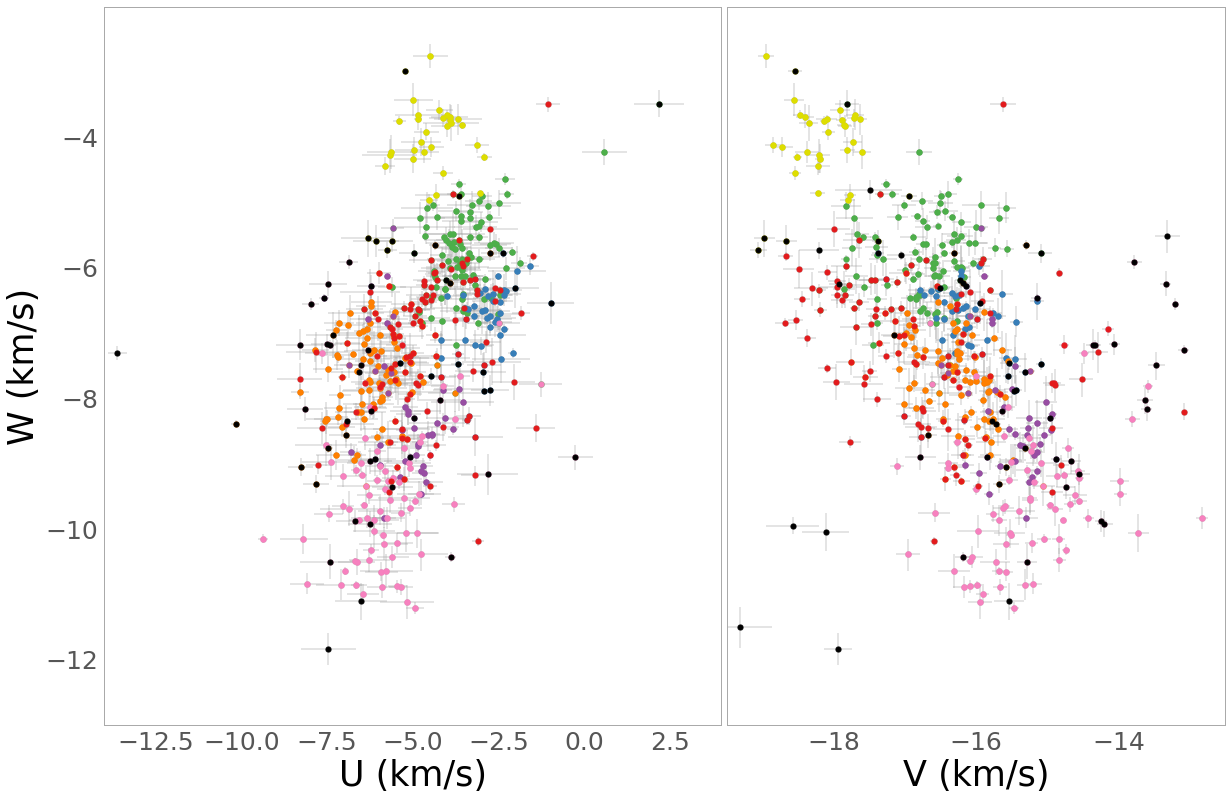}}}%
    \setlength{\belowcaptionskip}{-10pt}
    \setlength{\textfloatsep}{0pt}
    \caption{\textit{Top:} Heliocentric Cartesian coordinates (XYZ) and \textit{Bottom:} velocities (UVW) of confirmed YSOs coloured according to their kinematic groups.}
    \label{Cartesian_groups}%
\end{figure*}

In Fig.~\ref{Cartesian_groups} we show the Cartesian spatial (XYZ, \textit{top panel}) and velocity (UVW, \textit{lower panel}) structure of the 544 confirmed PMS stars with quality 6D information, colour-coded according to which subgroup they are assigned to. 

Notably, all subgroups apart from $\delta$ Sco have their largest spatial spread ($\sigma_{XYZ}$) in the Cartesian $X$ direction (Fig.~\ref{Cartesian_groups}). For both $\alpha$ Sco and $\pi$ Sco, $\sigma_{X}$ is more than twice as large as the next greatest, $\sigma_{Y}$, and is different from the spread in $\sigma_{Z}$ at the $>60\sigma$ significance level. We measure the significance of anisotropy by dividing the difference between the largest spatial spread and the smallest to their combined uncertainties for each subgroup. For $\beta$ Sco, spatial spreads are anisotropic at the 6$\sigma$ level, for $\delta$ Sco, at the $12\sigma$ level, for $\nu$ Sco, at the $16\sigma$ level, for $\rho$ Oph, at the $17\sigma$ level and for $\sigma$ Sco, at the $>30\sigma$ level. 

The Heliocentric Cartesian X coordinate is directed towards the Galactic center, so for Upper Sco this is the coordinate which contains the greatest component of observed parallax. Thus, parallax uncertainty, and uncertainty in line-of-sight distance, would likely inflate the spread of positions in this direction. However, as Upper Sco is relatively nearby (100 - 150 pc), the median uncertainty in LOS distance of our confirmed YSOs is relatively small, only $0.9$ pc, and thus cannot account for this spatial anisotropy. Rather, limitations in the coverage of observations, particularly the total sky area of observed fields, likely have a greater impact. In Fig.~\ref{MiretRoigFields} we plot the sky positions of all members of Upper Sco subgroups of \citet{miret-roig22} along with circles indicating the fields of observation. There are many members of $\alpha$ Sco, $\pi$ Sco and $\sigma$ Sco in particular that were not covered in our observations, which has the effect of reducing the spatial spread in Y and Z relative to X. 

Similarly, all subgroups apart from $\nu$ Sco have their largest velocity dispersion ($\sigma_{UVW}$) in the Cartesian $X$ direction ($U$). This could be due to the effect of unresolved binary orbits contributing to the observed RVs. We analyse this in greater detail in Section~\ref{Velocity dispersions}.


\begin{figure} 
    {\includegraphics[width=240pt]{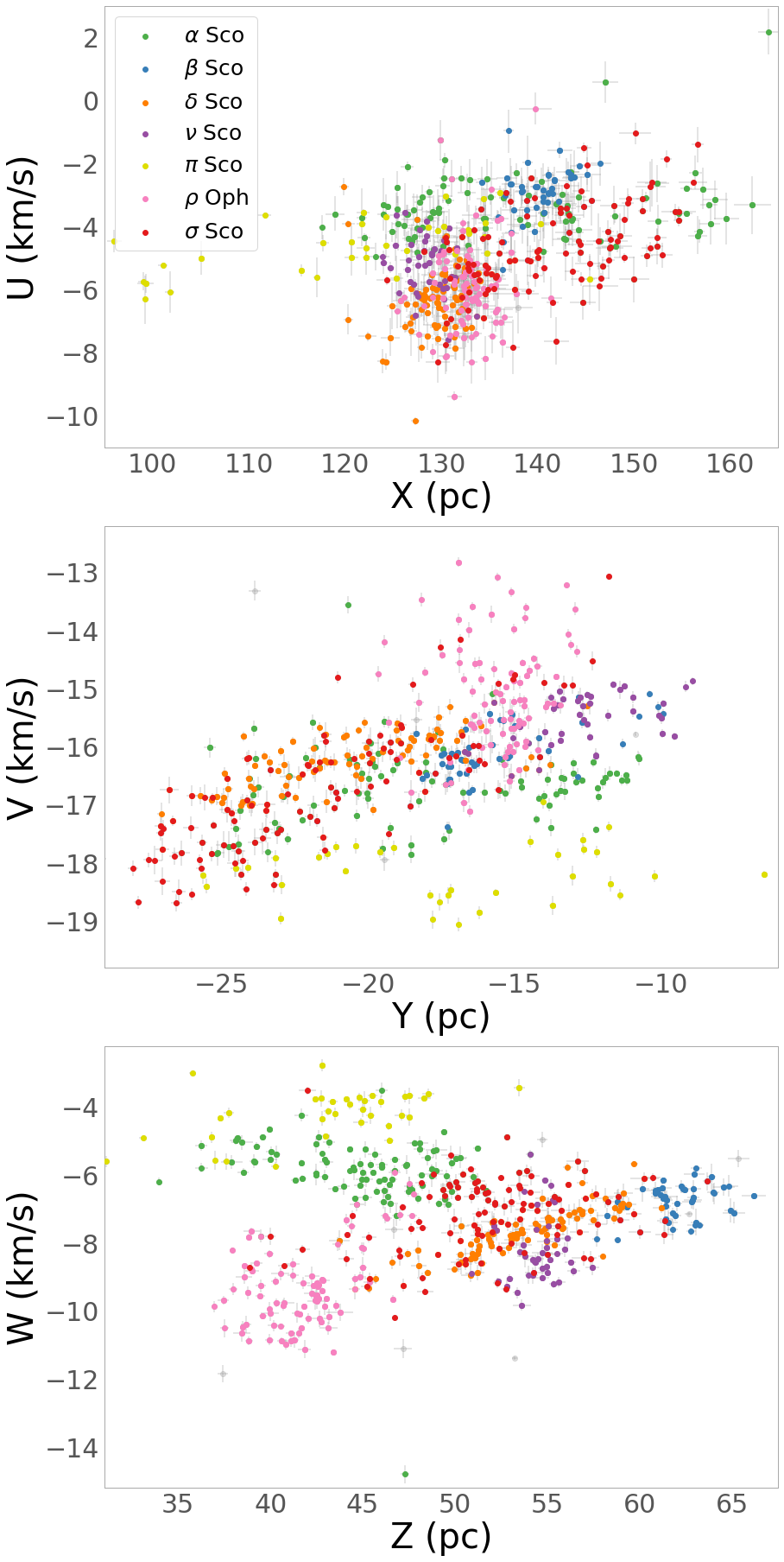}}
    \setlength{\belowcaptionskip}{-10pt}
    \setlength{\textfloatsep}{0pt}
    \caption{Heliocentric Cartesian coordinates (XYZ) and velocities (UVW) of confirmed YSOs coloured according to their kinematic groups. }
    \label{Cartesian_groups_position-velocity}%
\end{figure}

\begin{figure*} 
    {\includegraphics[width=490pt]{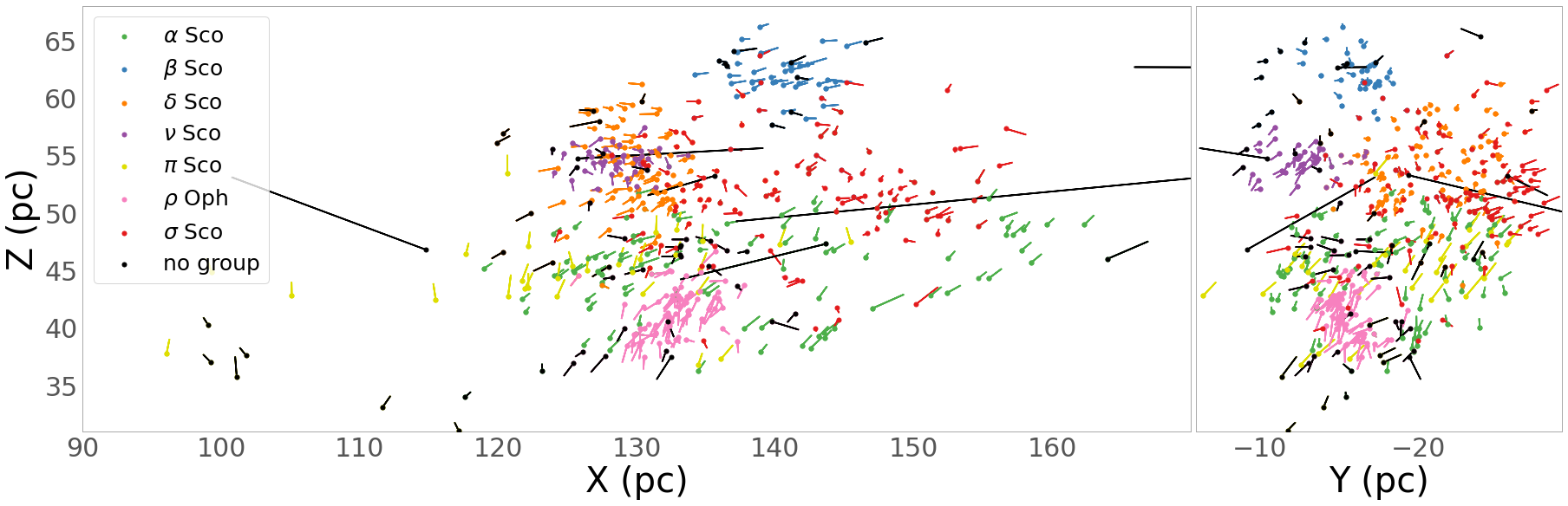}}
    \setlength{\belowcaptionskip}{-10pt}
    \setlength{\textfloatsep}{0pt}
    \caption{Heliocentric Cartesian coordinates of confirmed YSOs coloured according to their kinematic groups, with vectors indicating their velocities relative to the sample mean. }
    \label{Cartesian_groups_vectors}%
\end{figure*}

\begin{table*}
\begin{center}
{\renewcommand{\arraystretch}{1.5}
\begin{tabular}{|p{2.6cm}|p{1.6cm}|p{1.6cm}|p{1.6cm}|p{1.6cm}|p{1.6cm}|p{1.6cm}|p{1.6cm}| }
\hline
Group & $\alpha$ & $\beta$ & $\delta$ & $\nu$ & $\pi$ & $\rho$ & $\sigma$ \\
\hline
YSOs (new) & 280 (133) & 144 (70) & 263 (112) & 105 (48) & 117 (64) & 271 (126) &  174 (104)   \\
6D YSOs (new) & 201 (91) & 88 (35) & 184 (74) & 76 (33) & 63 (29) & 162 (69) &  111 (63)   \\ 
median X (pc) & $136.15^{+0.43}_{-0.43}$ & $140.27^{+0.21}_{-0.22}$ & $130.42^{+0.14}_{-0.14}$ & $128.44^{+0.13}_{-0.13}$ & $125.21^{+0.39}_{-0.37}$ & $132.92^{+0.16}_{-0.16}$ & $140.95^{+0.39}_{-0.40}$ \\
$\sigma_X$ (pc) & $10.94^{+0.10}_{-0.09}$ & $2.91^{+0.10}_{-0.10}$ & $3.27^{+0.07}_{-0.07}$ & $2.11^{+0.05}_{-0.05}$ & $11.96^{+0.08}_{-0.08}$ & $2.85^{+0.09}_{-0.09}$ & $7.53^{+0.10}_{-0.10}$ \\
median Y (pc) & $-19.16^{+0.09}_{-0.08}$ & $-16.59^{+0.04}_{-0.03}$ & $-21.00^{+0.06}_{-0.07}$ & $-12.61^{+0.02}_{-0.02}$ & $-17.33^{+0.07}_{-0.07}$ & $-15.33^{+0.03}_{-0.04}$ & $-21.93^{+0.09}_{-0.10}$ \\
$\sigma_Y$ (pc) & $4.42^{+0.01}_{-0.01}$ & $2.22^{+0.01}_{-0.01}$ & $3.10^{+0.01}_{-0.01}$ & $1.74^{+0.01}_{-0.00}$ & $4.96^{+0.01}_{-0.01}$ & $1.30^{+0.01}_{-0.01}$ & $3.98^{+0.02}_{-0.02}$ \\
median Z (pc) & $46.09^{+0.10}_{-0.09}$ & $61.78^{+0.10}_{-0.10}$ & $53.72^{+0.11}_{-0.11}$ & $54.63^{+0.06}_{-0.06}$ & $44.63^{+0.12}_{-0.12}$ & $41.81^{+0.09}_{-0.09}$ & $51.79^{+0.12}_{-0.12}$ \\
$\sigma_Z$ (pc) & $4.33^{+0.03}_{-0.03}$ & $2.49^{+0.04}_{-0.04}$ & $3.48^{+0.03}_{-0.03}$ & $1.23^{+0.02}_{-0.02}$ & $4.17^{+0.03}_{-0.03}$ & $2.39^{+0.03}_{-0.03}$ & $4.62^{+0.03}_{-0.03}$ \\
median U (kms$^{-1}$) & $-3.62^{+0.07}_{-0.07}$ & $-2.91^{+0.13}_{-0.12}$ & $-6.14^{+0.03}_{-0.03}$ & $-5.06^{+0.08}_{-0.08}$ & $-4.46^{+0.12}_{-0.12}$ & $-5.97^{+0.07}_{-0.07}$ & $-4.76^{+0.08}_{-0.07}$ \\
$\sigma_U$ (kms$^{-1}$) & $2.24^{+0.04}_{-0.04}$ & $1.11^{+0.08}_{-0.08}$ & $0.94^{+0.01}_{-0.01}$ & $0.90^{+0.04}_{-0.04}$ & $0.98^{+0.07}_{-0.07}$ & $1.36^{+0.04}_{-0.04}$ & $1.42^{+0.03}_{-0.03}$ \\
median V (kms$^{-1}$) & $-16.67^{+0.03}_{-0.03}$ & $-16.19^{+0.03}_{-0.03}$ & $-16.24^{+0.02}_{-0.02}$ & $-15.40^{+0.02}_{-0.02}$ & $-18.13^{+0.04}_{-0.04}$ & $-15.29^{+0.03}_{-0.03}$ & $-16.80^{+0.05}_{-0.05}$ \\
$\sigma_V$ (kms$^{-1}$) & $0.64^{+0.01}_{-0.01}$ & $0.48^{+0.02}_{-0.02}$ & $0.51^{+0.01}_{-0.01}$ & $0.39^{+0.01}_{-0.01}$ & $0.44^{+0.02}_{-0.02}$ & $0.89^{+0.01}_{-0.01}$ & $1.09^{+0.01}_{-0.01}$ \\
median W (kms$^{-1}$) & $-5.90^{+0.03}_{-0.03}$ & $-6.77^{+0.06}_{-0.06}$ & $-7.62^{+0.02}_{-0.02}$ & $-8.49^{+0.04}_{-0.04}$ & $-4.02^{+0.05}_{-0.05}$ & $-9.56^{+0.04}_{-0.04}$ & $-7.10^{+0.05}_{-0.05}$ \\
$\sigma_W$ (kms$^{-1}$) & $0.93^{+0.01}_{-0.01}$ & $0.66^{+0.04}_{-0.04}$ & $0.70^{+0.01}_{-0.01}$ & $0.91^{+0.02}_{-0.02}$ & $0.63^{+0.03}_{-0.03}$ & $1.12^{+0.01}_{-0.01}$ & $1.09^{+0.01}_{-0.01}$ \\
$\bar{v}_{out}$ (kms$^{-1}$) & $0.11^{+0.04}_{-0.04}$ & $0.13^{+0.06}_{-0.06}$ & $0.46^{+0.05}_{-0.05}$ & $0.16^{+0.09}_{-0.09}$ & $0.10^{+0.04}_{-0.04}$ & $0.24^{+0.07}_{-0.07}$ & $0.45^{+0.10}_{-0.10}$  \\
$v_{out}>0$ (\%) & 59 & 69 & 83 & 59 & 64 & 72 & 73 \\
$\mu_{pmRA}$ (masyr$^{-1}$) & $-23.31^{+0.14}_{-0.14}$ & $-21.77^{+0.06}_{-0.06}$ & $-24.01^{+0.08}_{-0.09}$ & $-24.12^{+0.11}_{-0.11}$ & $-25.47^{+0.43}_{-0.41}$ & $-25.61^{+0.17}_{-0.16}$ & $-23.23^{+0.16}_{-0.17}$ \\
$\sigma_{pmRA}$ (kms$^{-1}$) & $0.51^{+0.03}_{-0.03}$ & $0.55^{+0.05}_{-0.05}$ & $0.89^{+0.06}_{-0.05}$ & $0.76^{+0.08}_{-0.07}$ & $1.52^{+0.16}_{-0.14}$ & $0.80^{+0.05}_{-0.05}$ & $1.19^{+0.10}_{-0.08}$ \\
$\mu_{pmDec}$ (masyr$^{-1}$) & $-11.03^{+0.06}_{-0.06}$ & $-9.54^{+0.10}_{-0.10}$ & $-11.53^{+0.11}_{-0.11}$ & $-9.05^{+0.15}_{-0.16}$ & $-16.58^{+0.33}_{-0.34}$ & $-6.88^{+0.11}_{-0.11}$ & $-10.37^{+0.18}_{-0.18}$ \\
$\sigma_{pmDec}$ (kms$^{-1}$) & $1.22^{+0.08}_{-0.07}$ & $0.37^{+0.04}_{-0.03}$ & $0.68^{+0.04}_{-0.04}$ & $0.56^{+0.06}_{-0.05}$ & $1.85^{+0.21}_{-0.17}$ & $1.23^{+0.08}_{-0.07}$ & $1.07^{+0.09}_{-0.08}$ \\
$\mu_{RV}$ (kms$^{-1}$) & $-3.22^{+0.06}_{-0.05}$ & $-3.54^{+0.06}_{-0.06}$ & $-6.12^{+0.05}_{-0.06}$ & $-6.49^{+0.07}_{-0.07}$ & $-3.04^{+0.06}_{-0.06}$ & $-6.76^{+0.05}_{-0.07}$ & $-4.45^{+0.08}_{-0.06}$ \\
$\sigma_{RV}$ (kms$^{-1}$) & $0.93^{+0.05}_{-0.09}$ & $0.73^{+0.05}_{-0.11}$ & $0.90^{+0.04}_{-0.09}$ & $0.77^{+0.06}_{-0.11}$ & $0.98^{+0.06}_{-0.08}$ & $1.15^{+0.04}_{-0.08}$ & $1.55^{+0.06}_{-0.10}$ \\
$\sigma_{3D}$ (kms$^{-1}$) & $1.61^{+0.09}_{-0.11}$ & $0.99^{+0.08}_{-0.12}$ & $1.44^{+0.08}_{-0.11}$ & $1.21^{+0.11}_{-0.14}$ & $2.59^{+0.26}_{-0.23}$ & $1.86^{+0.10}_{-0.12}$ & $2.22^{+0.13}_{-0.15}$ \\
Virial Mass ($M\odot$) & $2150^{+250}_{-290}$ & $310^{+50}_{-70}$ & $1370^{+150}_{-210}$ & $345^{+63}_{-76}$ & $5600^{+1100}_{-1000}$ & $1020^{+110}_{-130}$ & $4050^{+490}_{-550}$ \\
Stellar Mass ($M\odot$) & $480\pm16$ & $150\pm9$ & $310\pm13$ & $110\pm8$ & $330\pm12$ & $480\pm19$ & $560\pm19$ \\
$\tau_\text{MST}$ (Myr) & $0.4_{-0.6}^{+0.5}$ & $-0.4_{-0.5}^{+0.4}$ & $-2.3_{-0.5}^{+0.4}$ & $0.2_{-0.3}^{+0.3}$ & $0.1_{-0.5}^{+0.5}$ & $-0.7_{-0.3}^{+0.4}$ & $-1.9_{-0.3}^{+0.3}$  \\
$\tau_\text{DSum}$ (Myr) & $1.8_{-0.4}^{+0.2}$ & $-1.0_{-0.4}^{+0.4}$ & $-2.3_{-0.3}^{+0.3}$ & $0.1_{-0.3}^{+0.3}$ & $-4.2_{-0.8}^{+0.8}$ & $-0.5_{-0.2}^{+0.2}$ & $-2.3_{-0.4}^{+0.5}$  \\
$r_{50,\tau_\text{MST}}$ (pc) & $10.5_{-0.1}^{+0.2}$ & $3.3_{-0.1}^{+0.1}$ & $4.2_{-0.1}^{+0.1}$ & $2.6_{-0.1}^{+0.1}$ & $9.8_{-0.3}^{+0.2}$ & $3.0_{-0.1}^{+0.1}$ & $8.4_{-0.1}^{+0.1}$  \\
$r_{90,\tau_\text{MST}}$ (pc) & $20.3_{-0.2}^{+0.4}$ & $6.8_{-0.1}^{+0.1}$ & $7.3_{-0.1}^{+0.1}$ & $4.7_{-0.1}^{+0.1}$ & $24.9_{-0.4}^{+0.5}$ & $5.7_{-0.1}^{+0.2}$ & $12.8_{-0.1}^{+0.1}$  \\
$r_{50,\tau_\text{DSum}}$ (pc) & $10.4_{-0.1}^{+0.1}$ & $3.2_{-0.1}^{+0.1}$ & $4.2_{-0.1}^{+0.1}$ & $2.6_{-0.1}^{+0.1}$ & $8.3_{-0.1}^{+0.2}$ & $3.0_{-0.1}^{+0.1}$ & $8.4_{-0.1}^{+0.1}$  \\
$r_{90,\tau_\text{DSum}}$ (pc) & $20.6_{-0.1}^{+0.2}$ & $6.8_{-0.1}^{+0.1}$ & $7.3_{-0.1}^{+0.1}$ & $4.7_{-0.1}^{+0.1}$ & $22.1_{-0.2}^{+0.5}$ & $5.7_{-0.1}^{+0.1}$ & $12.8_{-0.1}^{+0.1}$  \\

\hline
\end{tabular}}
\end{center}
\setlength{\belowcaptionskip}{-10pt}
\setlength{\textfloatsep}{0pt}
\caption{Properties of the Upper Scorpius subgroups based on our sample of confirmed YSOs with 6D kinematic data.}
\label{groups_table}
\end{table*}

The $\beta$ Sco (blue) and $\nu$ Sco (purple) subgroups appear to be the most spatially distinct, having little to no overlap with other subgroups in YZ space, though $\nu$ Sco overlaps with $\delta$ Sco (orange) in XZ space. However, there is much greater overlap between these subgroups and others in velocity space. There is also an apparent gap between the two subgroups, particularly noticeable in XZ space, with an approximate separation of $\sim$12 pc. Around these subgroups are distributed a number of YSOs with kinematics not closely matching any group, which may be candidate ejected stars with high velocities relative to the bulk of Upper Sco, or else may be sparse fringes of the $\beta$ Sco and $\nu$ Sco subgroups. 

$\rho$ Oph (pink) is the most densely concentrated subgroup spatially, but then appears to be one of the sparser subgroups in velocity space. It has significant spatial overlap with $\alpha$ Sco (green) and some with $\pi$ Sco (yellow), but these groups are then the most dissimilar to it in velocity space. 

$\sigma$ Sco (red) is perhaps the most sparsely distributed subgroup in both spatial and velocity spaces. Spatially the bulk of $\sigma$ Sco is located behind $\delta$ Sco in the line of sight, but its fringes also overlap with $\alpha$, $\beta$, $\pi$ and $\nu$ Sco. Only $\rho$ Oph does not have any members of $\sigma$ Sco overlapping in 3D.

In Fig.~\ref{Cartesian_groups_position-velocity} we show the Cartesian spatial - velocity coordinate pairs in each panel (X-U: \textit{top}, Y-V: \textit{middle}, Z-W: \textit{bottom}). In X-U and Y-V there are visible positive correlations between spatial and velocity coordinates across multiple subgroups, which suggests large-scale expansion across Upper Sco. Again, $\pi$ Sco (yellow) is a notable exception to these trends, with its members located well outside of the multi-subgroup linear trends. 

Notably, in Z-W space (Fig.~\ref{Cartesian_groups_position-velocity}, \textit{bottom}) $\rho$ Oph (pink) is clearly separated from the majority of other subgroups, overlapping with only a dozen members of $\sigma$ Sco and $\alpha$ Sco.




\section{Kinematics}
\label{section_kinematics}

In Fig.~\ref{Cartesian_groups_vectors} we show the Cartesian spatial coordinates of confirmed YSOs with quality 6D kinematic information, colour-coded according to which subgroup they have been allocated to, with vectors indicating their Cartesian velocities (in km~s$^{-1}$) relative to the sample mean. 

There is significant variation in the orientation of vectors for members of the sparse subgroups $\delta$ Sco (orange) and $\sigma$ Sco (red), which is a result of their large scatter in velocity space (as can be seen in Fig.~\ref{Cartesian_groups} \textit{bottom panels}) as well as proximity to the sample mean velocity. 

The more spatially concentrated subgroups $\beta$ Sco (blue), $\nu$ Sco (purple) and $\rho$ Oph (pink) tend to have less variation in the orientation of their velocity vectors, as they have less scatter in velocity space. $\rho$ Oph as a whole is clearly moving away from the center of Upper Sco, while $\beta$ and $\nu$ Sco appear to be moving primarily away from each other rather than from the center of Upper Sco. 

It can also be seen in Fig.~\ref{Cartesian_groups_vectors} that the majority of members of $\alpha$ Sco (green) and $\pi$ Sco (yellow) have vectors directed toward the center of Upper Sco, defying the outward trend followed by the other subgroups, which is also evidenced by the lack of positive correlation in Cartesian position - velocity space as shown in Fig.~\ref{Cartesian_groups_position-velocity} (particularly in Z). This indicates that $\alpha$ Sco and $\pi$ Sco likely have distinct origins from the other Upper Sco subgroups.

In the following analysis we explore the kinematics of individual subgroups in greater detail. 

\subsection{Velocity dispersions}
\label{Velocity dispersions}

\begin{figure} 
    \subfloat{{\includegraphics[width=240pt]{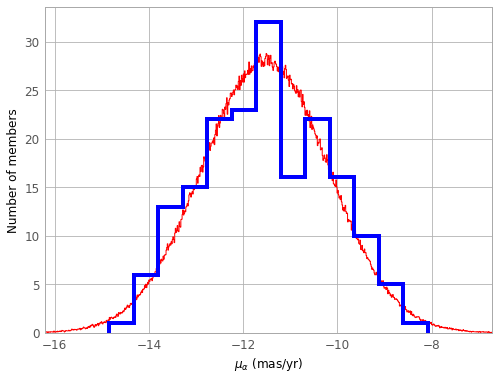}}}%
    \qquad
    \subfloat{{\includegraphics[width=240pt]{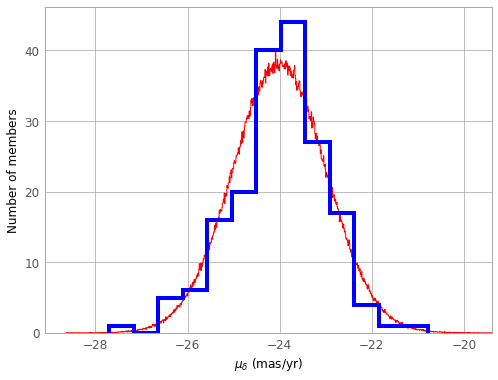}}}%
    \qquad
    \subfloat{{\includegraphics[width=240pt]{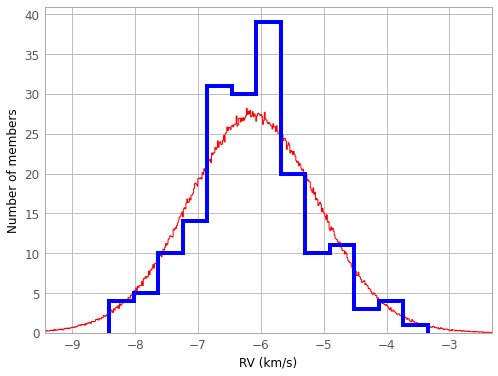}}}%
    \setlength{\belowcaptionskip}{-10pt}
    \setlength{\textfloatsep}{0pt}
    \caption{\textit{Top:} Histogram of proper motion in RA for confirmed YSO members of $\delta$ Sco (blue), with MCMC best-fit Gaussian distribution overplotted (red). \textit{Middle:} Histogram of proper motion in Dec for confirmed YSO members of $\delta$ Sco (blue), with MCMC best-fit Gaussian distribution overplotted (red). \textit{Bottom:} Histogram of RVs for confirmed YSO members of $\delta$ Sco (blue), with MCMC best-fit Gaussian distribution overplotted (red).}
    \label{Vel_dispersion_histograms}%
\end{figure}

Velocity dispersions provide crucial information related to the dynamical state of star clusters and associations. We estimate the velocity dispersions for each group in our sample using a Bayesian approach similar to that described in \citet{Armstrong22}. Where we model the velocity distributions as 3-dimensional Gaussians with the central velocity ($\mu$) and velocity dispersion ($\sigma$) in each proper motion and RV as free parameters. We add uncertainties randomly sampled from the observed uncertainty distribution in each dimension for each star.

As in \citet{Armstrong22}, we account for the possible inflation of our RV dispersion due to unresolved binaries by adding velocity offsets to our modeled RVs for 46\% of our modeled stars, according to the expected binary fraction \citep{raghavan10}. We create a synthetic population of binaries with primary star masses in the range of our observed sample, taken to be 0.65 - 0.1 $M_{\odot}$, following a \citet{maschberger13} IMF, with secondary star masses randomly sampled with uniform probability from between 0.1 -1.0 of their primaries' mass. We sample orbital periods from a log-normal distribution with mean period $log_{10}(5.03)$ and dispersion $log_{10}(2.28)$ days \citep{raghavan10} and we sample eccentricities from a flat distribution between a minimum of $e=0$ to a maximum scaled with orbital period \citep{parker09}. We calculate velocities along the line of sight for each star at random points in their binary orbits at random inclinations in 3D and then weight the velocities of each star per binary by their luminosities. 

We do not consider triple systems for the same reasons as given in \citet{Armstrong22}, i.e., their properties are not well constrained and that the RV contribution of third stars is likely to be negligible.

The posterior distribution function is sampled using \textit{emcee} (Section~\ref{cartesian}) and using an unbinned maximum likelihood test. We use uniform priors  of $-100$ to $+100$ kms$^{-1}$ for central velocities and 0 to 100 kms$^{-1}$ for velocity dispersions in each direction. We perform 2000 iterations with 1000 walkers and take the median, 16th and 84th percentile values as the best fit and 1 $\sigma$ uncertainties respectively, after discarding the first 1000 iterations as burn-in.

In Table~\ref{groups_table} we list the best fit central velocities $\mu$ and velocity dispersions $\sigma$ for all the seven subgroups of Upper Sco, based on the 544 YSO members with quality filtered 6D kinematic information. Figure~\ref{Vel_dispersion_histograms} shows the 3D velocity distributions for stars in $\delta$ Sco as an example, with the best-fitting velocity dispersion models overplotted in red. We measure the significance of anisotropy by dividing the difference between the largest velocity dispersion and the smallest to their combined uncertainties for each subgroup. The best fitting velocity dispersions for most subgroups are significantly anisotropic, with a confidence of $9\sigma$ for $\alpha$ Sco, $3\sigma$ for $\beta$ Sco, $3\sigma$ for $\delta$ Sco, $2\sigma$ for $\nu$ Sco, $4\sigma$ for $\pi$ Sco, $4\sigma$ for $\rho$ Oph, $3\sigma$ for $\sigma$ Sco. 

OB associations are typically sparse and unbound and thus are believed to be relatively dynamically un-evolved \citep{wright20}, i.e., they retain their initial velocity substructure, which is indicated by velocity anisotropy. The evidence of anisotropy in these subgroups indicates that they have not undergone sufficient dynamical mixing to develop isotropy and thus likely did not form as compact clusters.


\subsection{Virial mass}
\label{Virial mass}
In their analysis, \citet{posch24} estimate the masses of groups by multiplying the total number of likely members with the peak mass ($0.42 M_\odot$) of a \citet{kroupa01b} IMF. Following this approach we estimate total masses of subgroups by taking the number of members of each from \citet{miret-roig22}, apply a scaling factor which is the ratio of total YSOs per subgroup in our sample to new YSOs (those previously unallocated to subgroups by \citealt{miret-roig22}) and multiply by the peak mass. We approximate the uncertainty with a Poisson error. The approximate masses of subgroups are $\sim 480\pm16 M_\odot$ for $\alpha$ Sco, $\sim 150\pm9 M_\odot$ for $\beta$ Sco, $\sim 310\pm13 M_\odot$ for $\delta$ Sco, $\sim 110\pm8 M_\odot$ for $\nu$ Sco, $\sim 330\pm12 M_\odot$ for $\pi$ Sco, $\sim 480\pm19 M_\odot$ for $\rho$ Oph and $\sim 560\pm19 M_\odot$ for $\sigma$ Sco (Table.~\ref{groups_table}). This then implies a total mass of $\sim 2420\pm96 M_\odot$ for Upper Sco, which is reasonable agreement with previous estimates \citep[$\sim 2060 M_\odot$][]{preibisch08}.

We combine our observed velocity dispersions to obtain 3D velocity dispersions ($\sigma_{3D}$) and using these we estimate virial masses for each subgroup according to
\begin{equation}
    M_{\rm vir} = \eta\frac{\sigma_{3D}^2r_{50}}{3G},
\end{equation}
where $\eta = 10$ \citep{portegieszwart10}, and $r_{50}$ is the smallest radius containing half of a subgroup’s member YSOs. These values are given in Table~\ref{groups_table}.

Our estimated virial masses for subgroups are all significantly larger than those estimated using the IMF, in some cases by an order of magnitude, which is evidence that these groups are gravitationally unbound. The lowest virial masses we obtain are for $\beta$ Sco and $\nu$ Sco, the youngest subgroups after $\rho$ Oph \citep{ratzenbock23}, but even then their virial masses are twice and three times larger than their IMF-based mass estimates respectively. For subgroups with many candidate YSO members outside of the area covered by our observations (see Fig.~\ref{MiretRoigFields}), the half-mass radius $r_{50}$ is likely underestimated, and thus so is our virial mass estimate.

The $\rho$ Oph cluster in particular has previously been considered gravitationally bound \citep{rigliaco16,miret-roig22,wright24} despite the large velocity dispersion among its member YSOs. In fact, our velocity dispersion in RVs for $\rho$ Oph is in excellent agreement with the dispersion of \citet{rigliaco16} of $1.14 \pm 0.35$ kms$^{-1}$. This is due to the binding mass of gas still in the region, which has been estimated to be $\sim1750$ M$_{\odot}$ \citep{loren89}, which well exceeds the virial mass we estimate.




\subsection{Expansion velocity}
\label{expansion_velocity}

We calculate expansion velocities $v_{\rm out}$ (km~s$^{-1}$) similarly to \citet{armstrong24}, which are the components of velocity for each subgroup members directed away from the subgroup center in 3D. The median $v_{\rm out}$ ($\bar{v}_{out}$) and uncertainties for each subgroup are reported in Table~\ref{groups_table}. A significantly positive $\bar{v}_{out}$ indicates overall expansion of the subgroup, while a significantly negative $\bar{v}_{out}$ indicates overall contraction \citep{kuhn19,wright24}. We also count the number of subgroup members with positive individual expansion velocities and report the percentage per subgroup in Table~\ref{groups_table}.

For all subgroups we find that $\bar{v}_{out}$ is positive, though with varying significance. Also, all subgroups have at least $59\%$ of their members with 6D kinematic information moving away from the subgroup center in 3D. The highest and also most significant $\bar{v}_{out}$ belongs to $\delta$ Sco ($\bar{v}_{out} = 0.46^{+0.05}_{-0.05}$ km~s$^{-1}$), which is also the subgroup with the highest proportion of its members moving away from the subgroup center ($83\%$). This is a reasonable value of $\bar{v}_{out}$ for a young cluster, falling near the median of values found by \citet{kuhn19} and \citet{wright24}.  

Interestingly the $\rho$ Oph cluster, which has previously been considered gravitationally bound \citep{rigliaco16,miret-roig22}, has both a positive $\bar{v}_{out}$ of $>3.5\sigma$ significance and $72\%$ of its members have positive individual $v_{\rm out}$, which is evidence that this group is expanding despite the binding mass of gas in the region. \citet{wright24} calculated an expansion velocity of $\bar{v}_{out} = -0.34^{+0.01}_{-0.12}$ km~s$^{-1}$ for $\rho$ Oph, implying contraction, though this was based on a sample of 38 member YSOs identified in the Gaia-ESO survey \citep{gilmore22,randich22} within a $\sim1^{\circ}$ diameter area centered on (RA=246.0$^\circ$, Dec=-23.8$^\circ$), whereas we have a sample of 162 YSOs with 6D kinematic information which are distributed across $\sim 4^{\circ}$. This discrepancy may indicate that there is a contracting core of YSOs in the center of $\rho$ Oph, surrounded by an expanding halo.

However, simulations of young clusters \citep[e.g.,][]{sills18,kuhn19} have indicated that bound clusters may still exhibit significantly positive median $v_{\rm out}$ up to $\sim0.5$ kms$^{-1}$ as they rebound from the initial collapse of hierarchical cluster assembly after $\sim1$ Myr, until reaching an equilibrium at $\sim6$ Myr. The $\rho$ Oph cluster is within the age range \citep[3-4 Myr; ][]{ratzenbock23} when this rebound would be expected to happen, though it should be noted that \citet{sills18} simulated much more massive clusters ($>1500$ M$_\odot$) than $\rho$ Oph is estimated to be ($\sim 480 M_\odot$), which become nearly spherical after $\sim1$ Myr following hierarchical cluster assembly, so it is unclear to what degree these simulations might represent the dynamical evolution of a sparser, low-mass cluster like $\rho$ Oph. 

\subsection{Linear Expansion Trends}\label{Expansion}
Linear expansion is a key indicator that a stellar association is unbound and in the process of dispersing into the Galactic field. The precise rates of expansion and potential anisotropy can also be useful in constraining an association's initial configuration and dynamical history.

\begin{figure*}
\begin{subfigure}{0.49\textwidth}
    \includegraphics[width=240pt]{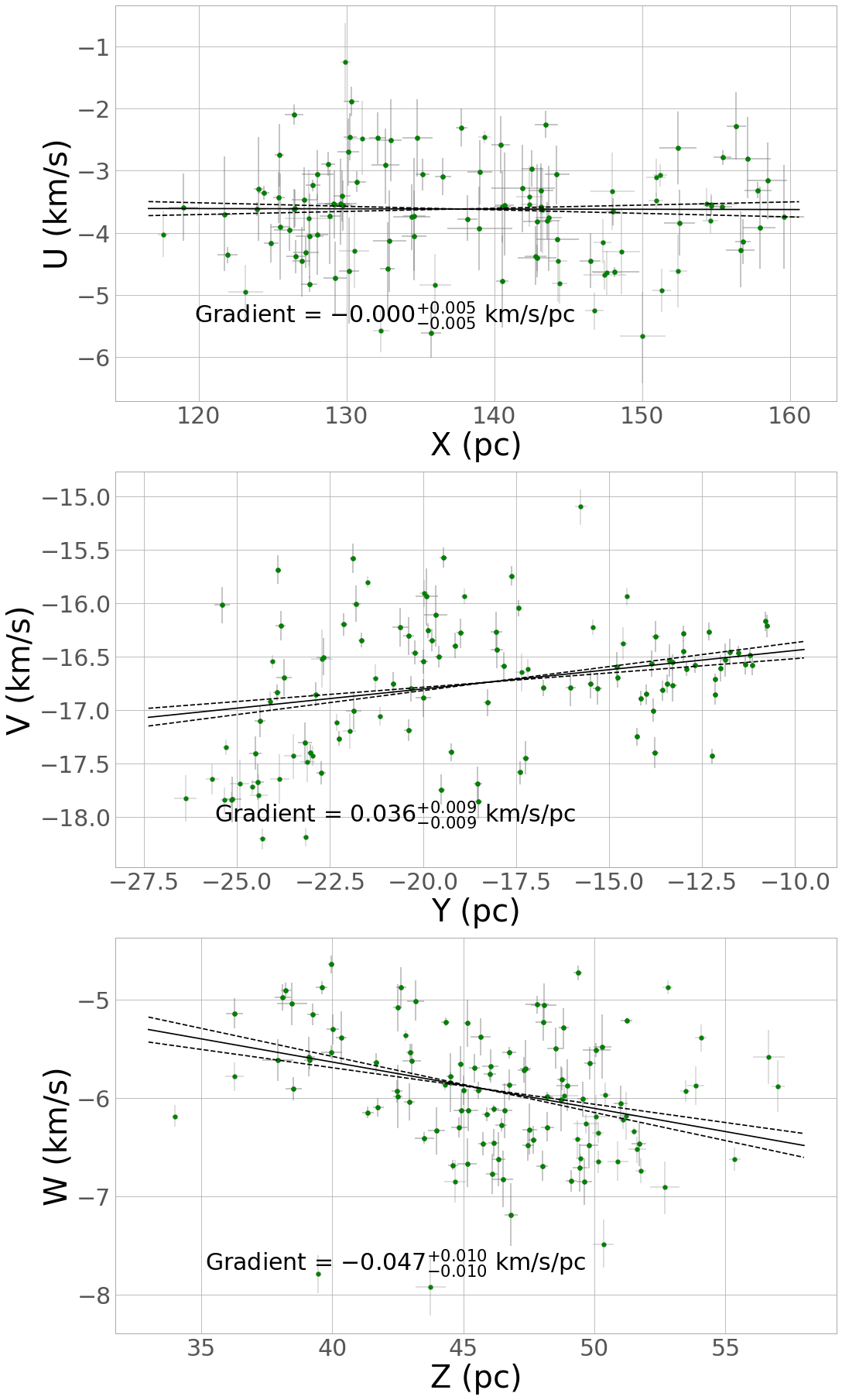}
\end{subfigure}
\begin{subfigure}{0.49\textwidth}
    \includegraphics[width=240pt]{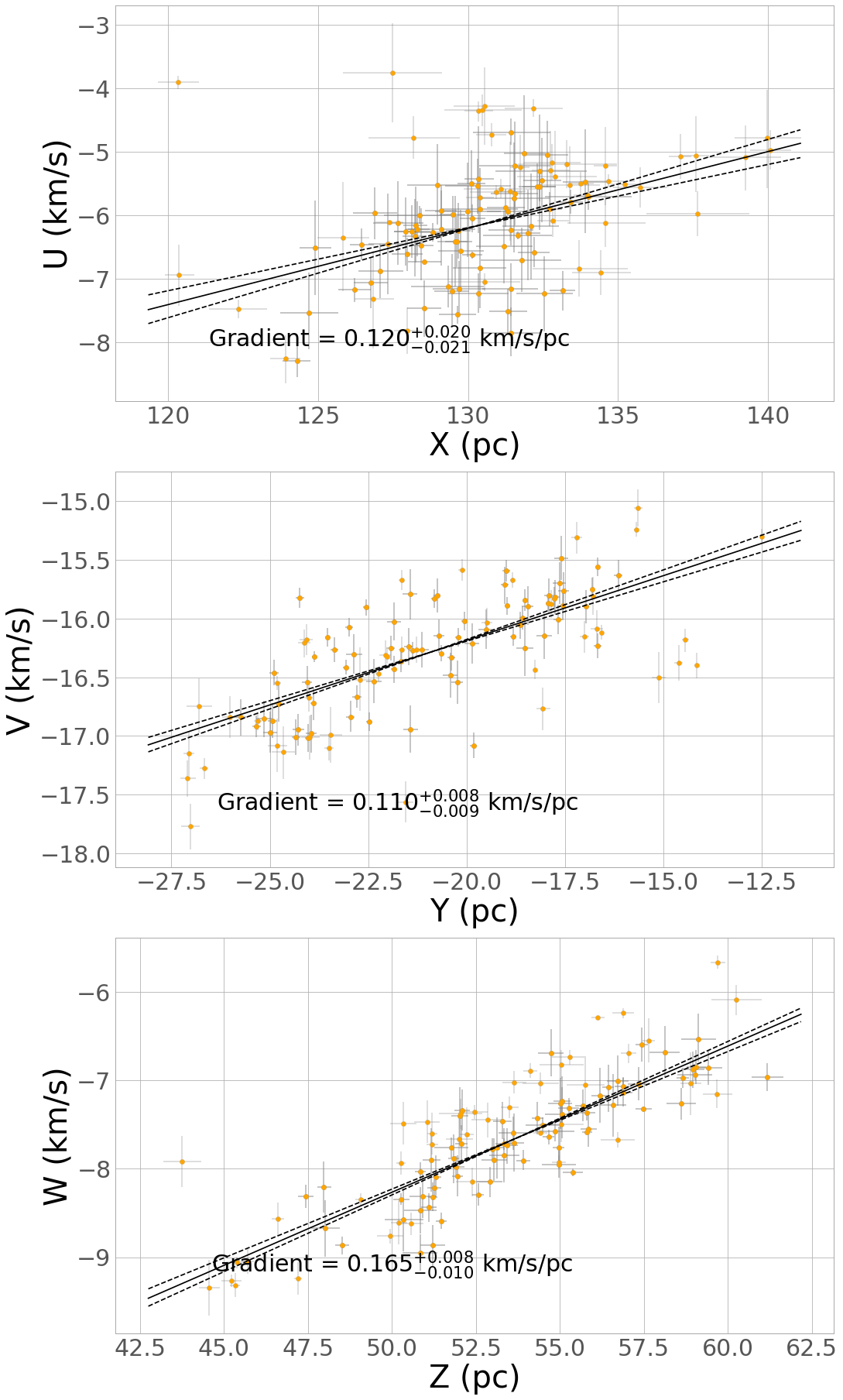}
\end{subfigure}

\caption{Cartesian velocity versus position in each of the three dimensions $XYZ$ for stars in $\alpha$ Sco (green) and $\delta$ Sco (orange) with uncertainties shown. The best-fitting gradients and the 16th and 84th percentiles values of the fit are shown as solid and dashed lines respectively in each panel.}
\label{figure_expansion}
\end{figure*}

We investigate evidence for expansion among the subgroups of Upper Sco by looking for linear correlations between Cartesian velocity and position in each direction, where positive or negative correlations will indicate either expansion or contraction, respectively. The best fitting parameters for the linear correlations, the gradient, intersection and the fractional underestimation of the variance ($m$, $b$, $f$), are determined following the same approach as described in \citet{Armstrong22}, using Bayesian inference, with $>3\sigma$ outliers in position and velocity removed. We also vary the positions of YSOs by randomly sampling from their uncertainties. We perform 2000 iterations with 200 walkers, half of which are discarded as burn in, and report the medians,16th and 84th percentiles from the posterior distribution as the linear best fit gradient and 1$\sigma$ uncertainties respectively. Our results are listed in Table~\ref{expansion_table}. In Fig.~\ref{figure_expansion} we plot linear expansion trends for $\alpha$ Sco (green) and $\delta$ Sco (orange) as examples.

All groups show expansion trends of at least 3$\sigma$ significance in at least one direction and $\beta$ Sco, $\delta$ Sco \& $\sigma$ Sco show trends of at least 3$\sigma$ significance in all directions, but the direction of greatest expansion is not the same for all groups. $\alpha$ Sco, $\beta$ Sco \& $\sigma$ Sco show the greatest rate of expansion in the direction of Galactic rotation Y, whereas $\delta$ Sco, $\nu$ Sco, $\pi$ Sco \& $\rho$ Oph show the greatest rate of expansion in the direction perpendicular to the Galactic plane Z. 

It is worth noting that the weak expansion trends in X may be due in part to the fact that this direction coincides most closely to the line-of-sight from the Sun, thus the errors on the X coordinate contain the largest component of parallax error, and the U velocity the largest component of RV error, including unresolved binarity,  which remain the greatest sources of uncertainty in our 6D kinematic information.

In $\alpha$ Sco, $\delta$ Sco \& $\sigma$ Sco the rates of expansion are significantly anisotropic (at least 4$\sigma$). This anisotropy may indicate a level of kinematic substructure in these groups that has survived dynamical mixing. Despite the relatively old ages of these groups \citep[$\sim10$ Myr;]{ratzenbock23}, they were likely formed too sparse to undergo significant mixing and will retain most of their substructure as they expand until they eventually disperse into the field, similar to what has been concluded for other nearby associations such as Vela OB2 \citep{wright20,Armstrong22}.

$\alpha$ Sco is the only subgroup to show evidence for contraction, at a rate of $-0.047^{+0.010}_{-0.010}$ km s$^{-1}$ pc$^{-1}$ in Z, the direction perpendicular to the Galactic plane.

We also fit linear gradients to the entire YSO sample, which are plotted in Fig.~\ref{figure_expansion_all}. We find significantly positive ($>6\sigma$) linear expansion rates in all directions, of $0.053^{+0.007}_{-0.007}$ km s$^{-1}$ in X versus U, $0.098^{+0.007}_{-0.008}$ km s$^{-1}$ in Y versus V and $0.054^{+0.008}_{-0.008}$ km s$^{-1}$ in Z versus W. This provides strong evidence that Upper Sco as a whole is unbound and expanding, with clear anisotropy ($>4\sigma$ difference in expansion rates between X and Y). However, the expansion rates of the whole Upper Sco sample are distinct from any of the internal rates for subgroups given in Table.~\ref{expansion_table}, which may indicate that the region-wide expansion trends are dominated by inter-subgroup kinematics rather than internal kinematics.  

\subsection{Expansion Timescales}\label{expansion_timescales}

We also derive expansion timescales in Myr by inverting the greatest expansion rates of each subgroup, and list these results in Table~\ref{expansion_table}. $\beta$ Sco, $\delta$ Sco, $\nu$ Sco, $\rho$ Oph \& $\sigma$ Sco all yield expansion timescales $< 10$ Myr, while $\alpha$ Sco \& $\pi$ Sco yield timescales of $28.4_{-5.7}^{+9.5}$ Myr and $16.0_{-3.4}^{+5.8}$ Myr, respectively, which are much larger than previous age estimates for these groups. On the other hand, the expansion timescale of $6.2_{-0.3}^{+0.4}$ Myr for $\delta$ Sco agrees reasonably well with the kinematic ages of 4-6 Myr given in \citet{miret-roig22}.

The expansion timescales of $7.8_{-0.9}^{+1.3}$ Myr, $12.2_{-5.7}^{+90.1}$ Myr, $6.0_{-1.1}^{+1.7}$ Myr \& $5.0_{-0.4}^{+0.5}$ Myr for $\beta$ Sco, $\nu$ Sco, $\rho$ Oph \& $\sigma$ Sco, respectively, are all significantly greater than the kinematic ages of 1-4 Myr, $>1$ Myr, $>1$ Myr and 1-3 Myr given in \citet{miret-roig22} for these groups. In particular, \citet{miret-roig22} found no evidence that $\rho$ Oph had been more compact in the past, and so estimated kinematic ages consistent with $0$. However, since timescales derived from expansion trends implicitly assume that the group of stars has been expanding from a point, they should be interpreted as upper limits. Therefore, there is no particular inconsistency between our expansion timescales and these literature kinematic ages.

However, considering expansion timescales as upper limit kinematic ages, it is then notable that both $\delta$ Sco and $\sigma$ Sco in particular have expansion timescales significantly lower than their isochronal age estimates, $9.8^{+1.2}_{-1.4}$ Myr and $10.0^{+1.0}_{-0.5}$ Myr from \citet{ratzenbock23}, respectively. Such a difference between kinematic and isochronal ages has been hypothesised by \citet{miret-roig24} to indicate a period in the early evolution of a cluster or association subgroup where the bulk of member YSOs remain gravitationally bound and possibly still embedded in their natal molecular cloud. The lower kinematic age then indicates when the subgroup becomes unbound and its members begin to disperse, possibly following residual gas expulsion. In the case of $\delta$ Sco and $\sigma$ Sco, our expansion timescales would imply embedded phases of $\sim3-5$ Myr, similar to the findings of \citet{miret-roig24}. However, the length of such a phase is highly uncertain as it depends on uncertainties in both kinematic and isochronal age methods. We discuss this further in Sect.~\ref{section_discussion}.

\subsection{Rotation}\label{rotation}

\begin{figure*}
\begin{subfigure}{0.49\textwidth}
    \includegraphics[width=240pt]{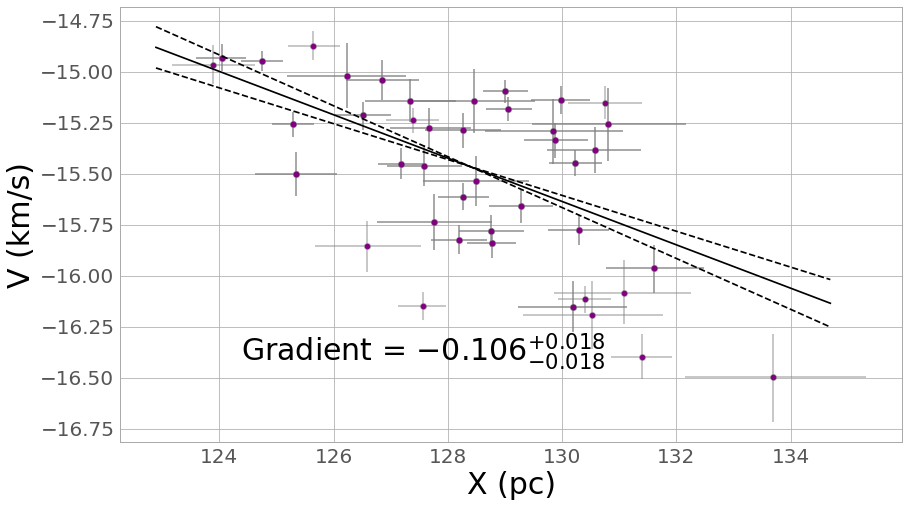}
\end{subfigure}
\begin{subfigure}{0.49\textwidth}
    \includegraphics[width=240pt]{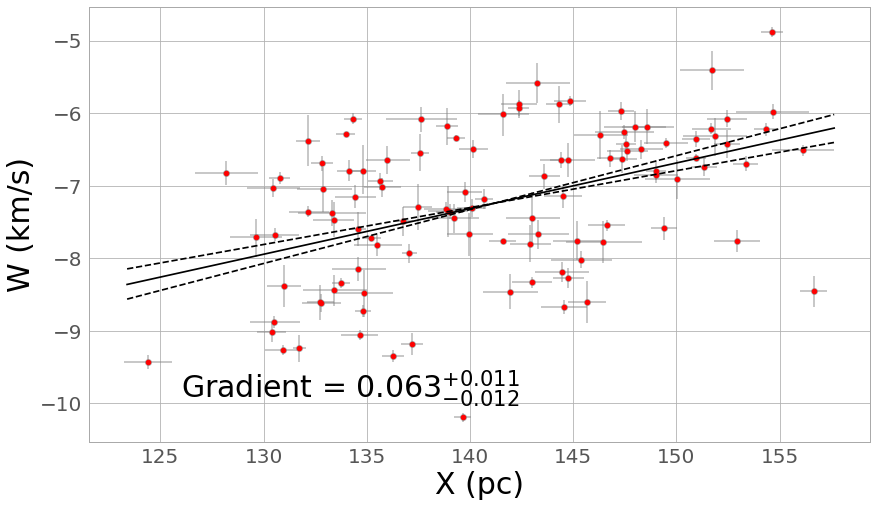}
\end{subfigure}

\caption{Cartesian V velocity versus X position for stars in $\nu$ Sco (purple) and W versus X for stars in $\sigma$ Sco (red) with uncertainties shown. The best-fitting gradients and the 16th and 84th percentiles values of the fit are shown as solid and dashed lines respectively in each panel.}
\label{figure_rotation}
\end{figure*}

We also investigate evidence for rotation in the subgroups of Upper Sco by fitting linear gradients to combinations of velocity and position in different directions. Note that since these subgroups are gravitationally unbound these gradients may not indicate true rotation, but rather residual angular momentum. Our results are listed in Table~\ref{table_rotation}. In Fig.~\ref{figure_rotation} we plot linear rotation trends for $\nu$ Sco (purple) in X versus V and $\sigma$ Sco (red) in X versus W as examples.

Rotation in young clusters has been hypothesised to arise from the merging of subclusters in the assembly phase of hierarchical cluster formation, the signature of which may even be stronger than that of cluster expansion in this scenario \citep[e.g.,][]{mapelli17}. It is therefore a potential way of discriminating between hierarchical and monolithic cluster formation scenarios.

In the case of rotation found in substructures within an extended population such as an OB association, it may be inherited from the turbulent motion of the natal gas from which the substructure forms.

Significant ($>3\sigma$) trends are found in at least one position - velocity pair in every subgroup. The subgroups with the greatest spatial extent ($\alpha$, $\delta$ and $\sigma$ Sco) exhibit significant rotation trends in multiple directions, though, apart from a rotation rate of $-0.132^{+0.024}_{-0.024}$ km~s$^{-1}$pc$^{-1}$ in W versus Y for $\sigma$ Sco, the rotation rates are relatively low in all directions in comparison to the expansion rates. These are among the older subgroups and so have had more time for expansion to suppress their rotation rates.

Meanwhile, $\rho$ Oph, which is the densest and possibly youngest subgroup, exhibits the greatest rotation rates of $-0.133^{+0.038}_{-0.037}$ kms$^{-1}$pc$^{-1}$ in U versus Z and of $-0.187^{+0.067}_{-0.062}$ kms$^{-1}$pc$^{-1}$ in U versus Y. This is expected given that $\rho$ Oph is still in a compact configuration and dynamically young, supporting the evidence from the asymmetry of its expansion trends (Sect.~\ref{Expansion}). That being said, these rates are still not significantly greater than the expansion rate in W versus Z, so there is not significant evidence for a hierarchical formation scenario for $\rho$ Oph.

\subsection{Traceback}\label{Traceback}

With precise kinematic information we can calculate the positions of stars in the past, and thus estimate when a cluster or association began expanding from its most initially compact configuration. This differs from a linear expansion age (Section \ref{expansion_timescales}) which, in effect, traces stars back to a point, and instead traces stars to an initial non-negligible volume. 

We calculate 3D positions as a function of time following the same approach as \citet{Armstrong22}, using the epicycle approximation and the orbital equations from \citet{fuchs06}, with the Oort $A$ and $B$ constants from \citet{feast97}, the local disc density from \citet{holmberg04}, the local standard of rest velocity from \citet{schonrich10} and a solar Z distance above the Galactic plane of 17 pc \citep{karim17}. 3D positions are calculated up to 10 Myrs in the past in 0.1 Myr steps. At each step we calculate the spatial coherence of the association using both the total length of the minimum spanning tree (MST) between YSO members of a subgroup \citep{squicciarini21}, which we denote as $\tau_\text{MSTlength}$, and the sum of 3D distances between YSOs and the median position of the subgroup \citep{quintana22}, which we denote as $\tau_\text{DSum}$. We estimate uncertainties on these using a Monte Carlo process with 1000 iterations, taking the 84th and 16th percentiles of the posterior distribution as the 1$\sigma$ uncertainties. We plot the resulting metrics as functions of traceback time $t$ in Fig.~\ref{MinAreaPlot} with different filters on the cluster members included, all members in red, 3$\sigma$ velocity outliers removed in green, 2$\sigma$ velocity outliers removed in blue, and the 32\% longest branches removed in black (Fig.~\ref{MinAreaPlot} \textit{left}). We also give the traceback times at which these metrics are minimised, with uncertainties, in Table~\ref{groups_table}.

For $\tau_\text{MSTlength}$ and $\tau_\text{DSum}$, negative values indicate that a subgroup would have been at its most compact in the past, while positive values indicate that a subgroup will be more compact in the future. Both $\alpha$ Sco and $\nu$ Sco, which exhibited evidence for contraction in Table~\ref{expansion_table}, have positive $\tau_\text{MSTlength}$ and $\tau_\text{DSum}$ values, and are thus likely to become more compact in the future, while the other groups are likely to have been more compact in the past. 

$\tau_\text{MSTlength}$ and $\tau_\text{DSum}$ estimates for each of $\beta$ Sco, $\delta$ Sco, $\rho$ Oph \& $\sigma$ Sco of $-0.4_{-0.5}^{+0.4}$ \& $-1.0_{-0.4}^{+0.5}$ Myr, $-2.3_{-0.5}^{+0.4}$ \& $-2.3_{-0.3}^{+0.3}$ Myr, $-0.7_{-0.3}^{+0.4}$ \& $-0.5_{-0.2}^{+0.2}$ Myr,  $-1.9_{-0.3}^{+0.3}$ \& $-2.3_{-0.4}^{+0.5}$ Myr, respectively, are consistent between size metrics. However, the $\tau_\text{MSTlength}$ and $\tau_\text{DSum}$ estimates for $\pi$ Sco of $0.1_{-0.5}^{+0.5}$ \& $-4.2_{-0.8}^{+0.8}$ Myr are significantly different, as are estimates for $\alpha$ Sco of $0.4_{-0.6}^{+0.5}$ \& $1.8_{-0.4}^{+0.2}$ Myr. This possibly indicates a certain amount of substructure remaining within these subgroups, making these different size metric inconsistent. Notably, these two subgroups are not recovered by the SIGMA algorithm \citep{ratzenbock23} in the same forms as by \citet{miret-roig22} with HDBSCAN, but their members are redistributed primarily among the Antares, $\rho$ Sco and US-foreground populations. 

\citet{miret-roig22} estimated dynamical traceback ages for these groups using a similar approach to ours, but used the determinant and trace of a covariance matrix consisting of radial, azimuthal and vertical sizes as the metrics minimised to determine the most compact configuration of groups in the past. According to the trace of the covariance matrix the subgroups have dynamical ages of, $1.0 \pm 1.2$ Myr ($\alpha$ Sco), $2.5 \pm 1.6$ Myr ($\beta$ Sco), $4.6 \pm 1.1$ Myr ($\delta$ Sco), $0.3 \pm 0.5$ Myr ($\nu$ Sco), $6.3 \pm 1.4$ Myr ($\pi$ Sco), $0.0 \pm 0.3$ Myr ($\rho$ Oph) and $2.1 \pm 0.7$ Myr ($\sigma$ Sco). While our dynamical age estimates $\tau_\text{MSTlength}$ and $\tau_\text{DSum}$ agree well with those of \citet{miret-roig22} for $\sigma$ Sco and within uncertainties for $\nu$ Sco, they are notably different for the majority of subgroups. For $\alpha$ Sco, $\beta$ Sco, $\delta$ Sco and $\pi$ Sco we obtain younger dynamical ages than \citet{miret-roig22} according to both of our size metrics, while for $\rho$ Oph we find larger dynamical ages according to both metrics.

The most likely causes of this difference in dynamical ages are the difference in samples of subgroup members and RVs available to enable dynamical traceback in 3D. The fact that we obtain younger dynamical ages for the majority of subgroups indicates that they may have become gravitationally unbound more recently than previously suggested. 

\subsubsection{Initial size}\label{size}

We also estimate the initial size of each subgroup via several metrics. We calculate half-mass radii and the radii containing $90\%$ of subgroup members at the times when they were most compact, $r_{50}(\tau_0)$ and $r_{90}(\tau_0)$, according to the dynamical ages $\tau_\text{MSTlength}$ and $\tau_\text{DSum}$. These radii are given in Table~\ref{groups_table}. It should be noted that since $\alpha$ and $\nu$ Sco have positive dynamical ages, we don't expect their initial configurations to be significantly more compact than the present day distributions.   


Cometary or pillar shaped cloud structures within which protostars are thought to form from triggering via feedback, such as the ``Pillars of Creation'' in the Eagle Nebula, have been observed to have typical diameters $\sim1$ pc \citep[e.g.,][]{miao06}. If the subgroups of Upper Sco had formed in such structures, as has been proposed in the ``cluster-chain'' triggered formation scenario \citep{posch24}, we would expect their initial sizes to be similar. However, we find that even the subgroups which become most compact at their initial configurations have half-mass radii $r_{50}(\tau_0)>2.5$ pc and $90\%$ radii $r_{90}(\tau_0)>4.5$ pc, an order of magnitude larger than such pillar structures. 

However, this estimate of initial size will be inflated by scatter due to observational uncertainties in parallax, proper motions and RVs which are propagated through the traceback calculation. We estimate the inflation of the radii $r_{50}(\tau_0)$ and $r_{90}(\tau_0)$ due to uncertainties by performing numerical simulations, where we create a synthetic population of $n$ stars equal to the number of confirmed YSOs with 6D kinematics per subgroup in our data, and place them at the origin of a 3D Cartesian space. We then add perturbations in 3D positions and velocities randomly sampled from the observed position and velocity uncertainties to the synthetic population of stars and calculate their radial distances from the origin at time $t$, equal to the larger of the dynamical traceback ages $\tau_\text{MSTlength}$ or $\tau_\text{DSum}$ for a given subgroup. We then calculate the radii $r_{50}(\tau_0)$ and $r_{90}(\tau_0)$ of the synthetic populations at time $t$. We perform 10 000 iterations of this simulation for each subgroup and take the median of the posterior distributions of $r_{50}(\tau_0)$ and $r_{90}(\tau_0)$ as the predicted inflation of initial size estimates, and the 16th and 84th percentiles values as the $1\sigma$ uncertainties. 

For the subgroups hypothesised to belong to the ``cluster-chains'' described by \citep{posch24} which we infer to have been more compact in the past via our traceback calculations ($\beta$ Sco, $\delta$ Sco, $\rho$ Oph and $\sigma$ Sco), we estimate that $r_{50}(\tau_0)$ for $\beta$ Sco is inflated by $1.26^{+0.10}_{-0.10}$ pc, $r_{90}(\tau_0)$ for $\beta$ Sco is inflated by $2.47^{+0.21}_{-0.20}$ pc, $r_{50}(\tau_0)$ for $\delta$ Sco is inflated by $0.42^{+0.02}_{-0.02}$ pc, $r_{90}(\tau_0)$ for $\delta$ Sco is inflated by $0.80^{+0.04}_{-0.04}$ pc, $r_{50}(\tau_0)$ for $\rho$ Oph is inflated by $0.56^{+0.04}_{-0.04}$ pc, $r_{90}(\tau_0)$ for $\rho$ Oph is inflated by $1.22^{+0.09}_{-0.08}$ pc, $r_{50}(\tau_0)$ for $\sigma$ Sco is inflated by  $2.03^{+0.15}_{-0.14}$ pc and $r_{90}(\tau_0)$ for $\sigma$ Sco is inflated by $4.04^{+0.31}_{-0.29}$ pc. 

In each of these cases, the estimated initial sizes of these subgroups can not be accounted for by scatter due to observational uncertainties alone. Subtracting the estimates of inflation from the estimated radii for these subgroups still results in $r_{50}(\tau_0)>2.0$ pc and $r_{90}(\tau_0)>4.0$ pc. 

We acknowledge the caveat that we are not accounting for unresolved binaries in our sample when making these estimates, which we would expect to contribute to scatter in the RVs especially. With further information about binarity among YSOs, more complete coverage of the region with spectroscopic observations for RVs and youth identification, and with greater precision of parallaxes and proper motions with the upcoming Gaia DR4, improved estimates of the initial size of subgroups in Upper Sco could provide strong constraints on the likely star formation scenario of the region.

\begin{table}
\begin{center}
{\renewcommand{\arraystretch}{1.5}
\begin{tabular}{|p{0.2cm}|p{1.2cm}|p{1.8cm}|p{0.4cm}|p{1.2cm}|p{0.8cm}|}
\hline
 & Velocity - Position pair & Gradient ($kms^{-1}pc^{-1}$) & Sig. & Timescale (Myr) & Aniso. \\
\hline
$\alpha$ & $U$  $X$ & $ -0.000_{-0.005}^{+0.005}$ & - & & \\
         & $V$  $Y$ & $ 0.036_{-0.009}^{+0.009}$ & 4$\sigma$ & $28.4_{-5.7}^{+9.8}$ & 6$\sigma$ \\
         & $W$  $Z$ & $ -0.047_{-0.010}^{+0.010}$ & 4$\sigma$ & & \\
\hline
$\beta$ & $U$  $X$ & $ 0.105_{-0.028}^{+0.026}$ & 3$\sigma$ & & \\
         & $V$  $Y$ & $ 0.131_{-0.019}^{+0.018}$ & 6$\sigma$ & $7.8_{-0.9}^{+1.3}$ & 1$\sigma$ \\
         & $W$  $Z$ & $ 0.099_{-0.023}^{+0.023}$ & 4$\sigma$ & &  \\
\hline
$\delta$ & $U$  $X$ & $ 0.120_{-0.021}^{+0.020}$ & 5$\sigma$ & & \\
         & $V$  $Y$ & $ 0.110_{-0.009}^{+0.008}$ & 12$\sigma$ & & \\
         & $W$  $Z$ & $ 0.165_{-0.010}^{+0.008}$ & 16$\sigma$ & $6.2_{-0.3}^{+0.4}$ & 4$\sigma$ \\
\hline
$\nu$ & $U$  $X$ & $ -0.079_{-0.044}^{+0.045}$ & 1$\sigma$ & & \\
         & $V$  $Y$ & $ 0.075_{-0.025}^{+0.024}$ & 3$\sigma$ & & \\
         & $W$  $Z$ & $ 0.084_{-0.074}^{+0.073}$ & 1$\sigma$ &  $12.2_{-5.7}^{+90.1}$ & 1$\sigma$  \\
\hline
$\pi$ & $U$  $X$ & $ 0.033_{-0.009}^{+0.008}$ & 3$\sigma$ & & \\
         & $V$  $Y$ & $ 0.006_{-0.011}^{+0.011}$ & - & & \\
         & $W$  $Z$ & $ 0.064_{-0.017}^{+0.017}$ & 3$\sigma$ & $16.0_{-3.4}^{+5.8}$ & 2$\sigma$  \\
\hline
$\rho$ & $U$  $X$ & $ 0.040_{-0.036}^{+0.034}$ & 1$\sigma$ & & \\
         & $V$  $Y$ & $ 0.004_{-0.054}^{+0.055}$ & - & & \\
         & $W$  $Z$ & $ 0.172_{-0.039}^{+0.038}$ & 4$\sigma$ & $6.0_{-1.1}^{+1.7}$ & 2$\sigma$ \\
\hline
$\sigma$ & $U$  $X$ & $ 0.083_{-0.015}^{+0.014}$ & 5$\sigma$ & & \\
         & $V$  $Y$ & $ 0.206_{-0.017}^{+0.016}$ & 12$\sigma$ & $5.0_{-0.4}^{+0.5}$ & 5$\sigma$ \\
         & $W$  $Z$ & $ 0.073_{-0.020}^{+0.020}$ & 3$\sigma$ & & \\
\hline
\end{tabular}}
\end{center}
\setlength{\belowcaptionskip}{-10pt}
\setlength{\textfloatsep}{0pt}
\caption{Expansion gradients fitted for each group and in each dimension. Significance values listed are calculated from the ratio of the gradient to the uncertainty on the gradient, rounded down to the nearest integer.}
\label{expansion_table}
\end{table}

\begin{table}
\begin{center}
{\renewcommand{\arraystretch}{1.3}
\begin{tabular}{|p{0.6cm}|p{0.8cm}|p{0.8cm}|p{3.0cm}|p{0.8cm}| }
\hline
Group & Velocity & Position & Gradient (km/s/pc) & Signif. \\
\hline
$\alpha$ & $V$ & $X$ & $ -0.022_{-0.004}^{+0.004}$ & 5$\sigma$ \\
 & $W$ & $X$ & $ 0.019_{-0.004}^{+0.004}$ & 4$\sigma$ \\
 & $U$ & $Y$ & $ 0.009_{-0.012}^{+0.013}$ & - \\
 & $W$ & $Y$ & $ -0.037_{-0.009}^{+0.010}$ & 3$\sigma$ \\
 & $U$ & $Z$ & $ -0.031_{-0.012}^{+0.012}$ & 2$\sigma$ \\
 & $V$ & $Z$ & $ -0.053_{-0.009}^{+0.009}$ & 5$\sigma$ \\
 \hline
$\beta$ & $V$ & $X$ & $ -0.081_{-0.017}^{+0.017}$ & 4$\sigma$ \\
 & $W$ & $X$ & $ 0.005_{-0.021}^{+0.020}$ & - \\
 & $U$ & $Y$ & $ 0.001_{-0.033}^{+0.032}$ & - \\
 & $W$ & $Y$ & $ -0.046_{-0.023}^{+0.024}$ & 1$\sigma$ \\
 & $U$ & $Z$ & $ 0.005_{-0.032}^{+0.032}$ & - \\
 & $V$ & $Z$ & $ 0.009_{-0.024}^{+0.025}$ & - \\
 \hline
$\delta$ & $V$ & $X$ & $ -0.036_{-0.012}^{+0.012}$ & 3$\sigma$ \\
 & $W$ & $X$ & $ 0.014_{-0.017}^{+0.017}$ & - \\
 & $U$ & $Y$ & $ 0.049_{-0.023}^{+0.023}$ & 2$\sigma$ \\
 & $W$ & $Y$ & $ -0.044_{-0.017}^{+0.018}$ & 2$\sigma$ \\
 & $U$ & $Z$ & $ -0.004_{-0.018}^{+0.018}$ & - \\
 & $V$ & $Z$ & $ -0.047_{-0.010}^{+0.010}$ & 4$\sigma$ \\
 \hline
$\nu$ & $V$ & $X$ & $ -0.106_{-0.018}^{+0.018}$ & 5$\sigma$ \\
 & $W$ & $X$ & $ 0.015_{-0.047}^{+0.046}$ & - \\
 & $U$ & $Y$ & $ -0.017_{-0.053}^{+0.054}$ & - \\
 & $W$ & $Y$ & $ -0.043_{-0.053}^{+0.056}$ & - \\
 & $U$ & $Z$ & $ -0.087_{-0.076}^{+0.075}$ & 1$\sigma$ \\
 & $V$ & $Z$ & $ -0.008_{-0.037}^{+0.036}$ & - \\
 \hline
$\pi$ & $V$ & $X$ & $ 0.005_{-0.005}^{+0.005}$ & 1$\sigma$ \\
 & $W$ & $X$ & $ 0.012_{-0.006}^{+0.006}$ & 2$\sigma$ \\
 & $U$ & $Y$ & $ 0.007_{-0.022}^{+0.021}$ & - \\
 & $W$ & $Y$ & $ -0.054_{-0.014}^{+0.014}$ & 3$\sigma$ \\
 & $U$ & $Z$ & $ -0.008_{-0.024}^{+0.025}$ & - \\
 & $V$ & $Z$ & $ -0.006_{-0.013}^{+0.014}$ & - \\
 \hline
$\rho$ & $V$ & $X$ & $ 0.011_{-0.029}^{+0.028}$ & - \\
 & $W$ & $X$ & $ 0.073_{-0.035}^{+0.033}$ & 2$\sigma$ \\
 & $U$ & $Y$ & $ -0.187_{-0.062}^{+0.067}$ & 2$\sigma$ \\
 & $W$ & $Y$ & $ -0.010_{-0.064}^{+0.065}$ & - \\
 & $U$ & $Z$ & $ -0.133_{-0.037}^{+0.038}$ & 3$\sigma$ \\
 & $V$ & $Z$ & $ 0.040_{-0.031}^{+0.030}$ & 1$\sigma$ \\
 \hline
$\sigma$ & $V$ & $X$ & $ -0.060_{-0.012}^{+0.013}$ & 4$\sigma$ \\
 & $W$ & $X$ & $ 0.063_{-0.012}^{+0.011}$ & 5$\sigma$ \\
 & $U$ & $Y$ & $ -0.044_{-0.030}^{+0.029}$ & 1$\sigma$ \\
 & $W$ & $Y$ & $ -0.132_{-0.024}^{+0.024}$ & 5$\sigma$ \\
 & $U$ & $Z$ & $ 0.046_{-0.025}^{+0.025}$ & 1$\sigma$ \\
 & $V$ & $Z$ & $ -0.038_{-0.022}^{+0.022}$ & 1$\sigma$ \\

\hline
\end{tabular}}
\end{center}
\setlength{\belowcaptionskip}{-10pt}
\setlength{\textfloatsep}{0pt}
\caption{Rotation gradients for each group derived from fitting linear gradients between position and velocity in different dimensions.}
\label{table_rotation}
\end{table}

\begin{figure*}
	\includegraphics[width=\textwidth]{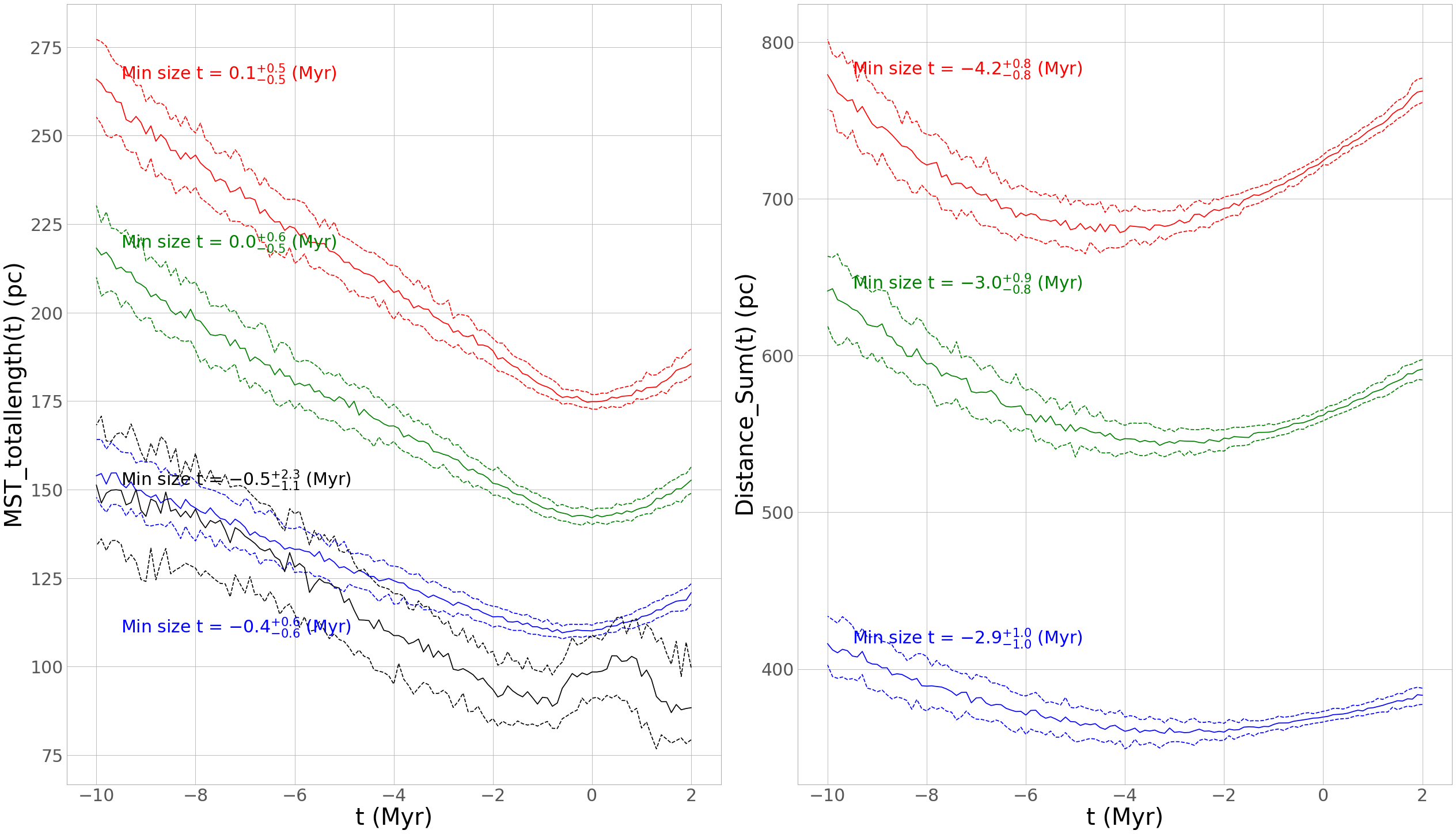}
	\setlength{\belowcaptionskip}{-10pt}
	\setlength{\textfloatsep}{0pt}
	\caption{\textit{Left:} Minimum spanning-tree total length as a function of trace-back time with no filter for outliers (red), 3$\sigma$ velocity outliers removed (green), 2$\sigma$ velocity outliers removed (blue) and 32\% longest branches removed (black) with their respective uncertainties. \textit{Right:} Sum of distances for each star to the association subgroup center as a function of trace-back time with no filter for outliers (red), 3$\sigma$ velocity outliers removed (green), 2$\sigma$ velocity outliers removed (blue). This example illustrates the traceback of the $\pi$ Sco subgroup. }
	\label{MinAreaPlot}
\end{figure*}



\subsection{Inter-group kinematics}

\begin{figure}
\includegraphics[width=240pt]{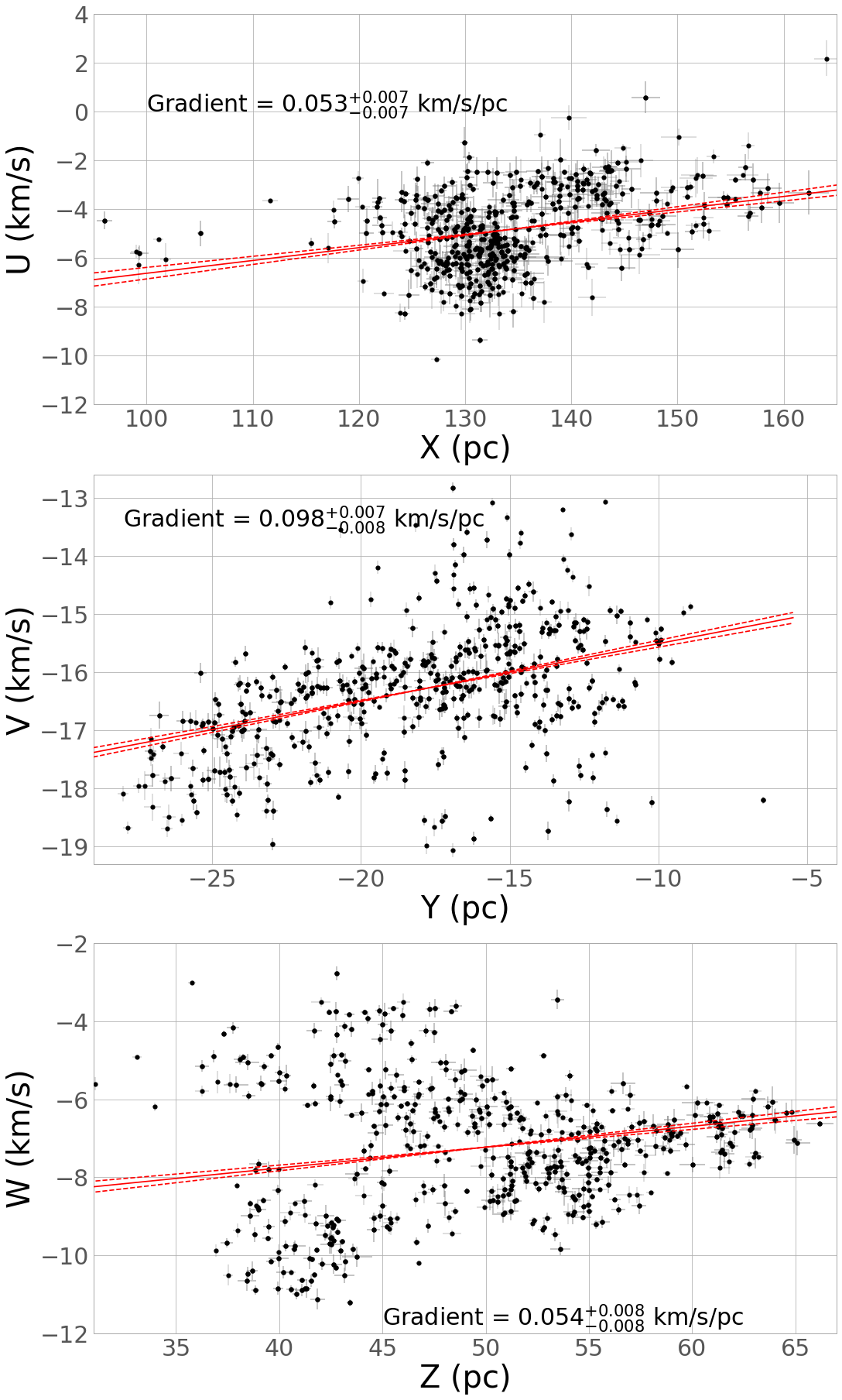}
\caption{Cartesian velocity versus position in each of the three dimensions $XYZ$ for all YSOs in our sample with uncertainties shown. The best-fitting gradients and the 16th and 84th percentiles values of the fit are shown as red solid and dashed lines respectively in each panel.}
\label{figure_expansion_all}
\end{figure}

\begin{figure*}
	\includegraphics[width=\textwidth]{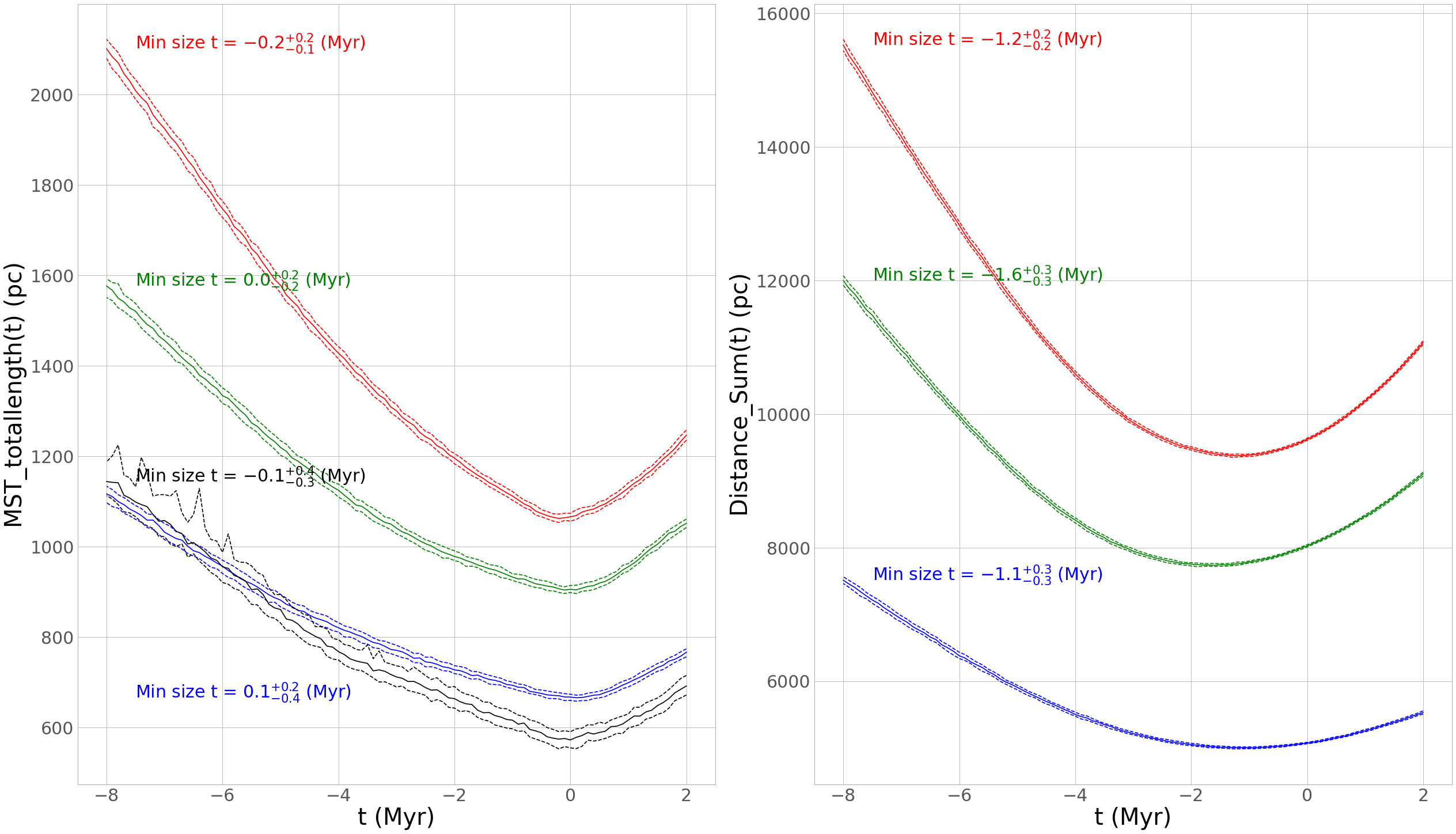}
	\setlength{\belowcaptionskip}{-10pt}
	\setlength{\textfloatsep}{0pt}
	\caption{Association size for all YSOs with quality 6D kinematic information in our Upper Sco sample. \textit{Left:} Minimum spanning-tree total length as a function of trace-back time with no filter for outliers (red), 3$\sigma$ velocity outliers removed (green), 2$\sigma$ velocity outliers removed (blue) and 32\% longest branches removed (black) with their respective uncertainties. \textit{Right:} Sum of distances for each star to the association subgroup center as a function of trace-back time with no filter for outliers (red), 3$\sigma$ velocity outliers removed (green), 2$\sigma$ velocity outliers removed (blue). }
	\label{MinAreaPlot_all}
\end{figure*}

As well as the internal kinematics of each separate subgroup we can also consider the kinematic trends between subgroups.

We calculate the component of 3D velocity directed away from the combined center of Upper Sco ($v_{\rm out}$) for each subgroup, using the median subgroup Cartesian positions and velocities and their respective uncertainties as given in Table.~\ref{groups_table}, and using the median Cartesian position and velocity of all YSOs in our sample as the combined center of Upper Sco.

We find $v_{\rm out}$ = $-1.26^{+0.04}_{-0.04}$ $kms^{-1}$ for $\alpha$ Sco, $v_{\rm out}$ = $1.95^{+0.11}_{-0.11}$ $kms^{-1}$ for $\beta$ Sco, $v_{\rm out}$ = $-0.54^{+0.03}_{-0.03}$ $kms^{-1}$ for $\delta$ Sco, $v_{\rm out}$ = $-0.31^{+0.04}_{-0.04}$ $kms^{-1}$ for $\nu$ Sco, $v_{\rm out}$ = $0.56^{+0.14}_{-0.14}$ $kms^{-1}$ for $\pi$ Sco, $v_{\rm out}$ = $1.60^{+3.00}_{-0.15}$ $kms^{-1}$ for $\rho$ Oph and $v_{\rm out}$ = $0.36^{+0.12}_{-0.12}$ $kms^{-1}$ for $\sigma$ Sco. 

Despite the evidence we have presented that most of these subgroups are expanding from their initial configurations, albeit anisotropically (see Section~\ref{Expansion}), we find that the large-scale motions of Upper Sco are not consistent with simple radial expansion of all subgroups relative to one another. Indeed, $\alpha$ Sco, $\delta$ Sco and $\nu$ Sco are moving toward the combined center of Upper Sco on average, as indicated by their significantly ($>7\sigma$ in all cases) negative $v_{\rm out}$ values.

We also calculate the inter-subgroup velocity dispersion, again using the median velocities of each subgroup as given in Table.~\ref{groups_table}. The inter-subgroup velocity dispersion in the X direction ($\sigma_U$) = $1.09^{+0.03}_{-0.03}$ kms$^{-1}$, in the Y direction ($\sigma_V$) = $0.89^{+0.01}_{-0.01}$ kms$^{-1}$ and in the Z direction ($\sigma_W$) = $1.66^{+0.01}_{-0.01}$ kms$^{-1}$.

If we calculate the inter-subgroup velocity dispersions with the velocity components attributed to radial expansion ($v_{\rm out}$) subtracted, we find $\sigma_U = 0.26^{+0.04}_{-0.04}$ kms$^{-1}$, $\sigma_V = 0.62^{+0.02}_{-0.02}$ kms$^{-1}$ and $\sigma_W = 1.01^{+0.04}_{-0.04}$ kms$^{-1}$. 

\begin{figure*}
\begin{subfigure}{0.99\textwidth}
    \includegraphics[width=490pt]{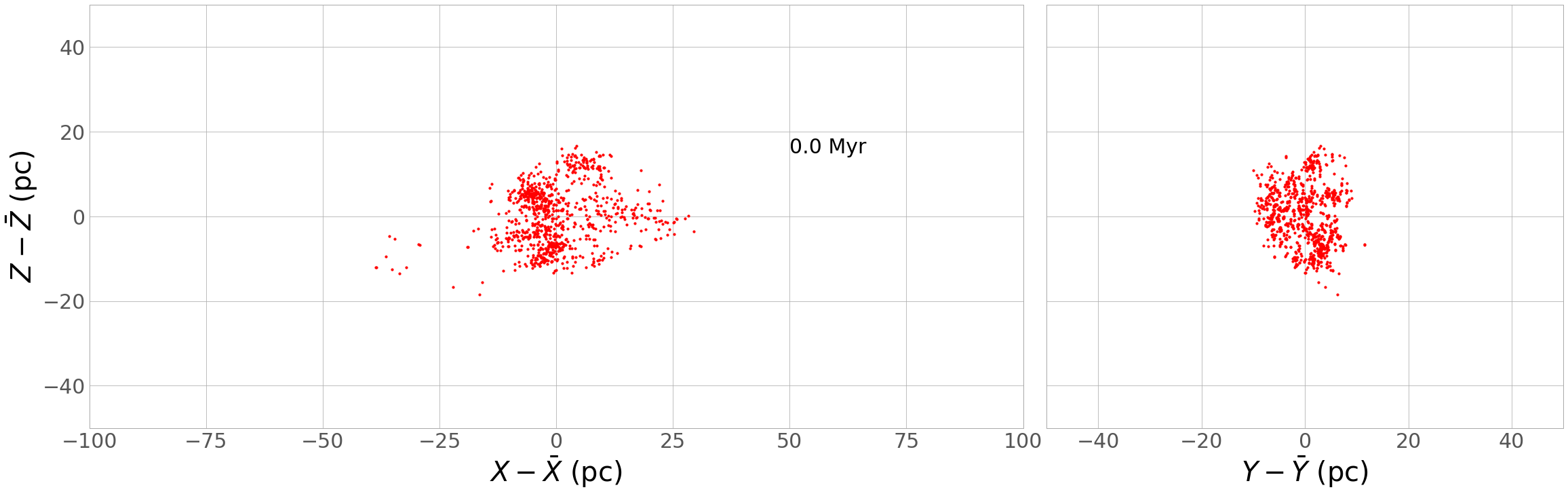}
\end{subfigure}
\begin{subfigure}{0.99\textwidth}
    \includegraphics[width=490pt]{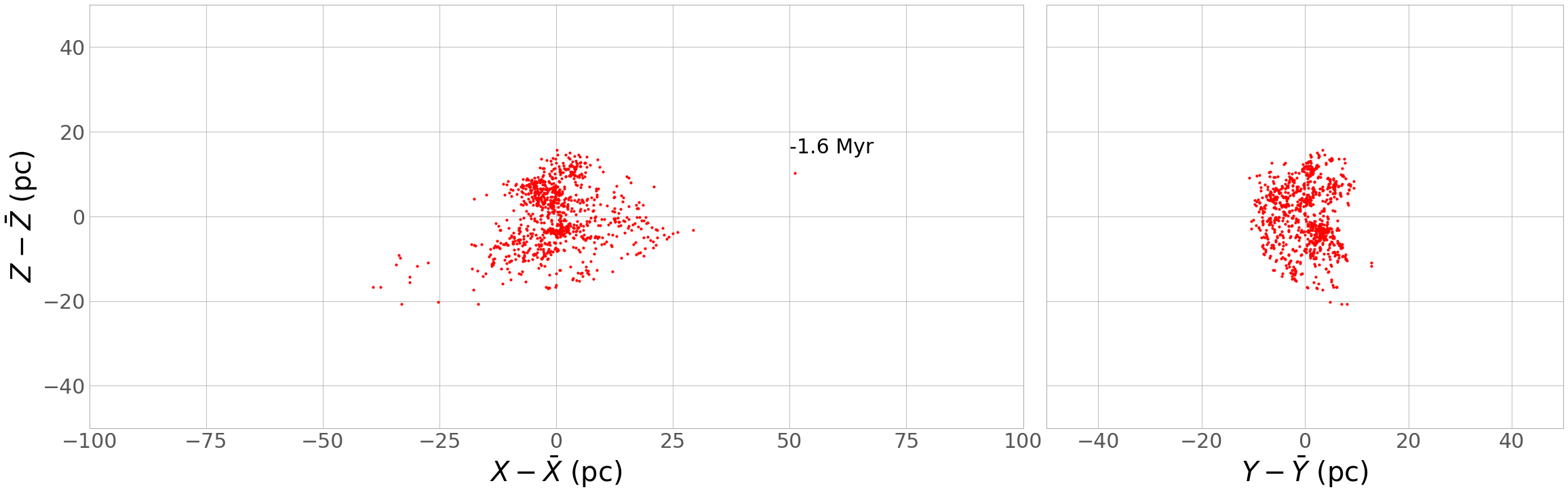}
\end{subfigure}
\begin{subfigure}{0.99\textwidth}
    \includegraphics[width=490pt]{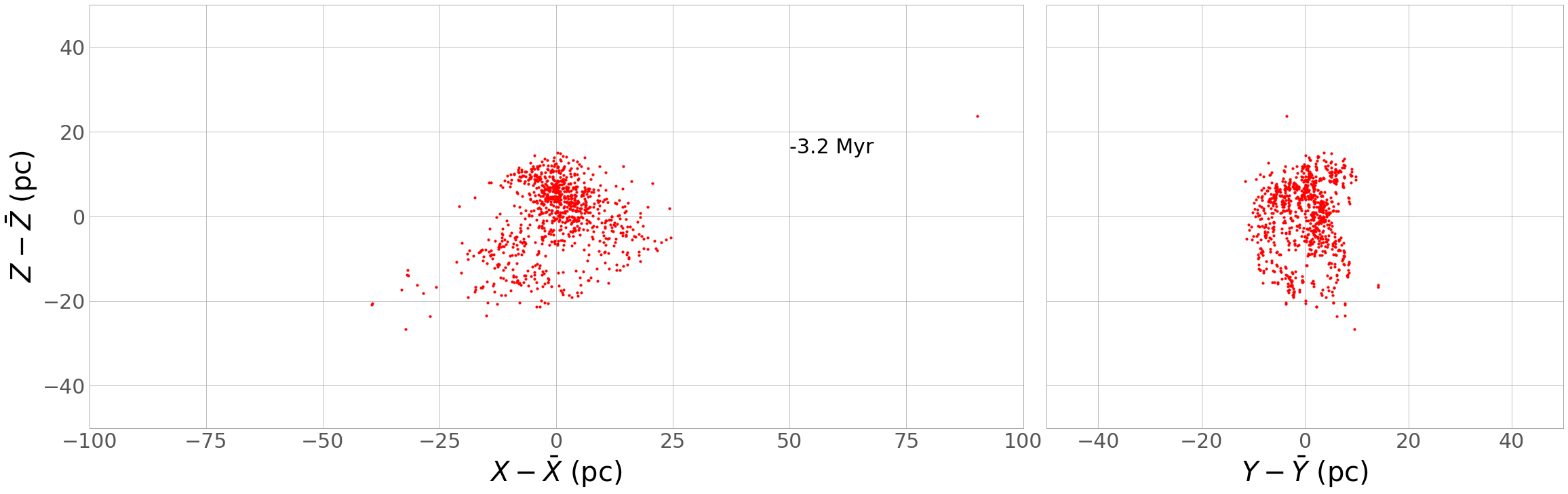}
\end{subfigure}
\caption{Relative Cartesian positions for all YSOs with quality 6D kinematic information in our sample at the present, 1.6 Myr in the past and 3.2 Myr in the past. }
\label{Past_configuration_all}
\end{figure*}

We also note that, despite the significant expansion trends seen across the entire Upper Sco YSO sample (Fig.~\ref{figure_expansion_all}) the greatest of which ($0.098^{+0.007}_{-0.008}$ kms$^{-1}$pc$^{-1}$) indicates an expansion timescale of $10.20^{+0.91}_{-0.68}$ Myr, traceback calculations performed for the entire sample following the approach described in Section.~\ref{Traceback} do not provide strong evidence that the region as a whole was significantly more compact in the past. According to the $\tau_\text{MSTlength}$ size metric the distribution of Upper Sco YSOs becomes sparser back in time, with the minimum MST total-length being found consistently in the present configuration, regardless of the level of filtering. According to the $\tau_\text{DSum}$ metric the distribution of Upper Sco YSOs may be considered to be more compact $\sim1.1 - 1.6$ Myr in the past, but then becomes sparser for earlier times. We also plot the relative positions of all YSOs in our sample at the present time, at $1.6$ Myr in the past and at $3.2$ Myr in the past in Fig.~\ref{Past_configuration_all} as a visualisation of this traceback.

A possible reason for this difference may be the sensitivity of the traceback to minimum configuration approach to kinematic outliers and, in this case, a potential interloper subgroup. The large negative $v_{\rm out}$ of $\alpha$ Sco, as well as its older isochronal age relative to most other subgroups, indicates its likely origin outside of Upper Sco as it is commonly defined, or at least that its formation was not directly related to the subgroups considered part of the Upper Sco 'cluster-chain' by \citet{posch24}. This is similar to the conclusion of \citet{Armstrong22} that the NGC 2547 cluster is presently interloping in the younger population of Vela OB2. The linear expansion analysis is less sensitive to kinematic outliers and so still produces significant expansion gradients for Upper Sco as a whole despite the inclusion of $\alpha$ Sco. 

If the traceback calculations are performed on the YSOs sample with $\alpha$ Sco removed, then the traceback time to the most compact configuration becomes $2.4^{+0.2}_{-0.2}$ Myr, $3.0^{+0.2}_{-0.3}$ Myr and $2.3^{+0.3}_{-0.3}$ Myr, according to the $\tau_\text{DSum}$ metric with no filter for outliers, 3$\sigma$ velocity outliers removed and 2$\sigma$ velocity outliers removed, respectively.

\section{Discussion}\label{section_discussion}
\subsection{Association membership}
The confirmation of 1204 YSOs in the Upper Scorpius region, 118 that were not included in the clustering of \citet{miret-roig22}, and 62 YSOs not included in \citet{ratzenbock23}, underlines the importance of spectroscopic indicators for accurate identification of members of young associations with complex kinematics. 

There are also 35 members of the \citet{miret-roig22} clusters in our sample that we cannot confirm as YSOs via either spectroscopic indicators or Gaia variability. Five of these have AAT-HERMES spectra of $SNR>20$, and so are unlikely to be YSOs, while the others have $SNR<5$ and so may not be confirmed due to large uncertainties in measured EW(Li).   

\subsection{The likely star formation scenario of the Upper Scorpius association}\label{star_formation}
\begin{figure*}
\begin{subfigure}{\textwidth}
    \includegraphics[width=\textwidth]{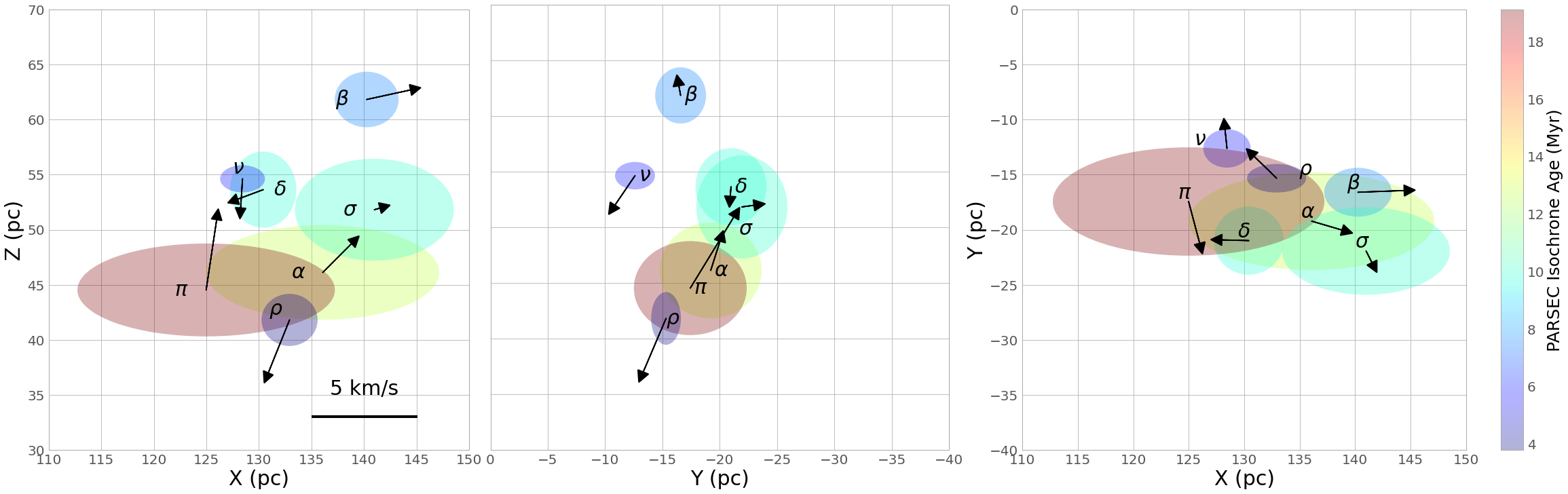}
\end{subfigure}
\begin{subfigure}{\textwidth}
    \includegraphics[width=\textwidth]{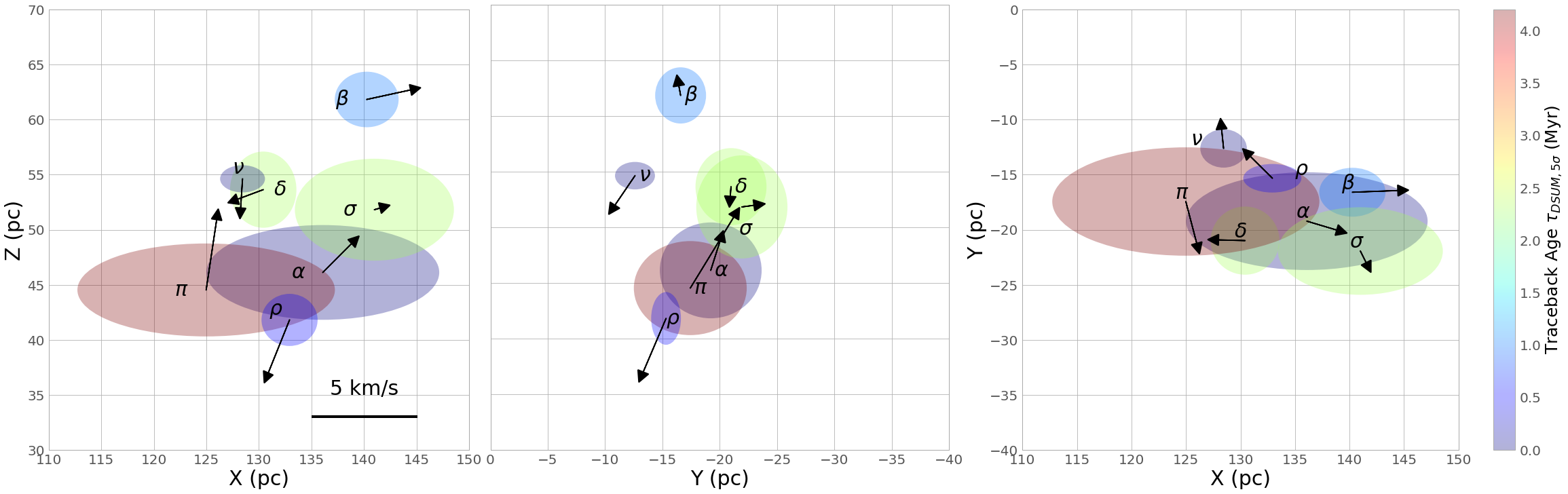}
\end{subfigure}
	\setlength{\belowcaptionskip}{-10pt}
	\setlength{\textfloatsep}{0pt}
	\caption{3D Cartesian positions of Upper Scorpius subgroups represented by ellipses, the central coordinates and length of the semi-major and semi-minor axes which are given as the median coordinates (XYZ) and twice the dispersions ($\sigma_X$,$\sigma_Y$,$\sigma_Z$), respectively. The ellipses are labeled with the greek letters pertaining to the names given in \protect\citet{miret-roig22} and colour-coded according to ages estimated in \protect\citet{ratzenbock23} using PARSEC stellar evolution models \citep{marigo17} with Gaia BP-RP colour in the \textit{top} panel, and traceback ages $\tau_\text{DSum}$ in the \textit{bottom} panel. Arrows indicate the average 3D velocities of each group relative to each other, with a scalebar in the rightmost panel. }
	\label{Ellipses}
\end{figure*}

Figure~\ref{Ellipses} presents a summary of the spatial distribution and kinematics that we have derived for the sub-groups of the Upper Sco OB association. Previous studies of this region have found evidence of an age gradient among the different subgroups present \citep{squicciarini21,ratzenbock23}, and thus infer scenarios of sequential star formation in this region. In particular, \citet{posch23} and \citet{posch24} recently proposed a star formation scenario in which feedback from massive stars in the oldest subgroups compresses molecular gas in an adjacent region, initiating a new generation of star formation while simultaneously pushing the star-forming gas away, similar to the triggered formation scenario of \citet{elmegreen76}. They invoke this scenario as an explanation for both the age gradient found in Upper Sco, with the older groups $\delta$ Sco and $\sigma$ Sco being centrally located and younger groups such as $\nu$ Sco being located toward the periphery of Upper Sco, as well as the apparent outward acceleration of the younger groups relative to the older groups. In this ``cluster-chain'' scenario, $\delta$ Sco and $\sigma$ Sco are proposed to have triggered star formation in $\beta$ Sco, which subsequently triggered the formation of $\nu$ Sco.

Our dynamical ages are generally younger than, and have a smaller range than, those of \citet{miret-roig22}, which does not necessarily indicate that the groups are younger, but rather that they began expanding (i.e., became free of gas or unbound) more recently than previously suggested \citep{miret-roig24}. 

In the case that residual gas expulsion is the mechanism by which these groups became unbound and initiated expansion, one would assume that feedback should expel gas within the cluster (and thus cause unbinding of the subgroup) before triggering star formation at a more distant region of the molecular cloud. Then, the smaller differences between dynamical ages of these groups also indicates shorter intervals between subgroup formation for the triggering via feedback to occur.

In fact, such a scenario requires that the unbinding of the first cluster ($C1$) precedes the formation of the second in the sequence ($C2$). Thus, assuming that an accurate isochronal cluster age ($\tau_{iso}$) indicates the formation time of a cluster and a dynamical age ($\tau_{dyn}$) indicates the time of unbinding, sequential cluster formation requires that 
\begin{equation}
    \tau_{dyn,C1} > \tau_{iso,C2}.
\end{equation}

In the case of the proposed cluster chain $\delta$ Sco $\rightarrow$ $\beta$ Sco $\rightarrow$ $\nu$ Sco \citep{posch24}, isochronal ages are estimated to be $9.8^{+1.2}_{-1.4}$ Myr, $7.6^{+0.8}_{-0.7}$ Myr and $5.8^{+1.8}_{-0.5}$ Myr respectively \citep[][according to PARSEC models with BP-RP colour]{ratzenbock23}, but have dynamical ages of $2.3^{+0.3}_{-0.3}$ Myr, $1.0^{+0.4}_{-0.4}$ Myr and $-0.1^{+0.3}_{-0.3}$ Myr respectively by our estimation of $\tau_\text{DSum}$, and dynamical ages of $4.6^{+0.6}_{-0.6}$ Myr, $2.4^{+1.7}_{-1.7}$ Myr and $0.3^{+0.5}_{-0.5}$ Myr respectively according to \citet{miret-roig24}. Notably, the oldest dynamical age of a cluster in this chain ($\delta$ Sco) is less than the lowest isochronal age ($\nu$ Sco), indicating that no cluster in this proposed chain had expelled its surrounding gas and become unbound until star formation had already begun in all the other clusters. This brings into doubt a scenario where the star formation of the younger groups was triggered by feedback from the previous group in a chain.

A possible alternative explanation that preserves the sequential cluster-chain hypothesis is that the clusters do not become unbound immediately after gas expulsion and before triggering the formation of the next cluster in the sequence, but several Myr later via a different mechanism, perhaps from gradual mass loss as unbound stars escape or from tidal forces exerted by nearby molecular clouds. Thus $\tau_{dyn,C1} > \tau_{iso,C2}$ would no longer be required. However, in this case the dynamical age does not indicate when a cluster became gas free, and so the difference between isochronal ages and dynamical ages does not give us the timescale of an ``embedded phase'' as suggested by \citet{miret-roig24}. 

This seems unlikely, though, given that natal molecular gas makes up a significant proportion of an embedded cluster's binding mass, and so the infant mortality rate of clusters having undergone residual gas expulsion is expected to be at least 50\% \citep[e.g.,][]{goodwin06}. \citep{wright24} found that the clusters they studied were all gravitationally unbound when considering the stellar mass only, but were nearly all gravitationally bound when considering the gas and stellar mass. This is very strong evidence for residual gas expulsion being responsible for unbinding a cluster. Our traceback analysis of the subgroups of Upper Sco indicates that they were likely initially too sparse, as well as too low-mass, to survive gas expulsion. Thus we consider that the dynamical age likely does indicate when gas expulsion occurred for each subgroup.  
  
The anisotropy of velocity dispersions (Table \ref{groups_table}) and expansion rates (Table \ref{expansion_table}), as well as the low fractions of subgroup members consistent with moving away from their subgroup center in 3D (Table \ref{groups_table}), indicate that subgroups did not expand uniformly from initially compact configurations, but rather formed with significant kinematic substructure. The initial size estimates we make for subgroups (Section~\ref{size}) as well as the short dynamical ages relative to linear expansion rates ($1.0^{+0.4}_{-0.4}$ and $7.81^{+1.32}_{-0.94}$ Myr respectively for $\beta$ Sco, $-0.1^{+0.3}_{-0.3}$ and $12.18^{+90.12}_{-5.66}$ Myr respectively for $\nu$ Sco) also imply initially sparse distributions in large volumes, which may be incompatible with subgroup formation as ``clusters'' embedded within ``pillars'' of gas \citep{posch24}. This is closer to the picture of ``low density'' star formation advocated for OB associations such as Cygnus OB2 by e.g., \citet{wright14}.

Large initial volumes for subgroups also suggests a star-formation scenario where the triggering mechanism, such as via a cloud-cloud collision \citep[e.g.,][]{tan00,wu17} or via stellar feedback \citep[e.g.,][]{farias24}, impacts a large surface area of molecular cloud at a similar time, producing a sparse distribution of coeval young stars. This implies a greater distance between the feedback-driving source and the impacted region of molecular cloud.

Together, these results better support scenarios in which star formation in Upper Sco is either initiated spontaneously via development of instabilities in a GMC or else via large-scale external triggering, such as via a GMC-GMC collision or from external stellar feedback, or some combination of these modes.

For the GMC-GMC collision scenario, the range in ages of the subgroups would be expected to be similar to the GMC crossing time at the collision speed, i.e., $t_{\rm age\:spread}\sim D_{\rm GMC}/v_{\rm coll}\sim 100\:{\rm pc}/10\:{\rm km\:s}^{-1}\sim 10\:$Myr. Such a scenario could lead to a coherent large-scale gradient in ages, i.e., as different parts of the cloud are engulfed in the collision. It could also lead to disturbed (unbound) kinematics of individual sub-groups.

For the external stellar feedback scenario, triggering may be from sources in the adjacent Upper Centaurus-Lupus (UCL) region, such as multiple likely past supernovae \citep[see][]{krause18}. However, the formation of distinct subgroups within Upper Sco at different times due to this feedback impacting different regions of molecular gas at different times and with differing intensity, initiating star formation later in regions further from the feedback driving sources (e.g., $\nu$ Sco and $\rho$ Oph), rather than star formation being triggered by feedback from previous generations in sequence within Upper Sco itself. 

This scenario is similar to that recently proposed by \citet{kerr25} to explain the observed kinematic and age structure they identified in the Circinus star forming complex. In particular, they make comparison of their observations to the outputs of fiducial STARFORGE simulations \citep{grudic21,guszejnov22}, which have previously been shown to produce unbound and expanding populations similar to observed OB associations \citep{farias24}, and find that the scenario of 'inside-out' sequential star formation driven by feedback from a central cluster can produce the structures and age gradient found in Circinus and also in $\gamma$ Vel \& Vela OB2 \citep{pang21,Armstrong22}. 

However, the difference between Circinus, Vela OB2 and Upper Sco is the lack of a central dense cluster within Upper Sco which would have driven feedback across the region, as ASCC 79 and $\gamma$ Vel are hypothesised to have done in Circinus and Vela OB2 respectively. Rather, the trends seen in these regions align better when considering Upper Sco as a part of Sco-Cen, with the central feedback-driving cluster being UCL. 

In future work based on this survey data we will investigate the origins of the 84 YSOs with kinematics distinct from any major subgroup, particularly as candidate runaway stars. We will compare a variety of kinematic age estimates for subgroups to ages derived from fitting stellar evolution models as well as Li-depletion models. We will also calculate mass accretion rates for YSOs and investigate trends between accretion, stellar mass and age.

\section{Summary}
\begin{itemize}
        \item We have carried out spectroscopic observations of $>7000$ targets selected from the Gaia DR3 source catalogue within 25 2$^\circ$-diameter fields across the young nearby Upper Scorpius association using the 2dF/HERMES spectrograph at the AAT.
        \item We reduce and analyse the spectra in order to derive radial velocities and youth indicators such as EW(Li)s, which we obtain for 6727 and 6937 unique sources, respectively.
        \item We identify YSOs on the basis of EW(Li)s, H$\alpha$ emission and optical variability flagged in Gaia DR3. We identify 1204 confirmed YSOs.
        \item We filter our YSO sample on Gaia astrometric quality criteria, leaving us with a sample of 544 confirmed YSOs with full reliable 6D kinematic information.
        \item We cross-match our confirmed YSOs with recent literature samples of Upper Scorpius members from \citet{luhman20,miret-roig22,ratzenbock23} and match 1009, 1030 \& 1085 of their samples respectively, whilst also identifying $>$100 new YSOs in our sample.
        \item We derive 3D heliocentric Cartesian positions $XYZ$ and velocities $UVW$ for our confirmed YSOs from observed astrometry.
        \item We allocate our confirmed YSOs into the 7 kinematically distinct subgroups of Upper Sco as defined by \citet{miret-roig22}, but we are also left with 84 YSOs whose kinematics are distinct from any of these groups by $>3\sigma$ which we consider separately as candidate ejected stars.
        \item We estimate 3D velocity dispersions for each kinematic subgroup using a Bayesian approach with randomly sampled RV offsets from a simulated binary population. We find that velocity dispersions for most subgroups are significantly ($>3\sigma$) anisotropic, indicating that the subgroups retain much of their initial kinematic substructure and have not yet undergone sufficient dynamical mixing to develop isotropy. This is a commonly observed feature of OB associations and their subgroups \citep{wright20}.
        \item We then use the 3D velocity dispersions to estimate virial masses for the subgroups and compare them to masses estimated by extrapolation from a \citet{kroupa01b} IMF. Virial mass estimates for all subgroups are significantly larger than those estimated by IMF extrapolation, in some cases by an order of magnitude, indicating that the subgroups are gravitationally unbound, as expected for OB associations.
        \item We calculate expansion velocities v$_{out}$ and find that they are positive above uncertainties for all subgroups with varying levels of significance. In particular, the $\rho$ Oph subgroup, often considered to be a bound young cluster, has a positive v$_{out}$ of $>3.5\sigma$ significance and 72\% of its members moving outward from the central position. This is in contrast to the recent observation that Rho Oph is contracting \citep{wright24}.
        \item We find evidence for expansion ($>3\sigma$ positive gradients) in at least one direction in all 7 groups, and evidence for contraction in $\alpha$ Sco and $\nu$ Sco. $\beta$ Sco, $\delta$ Sco \& $\sigma$ Sco exhibit significant expansion trends in all directions, while the trends of $\alpha$ Sco, $\beta$ Sco \& $\sigma$ Sco are significantly ($>4\sigma$) anisotropic. For $\alpha$ Sco, $\beta$ Sco \& $\sigma$ Sco the greatest rate of expansion is in the direction of Galactic rotation $Y$, while for $\delta$ Sco, $\nu$ Sco, $\pi$ Sco \& $\rho$ Oph the greatest rate of expansion is in the direction perpendicular to the Galactic plane $Z$.
        \item From the linear expansion gradients we also derive expansion timescales for each group, of $28.42_{-5.68}^{+9.47}$, $7.81_{-0.94}^{+1.32}$, $6.20_{-0.29}^{+0.40}$, $12.18_{-5.66}^{+90.12}$, $15.98_{-3.35}^{+5.78}$, $5.95_{-1.08}^{+1.74}$ \& $4.97_{-0.36}^{+0.45}$ Myr respectively. For $\beta$ Sco, $\delta$ Sco, $\rho$ Oph \& $\sigma$ Sco these are comparable with literature age estimates.
        \item By fitting linear gradients to perpendicular position and velocity pairs we can also detect rotation trends. We find evidence ($>3\sigma$) of rotation in all 7 subgroups of Upper Sco.
        \item We use an epicycle approximation and orbital equations to estimate the positions of YSOs up to 10 Myr in the past and thus trace the past evolution of the Upper Sco subgroups. We also use MST total length \citep[$\tau_\text{MSTlength}$; ][]{squicciarini21} and summed 3D distance \citep[$\tau_\text{DSum}$; ][]{quintana22} metrics to estimate when subgroups would have been at their most compact configurations in the past, assumed to be approximately their initial configurations. $\alpha$ Sco and $\nu$ Sco, which previously showed evidence for contraction, are found to be approaching their most compact configurations in the near future. For the other subgroups which are expanding overall, we derive non-zero kinematic ages of $1.0^{+0.4}_{-0.4}$, $2.3^{+0.3}_{-0.3}$, $4.2^{+0.8}_{-0.8}$, $0.5^{+0.2}_{-0.2}$ and $2.3^{+0.5}_{-0.4}$ Myr respectively. Except for $\rho$ Oph, these are consistently significantly younger than the expansion timescales derived by \citet{miret-roig22}.
        \item We estimate the sizes of subgroups at the past times when they would have been in their most compact configurations. We find that all subgroups would likely have had initial radii $>4$ pc, which seems inconsistent with formation in 'cometary-shape' or 'pillars' of molecular gas of typical diameter $\sim1$ pc and is more consistent with the extended, low-density view of OB association formation advocated by \citet{wright14} for Cygnus OB2.
        \item We compare the results of our kinematic analyses with those in recent literature as well as literature ages for the subgroups estimated by fitting PMS isochrones. We find that our dynamical ages estimates for the subgroups, as well as those from recent literature, are significantly younger than isochronal age estimates, which is problematic for the ``cluster-chain'' sequential formation scenario.
        \item We assess the results of our analysis and we propose that a star formation scenario where feedback from the older UCL region of Sco-Cen initiated star formation at different times in different locations of Upper Sco, creating the age gradient seen among the distinct subgroups. The low intensity of feedback from the relatively distant UCL region, impacting a large surface area of molecular gas in Upper Sco over an extended period, resulted in initially sparse subgroups which have largely retained their initial kinematic structure.
\end{itemize}

\section{Acknowledgments}
JJA \& JCT acknowledge support from ERC Advanced Grant MSTAR (788829).
This project has also received funding from the European Union's Horizon 2020 research and innovation programme under grant agreement No 101004719. This material reflects only the authors views and the Commission is not liable for any use that may be made of the information contained therein. This work has made use of data from the ESA space mission Gaia (http://www.cosmos.esa.int/gaia), processed by the Gaia Data Processing and Analysis Consortium (DPAC, http://www.cosmos.esa.int/web/gaia/dpac/consortium). Funding for DPAC has been provided by national institutions, in particular the institutions participating in the Gaia Multilateral Agreement. This research made use of the Simbad and Vizier catalogue access tools (provided by CDS, Strasbourg, France), Astropy \citep{astr13} and TOPCAT \citep{tayl05}. 
	
\section{Data Availability}
The data underlying this article, both the spectroscopic parameters for observed targets and the astrometric parameters for confirmed PMS members will be available at CDS via https://cdsarc.unistra.fr/viz-bin/cat/J/MNRAS.

\bibliographystyle{mnras}
\bibliography{references2}

\begin{thebibliography}{}
\makeatletter
\relax
\def\mn@urlcharsother{\let\do\@makeother \do\$\do\&\do\#\do\^\do\_\do\%\do\~}
\def\mn@doi{\begingroup\mn@urlcharsother \@ifnextchar [ {\mn@doi@}
  {\mn@doi@[]}}
\def\mn@doi@[#1]#2{\def\@tempa{#1}\ifx\@tempa\@empty \href
  {http://dx.doi.org/#2} {doi:#2}\else \href {http://dx.doi.org/#2} {#1}\fi
  \endgroup}
\def\mn@eprint#1#2{\mn@eprint@#1:#2::\@nil}
\def\mn@eprint@arXiv#1{\href {http://arxiv.org/abs/#1} {{\tt arXiv:#1}}}
\def\mn@eprint@dblp#1{\href {http://dblp.uni-trier.de/rec/bibtex/#1.xml}
  {dblp:#1}}
\def\mn@eprint@#1:#2:#3:#4\@nil{\def\@tempa {#1}\def\@tempb {#2}\def\@tempc
  {#3}\ifx \@tempc \@empty \let \@tempc \@tempb \let \@tempb \@tempa \fi \ifx
  \@tempb \@empty \def\@tempb {arXiv}\fi \@ifundefined
  {mn@eprint@\@tempb}{\@tempb:\@tempc}{\expandafter \expandafter \csname
  mn@eprint@\@tempb\endcsname \expandafter{\@tempc}}}

\bibitem[\protect\citeauthoryear{{AAO Software Team}}{{AAO Software
  Team}}{2015}]{2dfdr}
{AAO Software Team} 2015, \href
  {https://ui.adsabs.harvard.edu/abs/2015ascl.soft05015A} {p. ascl:1505.015}

\bibitem[\protect\citeauthoryear{{Ahumada} et~al.,}{{Ahumada}
  et~al.}{2020}]{ahumada20}
{Ahumada} R.,  et~al., 2020, \mn@doi [\apjs] {10.3847/1538-4365/ab929e}, \href
  {https://ui.adsabs.harvard.edu/abs/2020ApJS..249....3A} {249, 3}

\bibitem[\protect\citeauthoryear{Ambartsumian}{Ambartsumian}{1947}]{ambartsumian47}
Ambartsumian V.,  1947, Acad. Sci. Armenian SSR, Yerevan

\bibitem[\protect\citeauthoryear{{Armstrong} \& {Tan}}{{Armstrong} \&
  {Tan}}{2024}]{armstrong24}
{Armstrong} J.~J.,  {Tan} J.~C.,  2024, \mn@doi [\aap]
  {10.1051/0004-6361/202451538}, \href
  {https://ui.adsabs.harvard.edu/abs/2024A&A...692A.166A} {692, A166}

\bibitem[\protect\citeauthoryear{{Armstrong}, {Wright}, {Jeffries}, {Jackson}
  \& {Cantat-Gaudin}}{{Armstrong} et~al.}{2022}]{Armstrong22}
{Armstrong} J.~J.,  {Wright} N.~J.,  {Jeffries} R.~D.,  {Jackson} R.~J.,
  {Cantat-Gaudin} T.,  2022, \mn@doi [\mnras] {10.1093/mnras/stac3101}, \href
  {https://ui.adsabs.harvard.edu/abs/2022MNRAS.517.5704A} {517, 5704}

\bibitem[\protect\citeauthoryear{{Astropy Collaboration} et~al.,}{{Astropy
  Collaboration} et~al.}{2013}]{astr13}
{Astropy Collaboration} et~al., 2013, \mn@doi [\aap]
  {10.1051/0004-6361/201322068}, \href
  {http://adsabs.harvard.edu/abs/2013A%26A...558A..33A} {558, A33}

\bibitem[\protect\citeauthoryear{{Bailer-Jones}, {Rybizki}, {Fouesneau},
  {Demleitner}  \& {Andrae}}{{Bailer-Jones} et~al.}{2021}]{bailerjones21}
{Bailer-Jones} C.~A.~L.,  {Rybizki} J.,  {Fouesneau} M.,  {Demleitner} M.,
  {Andrae} R.,  2021, \mn@doi [\aj] {10.3847/1538-3881/abd806}, \href
  {https://ui.adsabs.harvard.edu/abs/2021AJ....161..147B} {161, 147}

\bibitem[\protect\citeauthoryear{{Baraffe}, {Homeier}, {Allard}  \&
  {Chabrier}}{{Baraffe} et~al.}{2015}]{baraffe15}
{Baraffe} I.,  {Homeier} D.,  {Allard} F.,   {Chabrier} G.,  2015, \mn@doi
  [\aap] {10.1051/0004-6361/201425481}, \href
  {http://adsabs.harvard.edu/abs/2015A%26A...577A..42B} {577, A42}

\bibitem[\protect\citeauthoryear{{Blaauw}}{{Blaauw}}{1964}]{blaauw64}
{Blaauw} A.,  1964, \mn@doi [\araa] {10.1146/annurev.aa.02.090164.001241},
  \href {http://adsabs.harvard.edu/abs/1964ARA%26A...2..213B} {2, 213}

\bibitem[\protect\citeauthoryear{{Buder} et~al.,}{{Buder}
  et~al.}{2021}]{Buder21}
{Buder} S.,  et~al., 2021, \mn@doi [\mnras] {10.1093/mnras/stab1242}, \href
  {https://ui.adsabs.harvard.edu/abs/2021MNRAS.tmp.1259B} {}

\bibitem[\protect\citeauthoryear{{Buder} et~al.,}{{Buder}
  et~al.}{2024}]{galahdr4}
{Buder} S.,  et~al., 2024, \mn@doi [arXiv e-prints]
  {10.48550/arXiv.2409.19858}, \href
  {https://ui.adsabs.harvard.edu/abs/2024arXiv240919858B} {p. arXiv:2409.19858}

\bibitem[\protect\citeauthoryear{{Cantat-Gaudin}, {Mapelli},
  {Balaguer-N{\'u}{\~n}ez}, {Jordi}, {Sacco}  \& {Vallenari}}{{Cantat-Gaudin}
  et~al.}{2019a}]{cantatgaudin19a}
{Cantat-Gaudin} T.,  {Mapelli} M.,  {Balaguer-N{\'u}{\~n}ez} L.,  {Jordi} C.,
  {Sacco} G.,   {Vallenari} A.,  2019a, \mn@doi [\aap]
  {10.1051/0004-6361/201834003}, \href
  {https://ui.adsabs.harvard.edu/abs/2019A%26A...621A.115C} {621, A115}

\bibitem[\protect\citeauthoryear{{Cantat-Gaudin}, {Mapelli},
  {Balaguer-N{\'u}{\~n}ez}, {Jordi}, {Sacco}  \& {Vallenari}}{{Cantat-Gaudin}
  et~al.}{2019b}]{cantatgaudin19}
{Cantat-Gaudin} T.,  {Mapelli} M.,  {Balaguer-N{\'u}{\~n}ez} L.,  {Jordi} C.,
  {Sacco} G.,   {Vallenari} A.,  2019b, \mn@doi [\aap]
  {10.1051/0004-6361/201834003}, \href
  {https://ui.adsabs.harvard.edu/abs/2019A%26A...621A.115C} {621, A115}

\bibitem[\protect\citeauthoryear{{Chabrier}}{{Chabrier}}{2003}]{chabrier03}
{Chabrier} G.,  2003, \mn@doi [\pasp] {10.1086/376392}, \href
  {https://ui.adsabs.harvard.edu/abs/2003PASP..115..763C} {115, 763}

\bibitem[\protect\citeauthoryear{{Damiani} et~al.,}{{Damiani}
  et~al.}{2014}]{damiani14}
{Damiani} F.,  et~al., 2014, \mn@doi [\aap] {10.1051/0004-6361/201323306},
  \href {http://adsabs.harvard.edu/abs/2014A%26A...566A..50D} {566, A50}

\bibitem[\protect\citeauthoryear{{Damiani}, {Prisinzano}, {Pillitteri},
  {Micela}  \& {Sciortino}}{{Damiani} et~al.}{2019}]{damiani19}
{Damiani} F.,  {Prisinzano} L.,  {Pillitteri} I.,  {Micela} G.,   {Sciortino}
  S.,  2019, \mn@doi [\aap] {10.1051/0004-6361/201833994}, \href
  {http://adsabs.harvard.edu/abs/2019A%26A...623A.112D} {623, A112}

\bibitem[\protect\citeauthoryear{{El-Badry}, {Rix}  \& {Heintz}}{{El-Badry}
  et~al.}{2021}]{elbadry21}
{El-Badry} K.,  {Rix} H.-W.,   {Heintz} T.~M.,  2021, \mn@doi [\mnras]
  {10.1093/mnras/stab323}, \href
  {https://ui.adsabs.harvard.edu/abs/2021MNRAS.506.2269E} {506, 2269}

\bibitem[\protect\citeauthoryear{{Elmegreen} \& {Lada}}{{Elmegreen} \&
  {Lada}}{1976}]{elmegreen76}
{Elmegreen} B.~G.,  {Lada} C.~J.,  1976, in Bulletin of the American
  Astronomical Society. p.~334

\bibitem[\protect\citeauthoryear{{Elmegreen} \& {Lada}}{{Elmegreen} \&
  {Lada}}{1977}]{elmegreen77}
{Elmegreen} B.~G.,  {Lada} C.~J.,  1977, \mn@doi [\apj] {10.1086/155302}, \href
  {http://adsabs.harvard.edu/abs/1977ApJ...214..725E} {214, 725}

\bibitem[\protect\citeauthoryear{{Fabricius} et~al.,}{{Fabricius}
  et~al.}{2021}]{fabricius21}
{Fabricius} C.,  et~al., 2021, \mn@doi [\aap] {10.1051/0004-6361/202039834},
  \href {https://ui.adsabs.harvard.edu/abs/2021A&A...649A...5F} {649, A5}

\bibitem[\protect\citeauthoryear{{Farias}, {Offner}, {Grudi{\'c}}, {Guszejnov}
  \& {Rosen}}{{Farias} et~al.}{2024}]{farias24}
{Farias} J.~P.,  {Offner} S. S.~R.,  {Grudi{\'c}} M.~Y.,  {Guszejnov} D.,
  {Rosen} A.~L.,  2024, \mn@doi [\mnras] {10.1093/mnras/stad3609}, \href
  {https://ui.adsabs.harvard.edu/abs/2024MNRAS.527.6732F} {527, 6732}

\bibitem[\protect\citeauthoryear{{Feast} \& {Whitelock}}{{Feast} \&
  {Whitelock}}{1997}]{feast97}
{Feast} M.,  {Whitelock} P.,  1997, \mn@doi [\mnras] {10.1093/mnras/291.4.683},
  \href {https://ui.adsabs.harvard.edu/abs/1997MNRAS.291..683F} {291, 683}

\bibitem[\protect\citeauthoryear{{Fiorellino} et~al.,}{{Fiorellino}
  et~al.}{2024}]{fiorellino24}
{Fiorellino} E.,  et~al., 2024, \mn@doi [\aap] {10.1051/0004-6361/202347777},
  \href {https://ui.adsabs.harvard.edu/abs/2024A&A...686A.160F} {686, A160}

\bibitem[\protect\citeauthoryear{Foreman-Mackey, Hogg, Lang  \&
  Goodman}{Foreman-Mackey et~al.}{2013}]{emcee}
Foreman-Mackey D.,  Hogg D.~W.,  Lang D.,   Goodman J.,  2013, Publications of
  the Astronomical Society of the Pacific, 125, 306

\bibitem[\protect\citeauthoryear{{Fortune-Bashee}, {Sun}  \&
  {Tan}}{{Fortune-Bashee} et~al.}{2024}]{2024ApJ...977L...6F}
{Fortune-Bashee} X.,  {Sun} J.,   {Tan} J.~C.,  2024, \mn@doi [\apjl]
  {10.3847/2041-8213/ad91a3}, \href
  {https://ui.adsabs.harvard.edu/abs/2024ApJ...977L...6F} {977, L6}

\bibitem[\protect\citeauthoryear{{Frasca}, {Biazzo}, {Alcal{\'a}}, {Manara},
  {Stelzer}, {Covino}  \& {Antoniucci}}{{Frasca} et~al.}{2017}]{frasca17}
{Frasca} A.,  {Biazzo} K.,  {Alcal{\'a}} J.~M.,  {Manara} C.~F.,  {Stelzer} B.,
   {Covino} E.,   {Antoniucci} S.,  2017, \mn@doi [\aap]
  {10.1051/0004-6361/201630108}, \href
  {https://ui.adsabs.harvard.edu/abs/2017A&A...602A..33F} {602, A33}

\bibitem[\protect\citeauthoryear{{Fuchs}, {Breitschwerdt}, {de Avillez},
  {Dettbarn}  \& {Flynn}}{{Fuchs} et~al.}{2006}]{fuchs06}
{Fuchs} B.,  {Breitschwerdt} D.,  {de Avillez} M.~A.,  {Dettbarn} C.,   {Flynn}
  C.,  2006, \mn@doi [\mnras] {10.1111/j.1365-2966.2006.11044.x}, \href
  {https://ui.adsabs.harvard.edu/abs/2006MNRAS.373..993F} {373, 993}

\bibitem[\protect\citeauthoryear{{Gaia Collaboration} et~al.,}{{Gaia
  Collaboration} et~al.}{2016}]{gaia16}
{Gaia Collaboration} et~al., 2016, \mn@doi [\aap]
  {10.1051/0004-6361/201629272}, \href
  {https://ui.adsabs.harvard.edu/abs/2016A&A...595A...1G} {595, A1}

\bibitem[\protect\citeauthoryear{{Gaia Collaboration} et~al.,}{{Gaia
  Collaboration} et~al.}{2021}]{Gaiaedr3}
{Gaia Collaboration} et~al., 2021, \mn@doi [\aap]
  {10.1051/0004-6361/202039657}, \href
  {https://ui.adsabs.harvard.edu/abs/2021A&A...649A...1G} {649, A1}

\bibitem[\protect\citeauthoryear{{Gilmore} et~al.,}{{Gilmore}
  et~al.}{2012}]{gilmore12}
{Gilmore} G.,  et~al., 2012, The Messenger, \href
  {http://adsabs.harvard.edu/abs/2012Msngr.147...25G} {147, 25}

\bibitem[\protect\citeauthoryear{{Gilmore} et~al.,}{{Gilmore}
  et~al.}{2022}]{gilmore22}
{Gilmore} G.,  et~al., 2022, \mn@doi [\aap] {10.1051/0004-6361/202243134},
  \href {https://ui.adsabs.harvard.edu/abs/2022A&A...666A.120G} {666, A120}

\bibitem[\protect\citeauthoryear{{Goodwin} \& {Bastian}}{{Goodwin} \&
  {Bastian}}{2006}]{goodwin06}
{Goodwin} S.~P.,  {Bastian} N.,  2006, \mn@doi [\mnras]
  {10.1111/j.1365-2966.2006.11078.x}, \href
  {https://ui.adsabs.harvard.edu/abs/2006MNRAS.373..752G} {373, 752}

\bibitem[\protect\citeauthoryear{{Grudi{\'c}}, {Guszejnov}, {Hopkins}, {Offner}
   \& {Faucher-Gigu{\`e}re}}{{Grudi{\'c}} et~al.}{2021}]{grudic21}
{Grudi{\'c}} M.~Y.,  {Guszejnov} D.,  {Hopkins} P.~F.,  {Offner} S. S.~R.,
  {Faucher-Gigu{\`e}re} C.-A.,  2021, \mn@doi [\mnras]
  {10.1093/mnras/stab1347}, \href
  {https://ui.adsabs.harvard.edu/abs/2021MNRAS.506.2199G} {506, 2199}

\bibitem[\protect\citeauthoryear{{Guszejnov}, {Markey}, {Offner}, {Grudi{\'c}},
  {Faucher-Gigu{\`e}re}, {Rosen}  \& {Hopkins}}{{Guszejnov}
  et~al.}{2022}]{guszejnov22}
{Guszejnov} D.,  {Markey} C.,  {Offner} S. S.~R.,  {Grudi{\'c}} M.~Y.,
  {Faucher-Gigu{\`e}re} C.-A.,  {Rosen} A.~L.,   {Hopkins} P.~F.,  2022,
  \mn@doi [\mnras] {10.1093/mnras/stac1737}, \href
  {https://ui.adsabs.harvard.edu/abs/2022MNRAS.515..167G} {515, 167}

\bibitem[\protect\citeauthoryear{{Hennebelle} \& {Chabrier}}{{Hennebelle} \&
  {Chabrier}}{2011}]{2011ApJ...743L..29H}
{Hennebelle} P.,  {Chabrier} G.,  2011, \mn@doi [\apjl]
  {10.1088/2041-8205/743/2/L29}, \href
  {https://ui.adsabs.harvard.edu/abs/2011ApJ...743L..29H} {743, L29}

\bibitem[\protect\citeauthoryear{{Hills}}{{Hills}}{1980}]{hills80}
{Hills} J.~G.,  1980, \mn@doi [\apj] {10.1086/157703}, \href
  {http://adsabs.harvard.edu/abs/1980ApJ...235..986H} {235, 986}

\bibitem[\protect\citeauthoryear{{Holmberg} \& {Flynn}}{{Holmberg} \&
  {Flynn}}{2004}]{holmberg04}
{Holmberg} J.,  {Flynn} C.,  2004, \mn@doi [\mnras]
  {10.1111/j.1365-2966.2004.07931.x}, \href
  {https://ui.adsabs.harvard.edu/abs/2004MNRAS.352..440H} {352, 440}

\bibitem[\protect\citeauthoryear{{Inutsuka}, {Inoue}, {Iwasaki}  \&
  {Hosokawa}}{{Inutsuka} et~al.}{2015}]{2015A&A...580A..49I}
{Inutsuka} S.-i.,  {Inoue} T.,  {Iwasaki} K.,   {Hosokawa} T.,  2015, \mn@doi
  [\aap] {10.1051/0004-6361/201425584}, \href
  {https://ui.adsabs.harvard.edu/abs/2015A&A...580A..49I} {580, A49}

\bibitem[\protect\citeauthoryear{{Jackson}, {Deliyannis}  \&
  {Jeffries}}{{Jackson} et~al.}{2018}]{jackson18}
{Jackson} R.~J.,  {Deliyannis} C.~P.,   {Jeffries} R.~D.,  2018, \mn@doi
  [\mnras] {10.1093/mnras/sty374}, \href
  {http://adsabs.harvard.edu/abs/2018MNRAS.476.3245J} {476, 3245}

\bibitem[\protect\citeauthoryear{{Jeffries}, {Jackson}, {Sun}  \&
  {Deliyannis}}{{Jeffries} et~al.}{2021}]{jeffries21}
{Jeffries} R.~D.,  {Jackson} R.~J.,  {Sun} Q.,   {Deliyannis} C.~P.,  2021,
  \mn@doi [\mnras] {10.1093/mnras/staa3141}, \href
  {https://ui.adsabs.harvard.edu/abs/2021MNRAS.500.1158J} {500, 1158}

\bibitem[\protect\citeauthoryear{{Johnson} \& {Soderblom}}{{Johnson} \&
  {Soderblom}}{1987}]{johnson87}
{Johnson} D.~R.~H.,  {Soderblom} D.~R.,  1987, \mn@doi [\aj] {10.1086/114370},
  \href {http://adsabs.harvard.edu/abs/1987AJ.....93..864J} {93, 864}

\bibitem[\protect\citeauthoryear{{Karim} \& {Mamajek}}{{Karim} \&
  {Mamajek}}{2017}]{karim17}
{Karim} M.~T.,  {Mamajek} E.~E.,  2017, \mn@doi [\mnras]
  {10.1093/mnras/stw2772}, \href
  {https://ui.adsabs.harvard.edu/abs/2017MNRAS.465..472K} {465, 472}

\bibitem[\protect\citeauthoryear{{Kerr}, {Farias}, {Prato}, {Rector}, {Speagle}
   \& {Kraus}}{{Kerr} et~al.}{2025}]{kerr25}
{Kerr} R.,  {Farias} J.~P.,  {Prato} L.,  {Rector} T.~A.,  {Speagle} J.~S.,
  {Kraus} A.~L.,  2025, \mn@doi [arXiv e-prints] {10.48550/arXiv.2503.02002},
  \href {https://ui.adsabs.harvard.edu/abs/2025arXiv250302002K} {p.
  arXiv:2503.02002}

\bibitem[\protect\citeauthoryear{{Kos} et~al.,}{{Kos} et~al.}{2019}]{Kos19}
{Kos} J.,  et~al., 2019, \mn@doi [\aap] {10.1051/0004-6361/201834710}, \href
  {https://ui.adsabs.harvard.edu/abs/2019A&A...631A.166K} {631, A166}

\bibitem[\protect\citeauthoryear{{Kounkel} et~al.,}{{Kounkel}
  et~al.}{2018}]{kounkel18}
{Kounkel} M.,  et~al., 2018, \mn@doi [\aj] {10.3847/1538-3881/aad1f1}, \href
  {https://ui.adsabs.harvard.edu/abs/2018AJ....156...84K} {156, 84}

\bibitem[\protect\citeauthoryear{{Krause} et~al.,}{{Krause}
  et~al.}{2018}]{krause18}
{Krause} M.~G.~H.,  et~al., 2018, \mn@doi [\aap] {10.1051/0004-6361/201732416},
  \href {http://adsabs.harvard.edu/abs/2018A%26A...619A.120K} {619, A120}

\bibitem[\protect\citeauthoryear{{Kroupa}}{{Kroupa}}{2001}]{kroupa01}
{Kroupa} P.,  2001, \mn@doi [\mnras] {10.1046/j.1365-8711.2001.04022.x}, \href
  {https://ui.adsabs.harvard.edu/abs/2001MNRAS.322..231K} {322, 231}

\bibitem[\protect\citeauthoryear{{Kroupa}, {Aarseth}  \& {Hurley}}{{Kroupa}
  et~al.}{2001}]{kroupa01b}
{Kroupa} P.,  {Aarseth} S.,   {Hurley} J.,  2001, \mn@doi [\mnras]
  {10.1046/j.1365-8711.2001.04050.x}, \href
  {https://ui.adsabs.harvard.edu/abs/2001MNRAS.321..699K} {321, 699}

\bibitem[\protect\citeauthoryear{{Krumholz} \& {McKee}}{{Krumholz} \&
  {McKee}}{2005}]{2005ApJ...630..250K}
{Krumholz} M.~R.,  {McKee} C.~F.,  2005, \mn@doi [\apj] {10.1086/431734}, \href
  {https://ui.adsabs.harvard.edu/abs/2005ApJ...630..250K} {630, 250}

\bibitem[\protect\citeauthoryear{{Kuhn}, {Hillenbrand}, {Sills}, {Feigelson}
  \& {Getman}}{{Kuhn} et~al.}{2019}]{kuhn19}
{Kuhn} M.~A.,  {Hillenbrand} L.~A.,  {Sills} A.,  {Feigelson} E.~D.,   {Getman}
  K.~V.,  2019, \mn@doi [\apj] {10.3847/1538-4357/aaef8c}, \href
  {https://ui.adsabs.harvard.edu/abs/2019ApJ...870...32K} {870, 32}

\bibitem[\protect\citeauthoryear{{Kurucz}}{{Kurucz}}{1992}]{Kurucz1992a}
{Kurucz} R.~L.,  1992, in {Barbuy} B.,  {Renzini} A.,  eds,  IAU Symposium Vol.
  149, The Stellar Populations of Galaxies. p.~225

\bibitem[\protect\citeauthoryear{{Lada} \& {Lada}}{{Lada} \&
  {Lada}}{2003}]{lada03}
{Lada} C.~J.,  {Lada} E.~A.,  2003, \mn@doi [araa]
  {10.1146/annurev.astro.41.011802.094844}, \href
  {http://adsabs.harvard.edu/abs/2003ARA%26A..41...57L} {41, 57}

\bibitem[\protect\citeauthoryear{{Lewis} et~al.,}{{Lewis} et~al.}{2002}]{2dF}
{Lewis} I.~J.,  et~al., 2002, \mn@doi [\mnras]
  {10.1046/j.1365-8711.2002.05333.x}, \href
  {https://ui.adsabs.harvard.edu/abs/2002MNRAS.333..279L} {333, 279}

\bibitem[\protect\citeauthoryear{{Li}, {Tan}, {Christie}, {Bisbas}  \&
  {Wu}}{{Li} et~al.}{2018}]{2018PASJ...70S..56L}
{Li} Q.,  {Tan} J.~C.,  {Christie} D.,  {Bisbas} T.~G.,   {Wu} B.,  2018,
  \mn@doi [\pasj] {10.1093/pasj/psx136}, \href
  {https://ui.adsabs.harvard.edu/abs/2018PASJ...70S..56L} {70, S56}

\bibitem[\protect\citeauthoryear{{Lindegren} et~al.,}{{Lindegren}
  et~al.}{2021}]{lindegren21}
{Lindegren} L.,  et~al., 2021, \mn@doi [\aap] {10.1051/0004-6361/202039653},
  \href {https://ui.adsabs.harvard.edu/abs/2021A&A...649A...4L} {649, A4}

\bibitem[\protect\citeauthoryear{{Loren}}{{Loren}}{1989}]{loren89}
{Loren} R.~B.,  1989, \mn@doi [\apj] {10.1086/167244}, \href
  {https://ui.adsabs.harvard.edu/abs/1989ApJ...338..902L} {338, 902}

\bibitem[\protect\citeauthoryear{{Luhman}}{{Luhman}}{2022}]{luhman22}
{Luhman} K.~L.,  2022, \mn@doi [\aj] {10.3847/1538-3881/ac35e2}, \href
  {https://ui.adsabs.harvard.edu/abs/2022AJ....163...24L} {163, 24}

\bibitem[\protect\citeauthoryear{{Luhman} \& {Esplin}}{{Luhman} \&
  {Esplin}}{2020}]{luhman20}
{Luhman} K.~L.,  {Esplin} T.~L.,  2020, \mn@doi [\aj]
  {10.3847/1538-3881/ab9599}, \href
  {https://ui.adsabs.harvard.edu/abs/2020AJ....160...44L} {160, 44}

\bibitem[\protect\citeauthoryear{{Majewski} et~al.,}{{Majewski}
  et~al.}{2017}]{apogee}
{Majewski} S.~R.,  et~al., 2017, \mn@doi [\aj] {10.3847/1538-3881/aa784d},
  \href {https://ui.adsabs.harvard.edu/abs/2017AJ....154...94M} {154, 94}

\bibitem[\protect\citeauthoryear{{Mapelli}}{{Mapelli}}{2017}]{mapelli17}
{Mapelli} M.,  2017, \mn@doi [\mnras] {10.1093/mnras/stx304}, \href
  {https://ui.adsabs.harvard.edu/abs/2017MNRAS.467.3255M} {467, 3255}

\bibitem[\protect\citeauthoryear{{Marigo} et~al.,}{{Marigo}
  et~al.}{2017}]{marigo17}
{Marigo} P.,  et~al., 2017, \mn@doi [\apj] {10.3847/1538-4357/835/1/77}, \href
  {https://ui.adsabs.harvard.edu/abs/2017ApJ...835...77M} {835, 77}

\bibitem[\protect\citeauthoryear{{Marton} et~al.,}{{Marton}
  et~al.}{2023}]{marton23}
{Marton} G.,  et~al., 2023, \mn@doi [\aap] {10.1051/0004-6361/202244101}, \href
  {https://ui.adsabs.harvard.edu/abs/2023A&A...674A..21M} {674, A21}

\bibitem[\protect\citeauthoryear{{Maschberger}}{{Maschberger}}{2013}]{maschberger13}
{Maschberger} T.,  2013, \mn@doi [\mnras] {10.1093/mnras/sts479}, \href
  {http://adsabs.harvard.edu/abs/2013MNRAS.429.1725M} {429, 1725}

\bibitem[\protect\citeauthoryear{{Miao}, {White}, {Nelson}, {Thompson}  \&
  {Morgan}}{{Miao} et~al.}{2006}]{miao06}
{Miao} J.,  {White} G.~J.,  {Nelson} R.,  {Thompson} M.,   {Morgan} L.,  2006,
  \mn@doi [\mnras] {10.1111/j.1365-2966.2006.10260.x}, \href
  {https://ui.adsabs.harvard.edu/abs/2006MNRAS.369..143M} {369, 143}

\bibitem[\protect\citeauthoryear{{Miret-Roig} et~al.,}{{Miret-Roig}
  et~al.}{2022a}]{miret-roig22a}
{Miret-Roig} N.,  et~al., 2022a, \mn@doi [Nature Astronomy]
  {10.1038/s41550-021-01513-x}, \href
  {https://ui.adsabs.harvard.edu/abs/2022NatAs...6...89M} {6, 89}

\bibitem[\protect\citeauthoryear{{Miret-Roig}, {Galli}, {Olivares}, {Bouy},
  {Alves}  \& {Barrado}}{{Miret-Roig} et~al.}{2022b}]{miret-roig22}
{Miret-Roig} N.,  {Galli} P.~A.~B.,  {Olivares} J.,  {Bouy} H.,  {Alves} J.,
  {Barrado} D.,  2022b, \mn@doi [\aap] {10.1051/0004-6361/202244709}, \href
  {https://ui.adsabs.harvard.edu/abs/2022A&A...667A.163M} {667, A163}

\bibitem[\protect\citeauthoryear{{Miret-Roig}, {Alves}, {Barrado}, {Burkert},
  {Ratzenb{\"o}ck}  \& {Konietzka}}{{Miret-Roig} et~al.}{2024}]{miret-roig24}
{Miret-Roig} N.,  {Alves} J.,  {Barrado} D.,  {Burkert} A.,  {Ratzenb{\"o}ck}
  S.,   {Konietzka} R.,  2024, \mn@doi [Nature Astronomy]
  {10.1038/s41550-023-02132-4}, \href
  {https://ui.adsabs.harvard.edu/abs/2024NatAs...8..216M} {8, 216}

\bibitem[\protect\citeauthoryear{{Padoan} \& {Nordlund}}{{Padoan} \&
  {Nordlund}}{2011}]{2011ApJ...730...40P}
{Padoan} P.,  {Nordlund} {\r{A}}.,  2011, \mn@doi [\apj]
  {10.1088/0004-637X/730/1/40}, \href
  {https://ui.adsabs.harvard.edu/abs/2011ApJ...730...40P} {730, 40}

\bibitem[\protect\citeauthoryear{{Palla}, {Randich}, {Flaccomio}  \&
  {Pallavicini}}{{Palla} et~al.}{2005}]{palla05}
{Palla} F.,  {Randich} S.,  {Flaccomio} E.,   {Pallavicini} R.,  2005, \mn@doi
  [\apjl] {10.1086/431668}, \href
  {https://ui.adsabs.harvard.edu/abs/2005ApJ...626L..49P} {626, L49}

\bibitem[\protect\citeauthoryear{{Pang}, {Yu}, {Tang}, {Hong}, {Yuan},
  {Pasquato}  \& {Kouwenhoven}}{{Pang} et~al.}{2021}]{pang21}
{Pang} X.,  {Yu} Z.,  {Tang} S.-Y.,  {Hong} J.,  {Yuan} Z.,  {Pasquato} M.,
  {Kouwenhoven} M.~B.~N.,  2021, \mn@doi [\apj] {10.3847/1538-4357/ac2838},
  \href {https://ui.adsabs.harvard.edu/abs/2021ApJ...923...20P} {923, 20}

\bibitem[\protect\citeauthoryear{{Parker}, {Goodwin}, {Kroupa}  \&
  {Kouwenhoven}}{{Parker} et~al.}{2009}]{parker09}
{Parker} R.~J.,  {Goodwin} S.~P.,  {Kroupa} P.,   {Kouwenhoven} M.~B.~N.,
  2009, \mn@doi [\mnras] {10.1111/j.1365-2966.2009.15032.x}, \href
  {https://ui.adsabs.harvard.edu/abs/2009MNRAS.397.1577P} {397, 1577}

\bibitem[\protect\citeauthoryear{{Pecaut} \& {Mamajek}}{{Pecaut} \&
  {Mamajek}}{2016}]{pecaut16}
{Pecaut} M.~J.,  {Mamajek} E.~E.,  2016, \mn@doi [\mnras]
  {10.1093/mnras/stw1300}, \href
  {https://ui.adsabs.harvard.edu/abs/2016MNRAS.461..794P} {461, 794}

\bibitem[\protect\citeauthoryear{Posch, Miret-Roig, Alves, Ratzenböck,
  Großschedl, Meingast, Zucker  \& Burkert}{Posch et~al.}{2023}]{posch23}
Posch L.,  Miret-Roig N.,  Alves J.,  Ratzenböck S.,  Großschedl J.,
  Meingast S.,  Zucker C.,   Burkert A.,  2023, \mn@doi [Astronomy &amp;
  Astrophysics] {10.1051/0004-6361/202347186}, 679, L10

\bibitem[\protect\citeauthoryear{{Posch}, {Alves}, {Miret-Roig},
  {Ratzenb{\"o}ck}, {Gro{\ss}schedl}, {Meingast}, {Swiggum}  \&
  {Konietzka}}{{Posch} et~al.}{2025}]{posch24}
{Posch} L.,  {Alves} J.,  {Miret-Roig} N.,  {Ratzenb{\"o}ck} S.,
  {Gro{\ss}schedl} J.,  {Meingast} S.,  {Swiggum} C.,   {Konietzka} R.,  2025,
  \mn@doi [\aap] {10.1051/0004-6361/202451312}, \href
  {https://ui.adsabs.harvard.edu/abs/2025A&A...693A.175P} {693, A175}

\bibitem[\protect\citeauthoryear{{Preibisch} \& {Mamajek}}{{Preibisch} \&
  {Mamajek}}{2008}]{preibisch08}
{Preibisch} T.,  {Mamajek} E.,  2008, in {Reipurth} B.,  ed., , Vol.~5,
  Handbook of Star Forming Regions, Volume II.
p.~235, \mn@doi{10.48550/arXiv.0809.0407}

\bibitem[\protect\citeauthoryear{{Quintana} \& {Wright}}{{Quintana} \&
  {Wright}}{2022}]{quintana22}
{Quintana} A.~L.,  {Wright} N.~J.,  2022, \mn@doi [\mnras]
  {10.1093/mnras/stac1526}, \href
  {https://ui.adsabs.harvard.edu/abs/2022MNRAS.515..687Q} {515, 687}

\bibitem[\protect\citeauthoryear{{Raghavan} et~al.,}{{Raghavan}
  et~al.}{2010}]{raghavan10}
{Raghavan} D.,  et~al., 2010, \mn@doi [\apjs] {10.1088/0067-0049/190/1/1},
  \href {https://ui.adsabs.harvard.edu/abs/2010ApJS..190....1R} {190, 1}

\bibitem[\protect\citeauthoryear{{Randich} et~al.,}{{Randich}
  et~al.}{2022}]{randich22}
{Randich} S.,  et~al., 2022, \mn@doi [\aap] {10.1051/0004-6361/202243141},
  \href {https://ui.adsabs.harvard.edu/abs/2022A&A...666A.121R} {666, A121}

\bibitem[\protect\citeauthoryear{{Ratzenb{\"o}ck}, {Gro{\ss}schedl},
  {M{\"o}ller}, {Alves}, {Bomze}  \& {Meingast}}{{Ratzenb{\"o}ck}
  et~al.}{2023a}]{ratzenbock23a}
{Ratzenb{\"o}ck} S.,  {Gro{\ss}schedl} J.~E.,  {M{\"o}ller} T.,  {Alves} J.,
  {Bomze} I.,   {Meingast} S.,  2023a, \mn@doi [\aap]
  {10.1051/0004-6361/202243690}, \href
  {https://ui.adsabs.harvard.edu/abs/2023A&A...677A..59R} {677, A59}

\bibitem[\protect\citeauthoryear{{Ratzenb{\"o}ck} et~al.,}{{Ratzenb{\"o}ck}
  et~al.}{2023b}]{ratzenbock23}
{Ratzenb{\"o}ck} S.,  et~al., 2023b, \mn@doi [\aap]
  {10.1051/0004-6361/202346901}, \href
  {https://ui.adsabs.harvard.edu/abs/2023A&A...678A..71R} {678, A71}

\bibitem[\protect\citeauthoryear{{Rigliaco} et~al.,}{{Rigliaco}
  et~al.}{2016}]{rigliaco16}
{Rigliaco} E.,  et~al., 2016, \mn@doi [\aap] {10.1051/0004-6361/201527253},
  \href {https://ui.adsabs.harvard.edu/abs/2016A&A...588A.123R} {588, A123}

\bibitem[\protect\citeauthoryear{{Sch{\"o}nrich}, {Binney}  \&
  {Dehnen}}{{Sch{\"o}nrich} et~al.}{2010}]{schonrich10}
{Sch{\"o}nrich} R.,  {Binney} J.,   {Dehnen} W.,  2010, \mn@doi [\mnras]
  {10.1111/j.1365-2966.2010.16253.x}, \href
  {https://ui.adsabs.harvard.edu/abs/2010MNRAS.403.1829S} {403, 1829}

\bibitem[\protect\citeauthoryear{{Scoville}, {Sanders}  \&
  {Clemens}}{{Scoville} et~al.}{1986}]{1986ApJ...310L..77S}
{Scoville} N.~Z.,  {Sanders} D.~B.,   {Clemens} D.~P.,  1986, \mn@doi [\apjl]
  {10.1086/184785}, \href
  {https://ui.adsabs.harvard.edu/abs/1986ApJ...310L..77S} {310, L77}

\bibitem[\protect\citeauthoryear{{Sheinis} et~al.,}{{Sheinis}
  et~al.}{2015}]{HERMES}
{Sheinis} A.,  et~al., 2015, \mn@doi [Journal of Astronomical Telescopes,
  Instruments, and Systems] {10.1117/1.JATIS.1.3.035002}, \href
  {https://ui.adsabs.harvard.edu/abs/2015JATIS...1c5002S} {1, 035002}

\bibitem[\protect\citeauthoryear{{Sills}, {Rieder}, {Scora}, {McCloskey}  \&
  {Jaffa}}{{Sills} et~al.}{2018}]{sills18}
{Sills} A.,  {Rieder} S.,  {Scora} J.,  {McCloskey} J.,   {Jaffa} S.,  2018,
  \mn@doi [\mnras] {10.1093/mnras/sty681}, \href
  {https://ui.adsabs.harvard.edu/abs/2018MNRAS.477.1903S} {477, 1903}

\bibitem[\protect\citeauthoryear{{Sneden}, {Bean}, {Ivans}, {Lucatello}  \&
  {Sobeck}}{{Sneden} et~al.}{2012}]{Sneden2012a}
{Sneden} C.,  {Bean} J.,  {Ivans} I.,  {Lucatello} S.,   {Sobeck} J.,  2012,
  \href {https://ui.adsabs.harvard.edu/abs/2012ascl.soft02009S} {p.
  ascl:1202.009}

\bibitem[\protect\citeauthoryear{{Soderblom}}{{Soderblom}}{2010}]{soderblom10}
{Soderblom} D.~R.,  2010, \mn@doi [araa] {10.1146/annurev-astro-081309-130806},
  \href {http://adsabs.harvard.edu/abs/2010ARA%26A..48..581S} {48, 581}

\bibitem[\protect\citeauthoryear{{Squicciarini}, {Gratton}, {Bonavita}  \&
  {Mesa}}{{Squicciarini} et~al.}{2021}]{squicciarini21}
{Squicciarini} V.,  {Gratton} R.,  {Bonavita} M.,   {Mesa} D.,  2021, \mn@doi
  [\mnras] {10.1093/mnras/stab2079}, \href
  {https://ui.adsabs.harvard.edu/abs/2021MNRAS.507.1381S} {507, 1381}

\bibitem[\protect\citeauthoryear{{Suwannajak}, {Tan}  \& {Leroy}}{{Suwannajak}
  et~al.}{2014}]{2014ApJ...787...68S}
{Suwannajak} C.,  {Tan} J.~C.,   {Leroy} A.~K.,  2014, \mn@doi [\apj]
  {10.1088/0004-637X/787/1/68}, \href
  {https://ui.adsabs.harvard.edu/abs/2014ApJ...787...68S} {787, 68}

\bibitem[\protect\citeauthoryear{{Tan}}{{Tan}}{2000}]{tan00}
{Tan} J.~C.,  2000, \mn@doi [\apj] {10.1086/308905}, \href
  {https://ui.adsabs.harvard.edu/abs/2000ApJ...536..173T} {536, 173}

\bibitem[\protect\citeauthoryear{{Tan}}{{Tan}}{2010}]{2010ApJ...710L..88T}
{Tan} J.~C.,  2010, \mn@doi [\apjl] {10.1088/2041-8205/710/1/L88}, \href
  {https://ui.adsabs.harvard.edu/abs/2010ApJ...710L..88T} {710, L88}

\bibitem[\protect\citeauthoryear{{Tan}, {Shaske}  \& {Van Loo}}{{Tan}
  et~al.}{2013}]{2013IAUS..292...19T}
{Tan} J.~C.,  {Shaske} S.~N.,   {Van Loo} S.,  2013, in {Wong} T.,  {Ott} J.,
  eds,  IAU Symposium Vol. 292, Molecular Gas, Dust, and Star Formation in
  Galaxies. pp 19--28 (\mn@eprint {arXiv} {1211.0198}),
  \mn@doi{10.1017/S1743921313000173}

\bibitem[\protect\citeauthoryear{{Tasker} \& {Tan}}{{Tasker} \&
  {Tan}}{2009}]{2009ApJ...700..358T}
{Tasker} E.~J.,  {Tan} J.~C.,  2009, \mn@doi [\apj]
  {10.1088/0004-637X/700/1/358}, \href
  {https://ui.adsabs.harvard.edu/abs/2009ApJ...700..358T} {700, 358}

\bibitem[\protect\citeauthoryear{{Taylor}}{{Taylor}}{2005}]{tayl05}
{Taylor} M.~B.,  2005, in {Shopbell} P.,  {Britton} M.,   {Ebert} R.,  eds,
  Astronomical Society of the Pacific Conference Series Vol. 347, Astronomical
  Data Analysis Software and Systems XIV. p.~29

\bibitem[\protect\citeauthoryear{{Tsantaki} et~al.,}{{Tsantaki}
  et~al.}{2022}]{tsantaki22}
{Tsantaki} M.,  et~al., 2022, \mn@doi [\aap] {10.1051/0004-6361/202141702},
  \href {https://ui.adsabs.harvard.edu/abs/2022A&A...659A..95T} {659, A95}

\bibitem[\protect\citeauthoryear{{Tutukov}}{{Tutukov}}{1978}]{tutukov78}
{Tutukov} A.~V.,  1978, \aap, \href
  {http://adsabs.harvard.edu/abs/1978A%26A....70...57T} {70, 57}

\bibitem[\protect\citeauthoryear{{Wright}}{{Wright}}{2020}]{wright20}
{Wright} N.~J.,  2020, \mn@doi [\nar] {10.1016/j.newar.2020.101549}, \href
  {https://ui.adsabs.harvard.edu/abs/2020NewAR..9001549W} {90, 101549}

\bibitem[\protect\citeauthoryear{{Wright} \& {Mamajek}}{{Wright} \&
  {Mamajek}}{2018}]{wright18}
{Wright} N.~J.,  {Mamajek} E.~E.,  2018, \mn@doi [\mnras]
  {10.1093/mnras/sty207}, \href
  {http://adsabs.harvard.edu/abs/2018MNRAS.476..381W} {476, 381}

\bibitem[\protect\citeauthoryear{{Wright}, {Parker}, {Goodwin}  \&
  {Drake}}{{Wright} et~al.}{2014}]{wright14}
{Wright} N.~J.,  {Parker} R.~J.,  {Goodwin} S.~P.,   {Drake} J.~J.,  2014,
  \mn@doi [\mnras] {10.1093/mnras/stt2232}, \href
  {http://adsabs.harvard.edu/abs/2014MNRAS.438..639W} {438, 639}

\bibitem[\protect\citeauthoryear{{Wright}, {Bouy}, {Drew}, {Sarro}, {Bertin},
  {Cuillandre}  \& {Barrado}}{{Wright} et~al.}{2016}]{wright16}
{Wright} N.~J.,  {Bouy} H.,  {Drew} J.~E.,  {Sarro} L.~M.,  {Bertin} E.,
  {Cuillandre} J.-C.,   {Barrado} D.,  2016, \mn@doi [\mnras]
  {10.1093/mnras/stw1148}, \href
  {http://adsabs.harvard.edu/abs/2016MNRAS.460.2593W} {460, 2593}

\bibitem[\protect\citeauthoryear{{Wright}, {Kounkel}, {Zari}, {Goodwin}  \&
  {Jeffries}}{{Wright} et~al.}{2023}]{wright23}
{Wright} N.~J.,  {Kounkel} M.,  {Zari} E.,  {Goodwin} S.,   {Jeffries} R.~D.,
  2023, in {Inutsuka} S.,  {Aikawa} Y.,  {Muto} T.,  {Tomida} K.,   {Tamura}
  M.,  eds,  Astronomical Society of the Pacific Conference Series Vol. 534,
  Protostars and Planets VII. p.~129

\bibitem[\protect\citeauthoryear{{Wright} et~al.,}{{Wright}
  et~al.}{2024}]{wright24}
{Wright} N.~J.,  et~al., 2024, \mn@doi [\mnras] {10.1093/mnras/stae1806}, \href
  {https://ui.adsabs.harvard.edu/abs/2024MNRAS.533..705W} {533, 705}

\bibitem[\protect\citeauthoryear{{Wu}, {Tan}, {Christie}, {Nakamura}, {Van Loo}
   \& {Collins}}{{Wu} et~al.}{2017}]{wu17}
{Wu} B.,  {Tan} J.~C.,  {Christie} D.,  {Nakamura} F.,  {Van Loo} S.,
  {Collins} D.,  2017, \mn@doi [\apj] {10.3847/1538-4357/aa6ffa}, \href
  {https://ui.adsabs.harvard.edu/abs/2017ApJ...841...88W} {841, 88}

\bibitem[\protect\citeauthoryear{{Wu}, {Tan}, {Nakamura}, {Christie}  \&
  {Li}}{{Wu} et~al.}{2018}]{2018PASJ...70S..57W}
{Wu} B.,  {Tan} J.~C.,  {Nakamura} F.,  {Christie} D.,   {Li} Q.,  2018,
  \mn@doi [\pasj] {10.1093/pasj/psx140}, \href
  {https://ui.adsabs.harvard.edu/abs/2018PASJ...70S..57W} {70, S57}

\bibitem[\protect\citeauthoryear{{Wu}, {Tan}, {Christie}  \& {Nakamura}}{{Wu}
  et~al.}{2020}]{2020ApJ...891..168W}
{Wu} B.,  {Tan} J.~C.,  {Christie} D.,   {Nakamura} F.,  2020, \mn@doi [\apj]
  {10.3847/1538-4357/ab77b5}, \href
  {https://ui.adsabs.harvard.edu/abs/2020ApJ...891..168W} {891, 168}

\bibitem[\protect\citeauthoryear{{Zari}, {Brown}  \& {de Zeeuw}}{{Zari}
  et~al.}{2019}]{zari19}
{Zari} E.,  {Brown} A.~G.~A.,   {de Zeeuw} P.~T.,  2019, \mn@doi [\aap]
  {10.1051/0004-6361/201935781}, \href
  {https://ui.adsabs.harvard.edu/abs/2019A&A...628A.123Z} {628, A123}

\bibitem[\protect\citeauthoryear{Zwart, McMillan  \& Gieles}{Zwart
  et~al.}{2010}]{portegieszwart10}
Zwart S.~P.,  McMillan S.,   Gieles M.,  2010, arXiv preprint arXiv:1002.1961

\bibitem[\protect\citeauthoryear{{Zwitter} et~al.,}{{Zwitter}
  et~al.}{2018}]{zwitter18}
{Zwitter} T.,  et~al., 2018, \mn@doi [\mnras] {10.1093/mnras/sty2293}, \href
  {https://ui.adsabs.harvard.edu/abs/2018MNRAS.481..645Z} {481, 645}

\bibitem[\protect\citeauthoryear{{de Zeeuw}, {Hoogerwerf}, {de Bruijne},
  {Brown}  \& {Blaauw}}{{de Zeeuw} et~al.}{1999}]{dezeeuw99}
{de Zeeuw} P.~T.,  {Hoogerwerf} R.,  {de Bruijne} J.~H.~J.,  {Brown} A.~G.~A.,
   {Blaauw} A.,  1999, \mn@doi [\aj] {10.1086/300682}, \href
  {http://adsabs.harvard.edu/abs/1999AJ....117..354D} {117, 354}

\makeatother
\end{thebibliography}


\bsp

\end{document}